\documentclass[journal, romanappendicesim]{IEEEtran}
\usepackage{amsmath,amsfonts,amssymb,amsthm}
\usepackage{array}
\usepackage{acronym}
\usepackage[caption=false,font=normalsize,labelfont=sf,textfont=sf]{subfig}
\usepackage{todonotes}
\usepackage{textcomp}
\usepackage{stfloats}
\usepackage{url}
\usepackage{verbatim}
\usepackage{graphicx}
\usepackage{cite}
\usepackage{xcolor}
\usepackage{xr}
\usepackage[ruled,vlined]{algorithm2e}

\input{Acronym}
\usepackage{listings}
\usepackage{multirow}
\usepackage{lscape}
\usepackage{hyperref}
\usepackage{circledsteps}
\newcommand\myCircled[2][]{\ifmmode
\Circled[fill color=black,inner color=white,#1]{\textbf{#2}}
\else
\Circled[fill color=black,inner color=white,#1]{\textbf#2}
\fi
}
\definecolor{codegreen}{rgb}{0,0.6,0}
\definecolor{codegray}{rgb}{0.5,0.5,0.5}
\definecolor{codepurple}{rgb}{0.58,0,0.82}
\definecolor{backcolour}{rgb}{0.95,0.95,0.92}
\lstdefinestyle{mystyle}{
    backgroundcolor=\color{backcolour},   
    commentstyle=\color{codegreen},
    keywordstyle=\color{black},
    stringstyle=\color{codepurple},
    basicstyle=\ttfamily\footnotesize,
    breakatwhitespace=false,         
    breaklines=true,                 
    captionpos=b,                    
    keepspaces=true,                             
    showspaces=false,                
    showstringspaces=false,
    showtabs=false,                  
    tabsize=2
}

\begin{document}

\title{\textsc{Enwar 3.0}: An Agentic Multi-Modal LLM Orchestrator for Situation-Aware Beamforming, Blockage Prediction, and Handover Management}

\author{~Ahmad M. Nazar,~\IEEEmembership{Member, IEEE},~Abdulkadir~Celik,~\IEEEmembership{Senior Member, IEEE},\\Asmaa~Abdallah,~\IEEEmembership{Senior Member, IEEE}, ~Mohamed Y. Selim,~\IEEEmembership{Senior Member, IEEE}, \\Daji~Qiao,~\IEEEmembership{Senior Member, IEEE}, and Ahmed~M.~Eltawil,~\IEEEmembership{Senior Member, IEEE}

\thanks{A. M. Nazar is with the Department of Electrical and Computer Engineering, Iowa State University (ISU), Ames, IA, 50014, USA, and with Gladiolus Technological Institute, USA}

\thanks{M. Y. Selim, and D. Qiao are with the Department of Electrical and Computer Engineering, ISU, Ames, IA, 50014, USA.}

\thanks{A. Celik, is with the School of Electronics and Computer Science, University of Southampton, Southampton SO17 1BJ, U.K.}

\thanks{A. Abdallah, and A. M. Eltawil are with Computer, Electrical, and Mathematical Sciences \& Engineering (CEMSE) Division at King Abdullah University of Science and Technology (KAUST), Thuwal, 23955 KSA. }

\thanks{A conference version of this work is submitted to ASILOMAR Signal Processing Conference'25 \cite{asilomar2025}.}
}

\maketitle
\vspace{-0.4cm}
\begin{abstract}
    Maintaining robust millimeter-wave (mmWave) connectivity in vehicular networks requires real-time adaptation to environmental dynamics, sensor degradation, and link variability. This paper presents \textsc{Enwar 3.0}, an environment-aware reasoning framework that unifies multi-modal sensing, agentic large language models (LLMs), and context-driven model selection for predictive beamforming, blockage detection, and handover management. Building upon prior iterations of \textsc{Enwar}, the proposed architecture integrates a classifier-driven assessment of sensor health with a primed LLM that orchestrates multiple specialized agents through structured, task-aware prompting. A novel synthetic degradation pipeline enables the training of a sensor degradation classifier that detects real-time impairments across camera, radar, LiDAR, and GPS inputs, achieving over 99\% accuracy. The LLM, trained via chain-of-thought (CoT) priming and human-in-the-loop feedback, coordinates agent calls for beam selection, blockage forecasting, and environment perception while dynamically loading sensor-specific models based on environmental context. Extensive evaluations across 15 sensor combinations demonstrate that \textsc{Enwar 3.0} delivers state-of-the-art performance in both predictive accuracy and interpretability, with beam selection accuracy exceeding 88\%, blockage F1-scores surpassing 98\%, and reasoning correctness reaching 87\% on complex decision prompts. This work establishes a scalable foundation for LLM-integrated wireless systems that reason, perceive, and adapt in real-time.
\end{abstract}

\section{Introduction}
\Ac{mmWave} communication enables high-throughput, directional links for next-generation wireless systems, but remains highly sensitive to environmental dynamics such as blockage, sensor degradation, and mobility-induced channel variability. Maintaining reliable and low-latency \ac{I2V} connectivity, therefore, requires joint perception and adaptive decision-making under real-time constraints \cite{celik2024genai}.

Traditional network control mechanisms rely on rule-based heuristics, static beam tracking, or loosely coupled perception modules. Such approaches do not scale well to multi-modal sensing environments and cannot adapt decisions based on sensor reliability, environmental context, and historical system behavior \cite{zou2023wireless}. As \acp{RSU} integrate camera, LiDAR, radar, and GPS inputs, practicial system orchestration demands unified reasoning across heterogeneous modalities.

To address these challenges, we introduce \textbf{\textsc{Enwar 3.0}}, the latest evolution in the \underline{\textsc{EN}}vironment-a\underline{\textsc{WAR}}e, multi-modal \ac{LLM} framework series. \textsc{Enwar 3.0} extends \textsc{Enwar 2.0} \cite{nazar2025enwar2} to support system orchestration, fine-grained sensor reasoning, multi-agent coordination, and long-term memory management. \textsc{Enwar 3.0} replaces static heuristics with dynamic agent invocation guided by a \ac{DRL}-trained orchestrator and incorporates memory-assisted reasoning for temporal continuity in decision-making. Unlike conventional fusion architectures that couple modalities at the feature level, \textsc{Enwar 3.0} integrates perception, policy selection, and reasoning into a hierarchical control framework. The system combines multi-modal predictors, degradation-aware routing, and structured \ac{LLM}-based coordination to generate interpretable decisions while satisfying real-time operational constraints in \ac{I2V} networks.


\subsection{Main Contributions}
We present a framework that extends prior \textsc{Enwar} capabilities toward closed-loop, degradation-aware network control supplemented by \ac{XAI}. {
Building upon the perception-centric pipeline of \textsc{Enwar 1.0}~\cite{nazar2024enwar} and the \ac{RAG}-based environment perception and beam prediction introduced in \textsc{Enwar 2.0}~\cite{nazar2025enwar2}, \textsc{Enwar 3.0} transitions the framework from perception-assisted decision support to policy-driven, degradation-aware orchestration across beam prediction, blockage detection, and handover control within a unified real-time loop}. 

{
While \textsc{Enwar 2.0} demonstrated strong beam prediction performance (up to 90.0\% Top-3 accuracy) and situation-aware interpretation correctness (up to 89.7\%) under a 100ms sampling cycle, its design primarily emphasized reasoning-assisted beam tracking without enforcing a bounded, closed-loop control interval. In contrast, \textsc{Enwar 3.0} addresses a broader systems objective: unified beam, blockage, and handover orchestration under degradation-aware routing with strict real-time guarantees. Operating at a 300ms control interval, \textsc{Enwar 3.0} maintains a worst-case end-to-end control-path latency of 289.7ms, ensuring execution within the sampling window without inter-window accumulation. Importantly, this bounded-latency guarantee is achieved while preserving strong beam prediction performance (up to 88.5\% Top-3 accuracy) and extending the framework toward deployment-ready, real-time policy-driven wireless control under sensing uncertainty.

The environment perception and beam prediction agents are adopted from \textsc{Enwar 2.0} \cite{nazar2025enwar2, BPEncoBPPIMRC2025}, and refined with enhanced preprocessing, expanded temporal context windows, and tighter integration within a hierarchical orchestration layer. Additionally, \textsc{Enwar 3.0} integrates a blockage prediction component previously introduced in the conference version~\cite{asilomar2025}. We emphasize that~\cite{asilomar2025} documents only the standalone blockage prediction model; it does not include degradation-aware policy learning, memory-conditioned agent selection, \ac{DRL}-based routing, or unified beam--blockage--handover control. These capabilities are introduced in \textsc{Enwar 3.0}. To further clarify the framework's evolution, Table~\ref{tab: upgrades} summarizes the progression from retrieval-based reasoning (\textsc{Enwar 1.0}), to multi-modal beam tracking with \ac{RAG}-based explainability (\textsc{Enwar 2.0}), to degradation-aware, latency-bounded orchestration with unified control (\textsc{Enwar 3.0}).

As such, the main contributions are summarized as follows: 
\begin{itemize}
\item[\bf \checkmark] \textbf{Hierarchical Agent Orchestration for Wireless Control:} We reformulate multi-modal wireless control as a hierarchical agent orchestration problem, and propose a modular \ac{LLM}-driven reasoning architecture that employs \ac{CoT} priming with \ac{RLHF} and structured prompts to coordinate beam and blockage prediction, and handover decisions within a unified inference loop.

\item[\bf \checkmark] \textbf{Degradation-Aware Control Layer:}  We introduce a synthetic sensor degradation-aware pipeline and a multi-output environment status classifier (99.1\% detection accuracy) that enables dynamic exclusion of unreliable modalities during runtime. This replaces static fusion strategies with reliability-aware sensing.


\item[\bf \checkmark] \textbf{\ac{DRL}-Based Adaptive Model Routing:} 
We design a degradation-aware \ac{DRL} policy that dynamically selects among 15 pretrained modality combinations per agent based on sensor status and historical memory context for adaptive routing rather than fixed model invocation.
\item[\bf \checkmark]  \textbf{Persistence-Aware Blockage and Handover Control:} We integrate blockage duration tracking and historical long-term memory context into a policy-triggered handover mechanism, forming a closed-loop beam--blockage--handover control pipeline that was absent in prior \textsc{Enwar} iterations.

\item[\bf \checkmark] \textbf{Real-Time, Bounded-Latency Operation:}
Within a 300ms sampling interval, the full system maintains worst-case end-to-end latency (289.7ms), ensuring no temporal accumulation and confirming compatibility with real-time situation-aware \ac{I2V} beamforming and reasoning.
\end{itemize}

Across multiple \ac{LLM} architectures ranging from 3B to 70B parameters, \textsc{Enwar 3.0} achieves up to 89.3\% reasoning correctness under full modality inclusion within strict latency bounds. The beam and blockage agents achieve over 88\% Top-3 accuracy and 98\% F1-score, respectively, in ideal conditions, while gracefully degrading under partial sensor failures. These results demonstrate that \textsc{Enwar 3.0} advances beyond reasoning-enhanced prediction to deliver adaptive, policy-driven wireless control under real-world sensing uncertainty. }


\begin{table}[]
    \centering
    \caption{A summary of differences between each iteration of \textsc{Enwar}}
    \begin{tabular}{|c|c|c|c|}
    \hline
         \textbf{Feature} & \textsc{Enwar 1.0} \cite{nazar2024enwar} & \textsc{Enwar 2.0} \cite{nazar2025enwar2} & \textsc{Enwar 3.0} \\ \hline
         \begin{tabular}[c]{@{}c@{}}AI Agent\\ Orchestration\end{tabular} & $\times$ & \textbf{\checkmark} & \textbf{\checkmark} \\ \hline
         Beam Prediction & $\times$ & \textbf{\checkmark} & \textbf{\checkmark} \\ \hline
         Blockage Prediction & $\times$ & $\times$ & \textbf{\checkmark} \\ \hline
         \begin{tabular}[c]{@{}c@{}}Chain-of-Thought\\ Reasoning\end{tabular} & $\times$ & \textbf{\checkmark} & \textbf{\checkmark} \\ \hline
         \begin{tabular}[c]{@{}c@{}}DRL Policy\\ Selection\end{tabular} & $\times$ & $\times$ & \textbf{\checkmark} \\ \hline
         \begin{tabular}[c]{@{}c@{}}Focused Feature\\ Extraction\end{tabular} & $\times$ & $\times$ & \textbf{\checkmark} \\ \hline
         \begin{tabular}[c]{@{}c@{}}Environment\\ Classifier\end{tabular} & $\times$ & $\times$ & \textbf{\checkmark} \\ \hline
         \begin{tabular}[c]{@{}c@{}}Handover\\ Management\end{tabular} & $\times$ & $\times$ & \textbf{\checkmark} \\ \hline
         \begin{tabular}[c]{@{}c@{}}Human-in-the-Loop \\ \ac{LLM} Priming\end{tabular} & $\times$ & $\times$ & \textbf{\checkmark} \\ \hline
         Interpretation & \textbf{\checkmark}& \textbf{\checkmark} & \textbf{\checkmark} \\ \hline
         \begin{tabular}[c]{@{}c@{}}Long-Term\\ Memory\end{tabular} & $\times$ & $\times$ & \textbf{\checkmark} \\ \hline
         \begin{tabular}[c]{@{}c@{}}Real-Time\\ Inference\end{tabular} & $\times$  & $\times$  & \textbf{\checkmark} \\ \hline
         \begin{tabular}[c]{@{}c@{}}Situation-Aware\\ Grounding\end{tabular} & \textbf{\checkmark} & \textbf{\checkmark} & \textbf{\checkmark} \\ \hline
    \end{tabular}
    \label{tab: upgrades}
    \vspace{-0.26cm}
\end{table}

\vspace{-0.25cm}
\subsection{Paper Organizations}
Our work hereafter is organized as follows, Section \ref{sec: related_work} discuss related works, Section \ref{sec:overview_enwar3} presents an overview of \textsc{Enwar 3.0}, Section \ref{sec:preprocessing_enwar3} showcases our data preprocessing pipeline, Section \ref{sec: llm_priming} details the \ac{LLM} priming process, Section \ref{sec:env_classifier} showcases the environment classifier, Sections \ref{sec:agent_management}-\ref{sec:agent_repo} discuss the agentic structures, Section \ref{sec:response_delivery} provides the mechanisms on response delivery, Section \ref{sec: results} showcases our analysis and ablation study of the components of \textsc{Enwar 3.0}, and Section \ref{sec: conclusion} concludes our work.

\section{Related Works}
\label{sec: related_work}
Recent research \footnote{App. \ref{appendix: related_work_summary} shows a tabular summary of related works} has explored integrating \acp{LLM} into wireless communication and network control, spanning \ac{RAG}, \ac{QA} training, instruction tuning, and multi-agent coordination. \ac{RAG} enhances \ac{LLM} outputs by retrieving domain-specific knowledge from curated knowledge bases, boosting response precision in telecom contexts \cite{xu2024large, yilma2024telecomrag, nazar2024enwar, nazar2025nextggptleveraginggenaiadvancing}. Meanwhile, \ac{QA} fine-tuning has been applied to telecom domains for spectrum management and protocol understanding, while instruction tuning tailors \acp{LLM} to domain tasks but demands large, high-quality datasets and substantial computational overhead \cite{zou2024telecomgpt}. Multi-agent \ac{LLM} architectures have also supported task decomposition and decision-making in network orchestration \cite{shen2024large, zou2023wireless, jiang2024large, tong2025wirelessagentlargelanguagemodel}.

However, most \ac{LLM}-based solutions focus on static, query-based interactions and lack real-time decision-making capabilities crucial for dynamic wireless environments. Many remain fundamentally text-centric and fall short in multi-modal tasks requiring rapid interpretation of sensory data such as LiDAR, radar, and cameras \cite{xu2024large, shao2024wirelessLLM, 6GLLM, yang2025wirelessgptgenerativepretrainedmultitask}. Although vision-language models (VLMs) like LLaMa3.2-Vision, LLaVa \cite{llama3, llava} and cross-modal transformers have shown promise in bridging perception and reasoning \cite{zhang2024visionlanguagemodelsvisiontasks}, practical deployments for real-time wireless optimization remain rare.

Parallel advances in multi-modal sensing have emphasized the benefits of fusing diverse data streams, such as LiDAR, radar, camera, and GPS, to capture fine-grained spatial and temporal patterns essential for tasks like beamforming, blockage prediction, and environment perception \cite{blockage2, blockage3, abdallah2024multiagentbeamtraining, beam_lstm}. Notable examples include vision-aided frameworks that leverage RGB images to predict optimal beam indices and blockage events, reducing overhead from exhaustive beam searches \cite{alrabeiah2019millimeterwavebasestations, deepsense3, blockage1}, and feature-fusion networks that proactively improve link resilience in vehicle-to-infrastructure (V2I) communications.

Recent work has also combined multi-modal data, semantic retrieval, and \ac{RL} to enhance efficiency in vehicular and bandwidth-constrained networks \cite{liu2025wirelessagenticairetrievalaugmented, Du_2025}. Yet, even these systems often rely on black-box reasoning without transparent and explainable logic \cite{Nasir2024xai, khanexplainableICC25, khanTCOM}.

Furthermore, \acp{LLM} struggle with large-scale optimization challenges, such as beamforming or resource allocation, due to the lack of mathematical rigor in purely generative architectures \cite{zou2023wireless, xu2024large, shen2024large, jiang2024large}, however with sufficient data and defined tasks, \acp{LLM} can be fine-tuned for wireless communication and sensing tasks \cite{alikhani2025largewirelessmodellwm}. Another promising alternative includes combining \ac{LLM}-based semantic understanding with domain-specific and structured optimization solvers \cite{ding2025newparadigmusercentricwireless}.

Agentic \acp{LLM} show promise in their ability to perform structured, multi-step reasoning and dynamic orchestration, as seen in NetOrchLLM for network management \cite{netOrchLLM} and surveys highlighting how reasoning loops, \ac{RAG}, \ac{CoT} and few-shot prompting enhance decision-making in complex domains \cite{plaat2025agenticlargelanguagemodels, wei2022chain, brown2020language}. These trends highlight growing momentum toward integrating \acp{LLM} as language tools and intelligent orchestrators within multi-modal, multi-agent systems.

Despite these advancements, substantial gaps remain in designing seamless frameworks that blend multi-modal sensing, interpretable reasoning, and adaptive decision-making for wireless control. \textsc{Enwar 3.0} directly addresses these gaps with a unified system capable of high-level reasoning and fine-grained environmental adaptation under real-time constraints.

\section{An Overview of \textsc{Enwar 3.0}}
\label{sec:overview_enwar3}
This section outlines \textsc{Enwar 3.0}'s pipeline at a high level, spanning offline training and priming to real-time inference. We first formalize the \ac{I2V} communication problem, then describe the integration of multi-agent orchestration, modality-aware perception, and adaptive reasoning in the system.

\subsection{Problem Definition}
\begin{figure}
    \centering
    \includegraphics[width=0.9\linewidth]{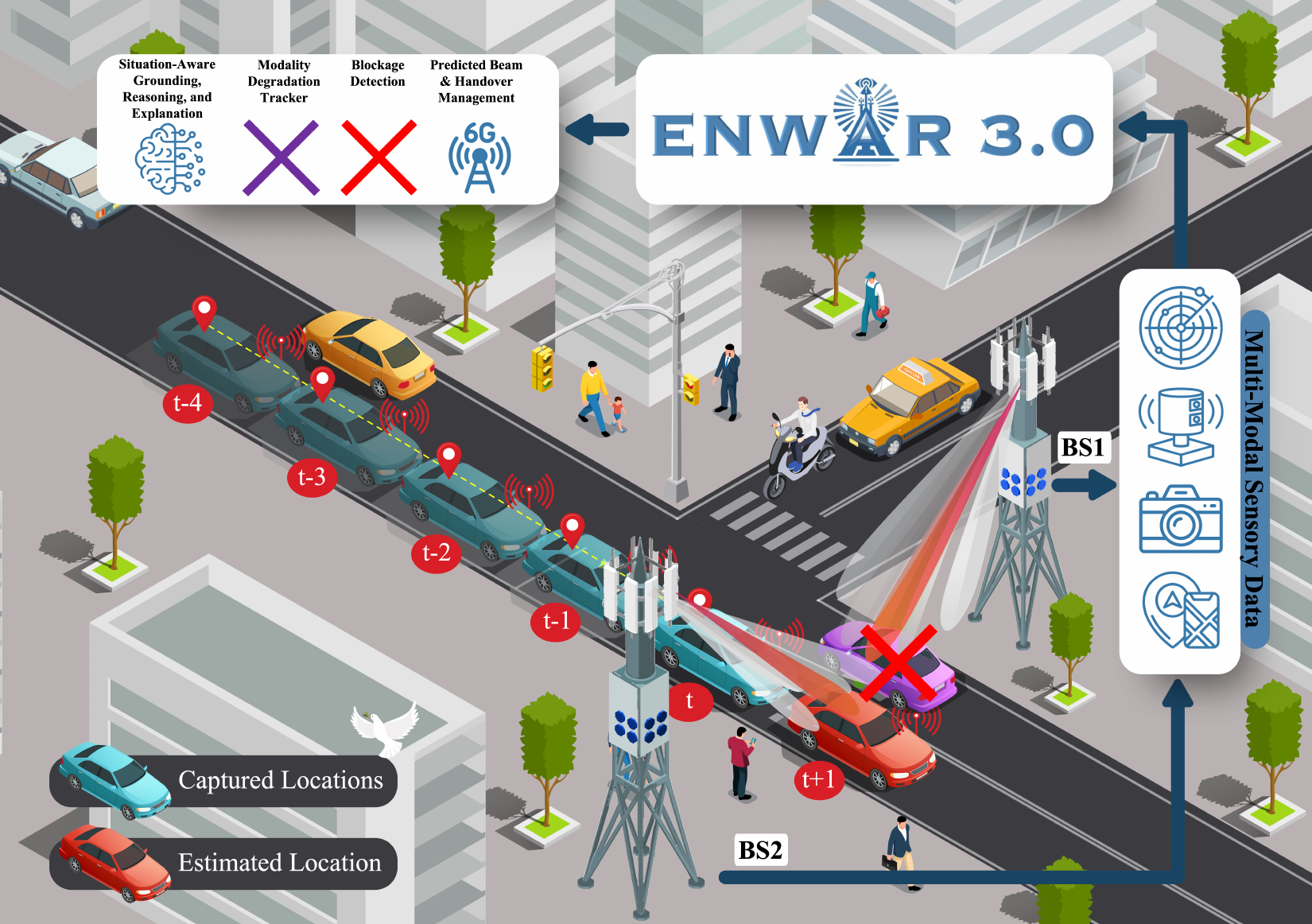}
    \caption{Illustration of \textsc{Enwar 3.0}'s system model.}
    \label{fig:enwar3_system_model}
    \vspace{-0.33cm}
\end{figure}

This work considers a \ac{mmWave} \ac{I2V} communication scenario composed of three geo-tagged units as seen in Fig. \ref{fig:enwar3_system_model}. The first unit (Unit 2) is a moving vehicle equipped with a \ac{ULA} of $M$ antennas and a GPS receiver. The second and third units are fixed roadside infrastructure elements, each consisting of a single-antenna \ac{BS} placed communication range of the vehicle. The primary \ac{BS} (Unit 1) is co-located with a time-synchronized sensor suite comprising a camera, radar, and LiDAR, forming the central \ac{RSU}. The secondary \ac{BS} (BS$_2$), also geo-tagged and within the communication range of the vehicle and has the same blockage and beam prediction capabilities, supports seamless handover by providing an alternative beamforming anchor point. This multi-BS setup enables the system to manage link blockages and mobility-induced coverage changes via environment- and sensor-aware handover decisions.

At discrete time steps $t\!\in\!\{t_1, t_2, \dots, t_n\}$ spaced by $T_s$, the \ac{RSU} collects synchronized multi-modal sensor data, forming the input to two learning agents: a blockage and beam prediction model. Each input sequence spans five consecutive observations ($\Delta T$) denoted as:
\begin{equation}
    \mathcal{X}_t = \left\{ \mathbf{x}_{t-4}, \mathbf{x}_{t-3}, \mathbf{x}_{t-2}, \mathbf{x}_{t-1}, \mathbf{x}_t \right\},
\end{equation}
where $\mathbf{x}_i \in \mathbb{R}^{C \times H \times W}$ represents the fused features extracted from the multi-modal sensors at time $i$. We model the wireless channel and pathloss between Unit 2 and the \acp{BS} using the standardized 3GPP Urban Microcell (UMi) model \cite{3gpp}, which captures realistic signal propagation in dense urban settings. All relevant simulation parameters, including pathloss coefficients and scenario configurations, are detailed in Section~\ref{sec: simulation_params}.

\vspace{0.3em}
\noindent\textbf{Blockage Prediction:} The blockage prediction agent estimates the likelihood of occlusion events—caused by large vehicles, pedestrians, or infrastructure—occurring up to $k$ steps into the future. The model outputs a vector of predicted probabilities:
\begin{equation}
    \mathbf{p}_t = \left[ p_{t+1}, p_{t+2}, \dots, p_{t+k} \right], \quad \text{where} \quad p_{t+i} \in [0, 1],
\end{equation}
with each $p_{t+i}$ indicating the likelihood of a blockage event at time $t+i$ given past sensor observations $\mathcal{X}_t$. The model is trained as a binary classifier using ground-truth labels $y_{t+i} \in \{0,1\}$, and defined by a deep fusion architecture:
\begin{equation}
    \mathbf{p}_t = f_{\theta}(\mathcal{X}_t),
\end{equation}
where $f_{\theta}$ denotes the parameterized model that captures multi-modal spatiotemporal dependencies for blockage prediction.

\vspace{0.3em}
\noindent\textbf{Beam Prediction:} Let $\mathbf{h}(t)\!\in\!\mathbb{C}^{M \times 1}$ denote the channel between the vehicle and the \ac{RSU} at time $t$. The transmitter employs a beamforming codebook $\boldsymbol{\mathcal{F}}\!=\!\{\mathbf{f}_i\}_{i=1}^{Q}$, where $\mathbf{f}_i \in \mathbb{C}^{M \times 1}$ is the $i$th codeword and $Q\!=\!OM\!=\!|\boldsymbol{\mathcal{F}}|$ defines the oversampled codebook size. The received signal is therefore:
\begin{equation}
    y(t) = \mathbf{f}_{\iota(t)}^H \mathbf{h}(t) \, x(t) + n(t),
\end{equation}
where $\iota(t)$ is the selected beam index, $x(t) \in \mathbb{C}$ is the transmitted symbol, and $n(t) \sim \mathcal{N}_{\mathbb{C}}(0, \sigma^2)$ is complex Gaussian noise. The optimal beam index maximizes received power:
\begin{equation}
    \overset{\star}{\iota}(t) = \operatorname*{argmax}_{i \in [1, Q]} \left| \mathbf{f}_i^H \mathbf{h}(t) \right|^2.
\end{equation}
However, exhaustive beam sweeping is computationally intensive in fast-changing environments. To overcome this challenge, a beam prediction model is trained to anticipate the optimal beam $\overset{\star}{\iota}(t+k)$ based on recent sensor trajectories $\mathcal{X}_t$.

\subsection{\textsc{Enwar 3.0}'s Flow}
\begin{figure*}
    \centering
    \includegraphics[width=0.9\linewidth]{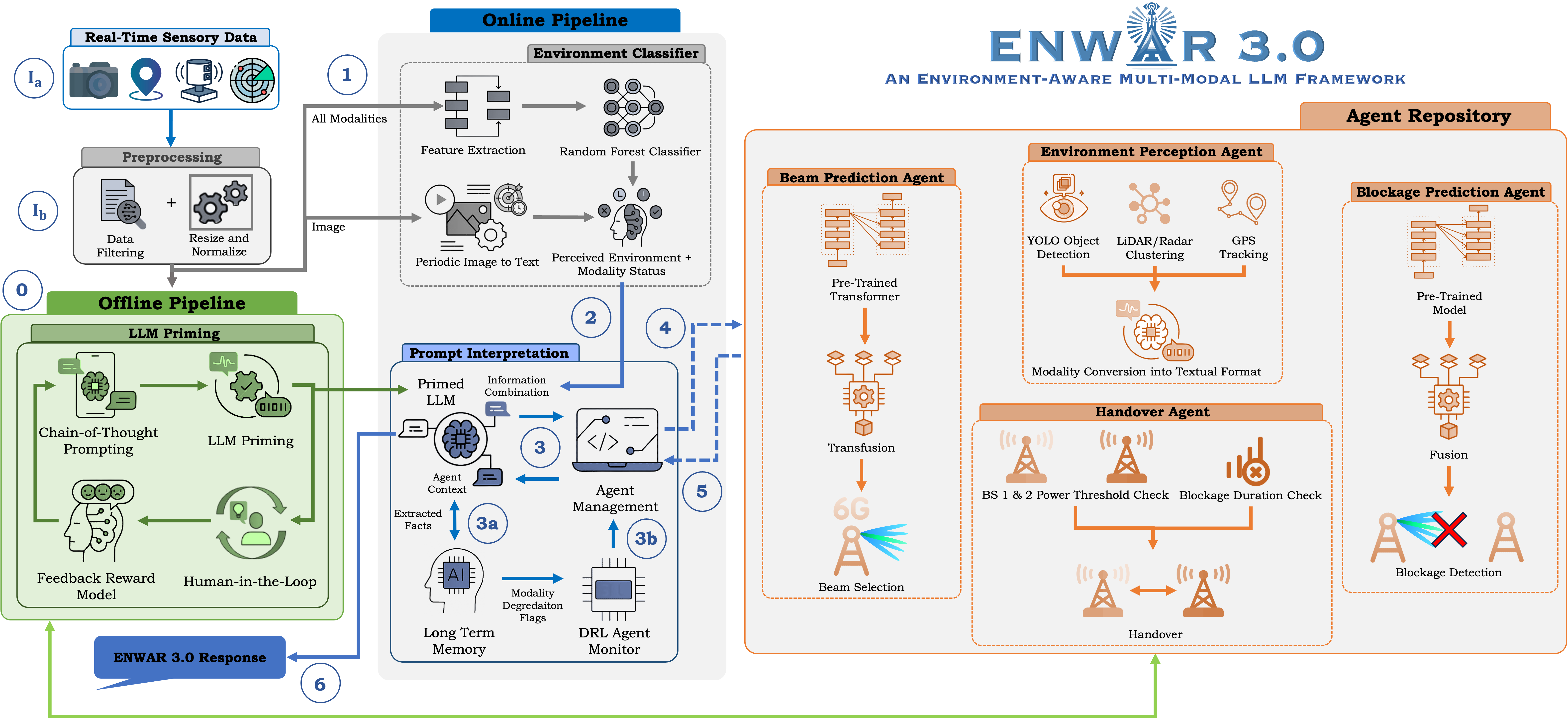}
    \caption{\textsc{Enwar 3.0}'s flow pipeline starting with preprocessing multi-modal inputs, LLM priming in the offline pipeline, and utilizing the primed LLM to process a detailed network perception and enhancement task in the online pipeline through an environment classifier, and available agents in the repository.}
    \label{fig:enwar3_flow}
    \vspace{-0.2cm}
\end{figure*}

As illustrated in Fig. \ref{fig:enwar3_flow}, \textsc{Enwar 3.0} begins with step \textbf{(I$_\text{a}$)}: real-time inputs from camera, GPS, LiDAR, and radar are acquired and processed via a unified preprocessing module in step \textbf{(I$_\text{b}$)} for normalization, filtering, and resizing. The pipeline then splits into two coupled tracks: an offline priming stage and an online inference path, unified through a central \ac{LLM}.

The offline priming pipeline in step \textbf{(0)} prepares the \ac{LLM} using \ac{CoT} prompting, few-shot learning, and human-in-the-loop feedback. Across multiple sessions, the \ac{LLM} is exposed to diverse sensing conditions and response expectations. A reward model scores outputs, and iteratively refines prompt-response alignment. This produces a primed \ac{LLM} capable of consistent multi-step reasoning without reliance on heavy prompt templating at inference time.

The online path begins at step \textbf{(1)} with an environment classifier that detects per-modality degradation to relay to downstream agents. In parallel, an image-to-text module produces periodic high-level summaries to contextualize decision-making. In step \textbf{(2)}, incoming inputs are framed within a structured instruction template. The primed \ac{LLM} interprets this context and determines whether agent invocation is required.

If needed, step \textbf{(3)} activates the agent manager, which orchestrates task-specific agents (e.g., beam and blockage prediction, handover, and environment perception) based on sensor status and task context. In parallel with structured input ingestion, two complementary modules enrich the agent selection process:
\begin{itemize}
	\item The long-term memory module (step \textbf{(3a)}) uses LlamaIndex’s fact extraction to persist meaningful patterns like modality degradation trends, blockage durations, and past agent decisions to enable context-aware reasoning and facilitate historical recall in future inferences.
	\item The \ac{DRL} policy module (step \textbf{(3b)} guides agent configuration using a \ac{PPO}-trained policy \cite{schulman2017proximalpolicyoptimizationalgorithms}, which prioritizes accurate, low-latency agent configurations while avoiding degraded modalities.
\end{itemize}

The long-term memory and \ac{DRL} policy modules operate concurrently with input processing and relay outputs to the agent manager. The \ac{LLM} combines this data with contextual priors and sensor cues for agents selection. {
Crucially, agent execution and network updates occur independently of full natural-language response generation, ensuring that time-critical decisions remain bounded within the sampling interval.}

Once agent outputs are obtained, they are returned to the \ac{LLM} (step \textbf{(7)}), which conducts multi-step reasoning and may re-engage additional agents as needed to complete the inference loop. Finally, step \textbf{(8)} executes a concluding \ac{LLM} reasoning pass, generating a structured and interpretable response that integrates perception, prediction, and control for robust \ac{I2V} orchestration.

\section{Data Preprocessing}
\label{sec:preprocessing_enwar3}
We utilize real-world multi-modal data from the DeepSense6G dataset~\cite{DeepSense1, DeepSense2} to support beam and blockage prediction. This dataset comprises 18,667 samples with synchronized sensor streams of an \ac{I2V} environment. We utilize Scenarios 31–34, which provide time-aligned measurements such as adjusted GPS traces from Units 1 and 2, high-resolution RGB images, LiDAR-based 3D point clouds, and radar signatures. Each training instance is constructed from five temporally ordered multi-modal samples to capture short-term spatiotemporal dynamics.

\subsection{Data Augmentation}
To improve generalization, modality-specific augmentations were applied during training. Images were randomly flipped, rotated, and blurred to emulate viewpoint and lighting variations. LiDAR point clouds underwent random flips, rotations, and spatial scaling to increase geometric diversity. Radar inputs were perturbed with Gaussian noise to model measurement uncertainty.

\subsection{Data Labeling}
Each sample is labeled at time $t{+}1$. Beam prediction uses the beam index with maximum received power as ground truth. Blockage labels are assigned when a sustained \ac{RSSI} drop aligns with visually confirmed obstruction events on Unit~2.

\subsection{Blockage Class Imbalance Handling}
The blockage classification task exhibits significant class imbalance, with 15,050 non-blocked and 3,617 blocked samples. To reduce bias toward the majority class, we employ a class-weighted loss that emphasizes blocked instances:
\begin{equation}
\label{eq: imbalance}
    w_{\text{pos}} = \alpha \cdot \left( \frac{N_\text{non-blocked}}{N_\text{blocked}} \right),
\end{equation}
where $N_{\text{non-blocked}}$ and $N_{\text{blocked}}$ denote the number of samples in each class, and $\alpha$ is a scaling factor that modulates the weighting intensity.

\subsection{Modality Preprocessing}
\subsubsection{Image Preprocessing}
RGB images are resized to $256 \times 256$ and normalized prior to feature extraction.

\subsubsection{GPS Preprocessing}
We derive motion features from the GPS trajectory by computing first- and second-order temporal derivatives, capturing vehicle displacement, velocity, acceleration, angular velocity, and curvature. These descriptors form an 18-dimensional vector and are min-max normalized using a pre-fitted training scaler to ensure consistency across samples.

\subsubsection{LiDAR Preprocessing}
LiDAR point clouds first undergo voxel grid downsampling to reduce point density while maintaining the underlying spatial geometry followed by ground plane extraction via \ac{RANSAC} \cite{voxel_lidar_ransac, pointnet}. As a noise filter, a statistical outlier filter computes the mean distance, $\epsilon$, of each point to its $k$ nearest neighbors and the minimum number of samples to construct an identifiable cluster. Points whose distances deviate from the local neighborhood distribution are discarded based on a threshold defined by the dataset's standard deviation. Remaining above-ground points are clustered using \ac{DBSCAN} ($\epsilon\!=\!0.75$, $\texttt{min\_samples}\!=\!5$), and small clusters are discarded.

The cleaned 3D point cloud is projected into a \ac{BEV} grid over $X,Y \in [-50, 50]$m and $Z \in [-2.5, 15]$m at $0.25$m resolution. Each \ac{BEV} cell encodes: (i) maximum height, (ii) log-scaled point count (density), and (iii) height variance. Temporal sequences are concatenated to yield a $(700, 1200, 15)$ tensor. Each channel is individually normalized: height via min-max scaling, density via log transform, and variance via global standardization.

\subsubsection{Radar Preprocessing}
Radar inputs comprise complex-valued tensors of shape $(4, 256, 250)$ across four virtual antennas. We decompose signals into magnitude and phase, and apply 1D \ac{FFT} along the Doppler axis to extract motion-related spectral features.

We augment these raw features with statistical descriptors: per-channel mean, standard deviation, entropy (for scene complexity), and Doppler-based motion metrics such as mean velocity and spectral spread. All feature maps are resized to $(256\!\times\!64)$ via padding or trimming, and single-value descriptors are broadcast as needed. The final radar input is an $(8, 256, 64)$ tensor capturing spatial, spectral, and dynamic cues relevant for downstream inference.

\section{LLM Priming}
\label{sec: llm_priming}
To enable structured multi-agent orchestration in \textsc{Enwar 3.0}, we prime a centralized \ac{LLM} in an offline pre-deployment phase. The objective of priming is to train the model to interpret structured environment states, invoke appropriate agents exactly once, and generate policy-consistent justifications aligned with sensor reliability and \ac{DRL}-recommended actions. Few-shot examples follow a fixed prompt schema (App.~\ref{appendix:priming_template}) that includes fields such as \textit{Environment Status}, \textit{Trajectory}, \textit{Blockage Status}, and \textit{Predicted Beam Properties}, ensuring a consistent reasoning structure.

Priming is guided through a lightweight human-in-the-loop reward mechanism inspired by \ac{RLHF}. Each structured input state $s_p$ consists of multi-modal environmental features (e.g., object presence, sensor degradation levels), long-term memory context, and \ac{DRL}-suggested agent selections. The \ac{LLM} generates an action $a_p$ consisting of selected agent calls and structured justification. Human evaluators then score each $a_p$ using a fixed rubric along three axes: (i) \textit{Correctness}, based on valid beam, blockage, and handover decisions; (ii) \textit{Agentic Justification}, assessing the alignment between \ac{DRL} output and modality health; and (iii) \textit{Explanation Clarity}, ensuring responses are well-structured, complete, and non-redundant. These criteria are aggregated into a scalar reward $R_p(a_p) \in [0,10]$. The reward function $R_{\text{priming},\theta}(s_p, a_p)$ is parameterized by rule-based scoring parameters $\theta$, ensuring evaluation consistency rather than learned reward modeling. If responses misuse agents, violate policy alignment, or exhibit structural redundancy, evaluators provide corrective feedback, and the prompt is iteratively refined.

This process defines a reward-guided prompt optimization loop over the induced \ac{LLM} policy $\pi(a_p \mid s_p)$:
\begin{equation}
\theta \leftarrow \theta + \nabla_\theta \mathbb{E}{a_p \sim \pi(\cdot | s_p)}[R{_{\text{priming},\theta}}(s_p, a_p)],
\end{equation}
where $\theta$ governs rubric parameters only; the \ac{LLM} weights remain unchanged. Instead of gradient-based fine-tuning, the model internalizes structured reasoning patterns through iterative \ac{CoT} prompting and feedback until responses satisfy $R_p(a_p) > \tau_{\text{reward}}$. The corresponding pseudocode is provided in App.~\ref{appendix:priming_code}.

At deployment, the primed \ac{LLM} receives real-time sensory inputs and \ac{DRL} outputs, determines whether agents are required, executes the necessary calls, and generates a structured decision report. For example, if radar is flagged as degraded and excluded by the \ac{DRL} policy, the \ac{LLM} must justify its omission and avoid redundant invocations. This behavior emerges from the structured reward-guided priming process.

The priming mechanism enforces policy-consistent, degradation-aware, and non-redundant agent orchestration under dynamic \ac{I2V} conditions. An example of the priming loop is provided in App.~\ref{appendix:priming_example} and summarized below.


\vspace{5pt}
\hrule
\vspace{1pt}
\hrule
\vspace{5pt}
\noindent \textbf{Prompt 1:}
Perceive the environment using the available modalities based on your sense of the environment status. Which beam and blockage models should be selected and justify your decision.
\vspace{3pt}
\hrule
\vspace{1pt}
\hrule
\vspace{3pt}
\noindent \textbf{Summarized Response 1:}
\begin{itemize}
    \item Modality Status: Radar degraded (42.1\%), all others reliable.
    \item Environment: Urban, 3 vehicles near stop sign.
    \item DRL Agent Recommendation: Use camera-gps-lidar for beam, camera-only for blockage
    \item Select camera-only for blockage due to radar degradation.
    \item Start handover to BS2. 
    \item Invoked blockage prediction and handover agents twice, beam prediction and perception agents once.
\end{itemize}
\vspace{5pt}
\hrule
\vspace{1pt}
\hrule
\vspace{5pt}
\noindent \textbf{Prompt 2:}
Going forward, focus more on the detailed justifications of why handover is necessary, and do not repeat agent invocations. Score: 2.4
\vspace{5pt}
\hrule
\vspace{1pt}
\hrule
\vspace{5pt}
\noindent \textbf{Summarized Response 2:}
...\\
Handover Status: The blockage condition threshold has been flagged and BS2 has a higher power than Unit 1, therefore initiate handover. Invoked all agents once.
\vspace{5pt}
\hrule
\vspace{1pt}
\hrule

\section{Perceived Environment Status}
In multi-modal sensor-equipped environments, sensors can degrade due to noise, weather, and obstructions. Early environment classification is crucial for decision-making. \textsc{Enwar 3.0} features an environment classifier and an image-to-text module that periodically converts images into a generalized environment description. This section details these modules.

\subsection{Environment  Classifier}
\label{sec:env_classifier}
\textsc{Enwar 3.0} includes a dedicated environment status classifier that outputs four binary flags indicating whether each modality (camera, GPS, LiDAR, radar) is degraded to support adaptive model selection. These real-time flags guide downstream agents to rely only on trustworthy sensors, improving robustness under dynamic conditions.

We developed a synthetic degradation pipeline because the original dataset lacked extensive coverage of adverse scenarios. Camera frames were blurred, oversaturated, or darkened to simulate fog, rain, or night. LiDAR point clouds were sparsified to mimic occlusion, radar tensors corrupted with Gaussian noise, and GPS traces jittered to emulate jamming or spoofing. Each modality was degraded independently and probabilistically to create diverse sensor failure states. In deployment, this classifier may be augmented with weather metadata (e.g., visibility, time of day) via real-time APIs.

The classifier uses a two-stage pipeline: handcrafted feature extraction and multi-output classification. Extracted features include blur entropy and brightness stats (camera), point dispersion and density (LiDAR), spectral entropy and SNR (radar), and variance and displacement (GPS). These are concatenated, normalized, and passed to a multi-output Random Forest with 100 estimators and max depth 10. Each output node predicts whether a modality is degraded.

Using modality-specific degradation metrics, a modality is flagged as degraded if $>\!\tau_{degradation}$ of its signal is impaired over five consecutive time steps. These scores are smoothed over time to avoid reacting to transient noise. Degraded modalities are temporarily excluded from model selection until their scores drop below the threshold, such that \textsc{Enwar 3.0} relies only on high-confidence inputs.


\subsection{Periodic Image-to-Text Environment Perception}
\textsc{Enwar 3.0} includes a low-frequency environment perception module that converts visual inputs into structured textual summaries. At periodic intervals (e.g., hourly), a representative camera frame is processed by a vision-language model to generate concise scene descriptors (e.g., \textit{``urban intersection with moderate traffic and clear weather''}). 

These summaries capture slowly varying environmental attributes such as lighting, weather, and scene type, providing high-level contextual information beyond instantaneous sensor measurements. Since these attributes evolve gradually, the module operates independently of the real-time inference loop, preserving low latency while providing stable contextual cues to downstream reasoning agents.

\section{Agent Management}
\label{sec:agent_management}
The agent management layer mediates interaction between the centralized \ac{LLM} and task-specific agents. When a subtask associated with a specialized agent is required, the \ac{LLM} issues a structured JSON request to the manager, which selects the appropriate agent based on task type and current sensor availability as determined by the environment classifier.

Agents are configured for different degradation profiles. For instance, if the camera is flagged as unreliable, camera-dependent models are bypassed in favor of radar- or LiDAR-based alternatives. This routing mechanism ensures inference relies on the most reliable modalities.

The manager is supported by a \textbf{memory module} and a \textbf{\ac{DRL} policy} (Sections~\ref{sec: memory_module}–\ref{sec: drl_agent}), which incorporate historical context, sensor degradation, and blockage persistence into agent selection. Agent outputs are returned to the \ac{LLM} as structured semantic blocks, which are intergated into the final \ac{LLM} reasoning chain's agent selection.


\vspace{-0.2cm}
\subsection{Long-Term Memory and Sensor Context Tracking}
\label{sec: memory_module}
The long-term memory module in \textsc{Enwar 3.0} serves as a persistent, queryable repository for encoding sensor reliability patterns and agent decision outcomes over time. It complements the real-time behavior of the \ac{DRL}-based agent selection module by supplying historical context, enabling the system to reason reactively with temporal foresight.

This module tracks two critical forms of long-range metadata: agent context histories and modality degradation. Agent contexts include past beam and blockage model selections, blockage durations, environment labels and conditions, and associated \ac{DRL} rewards. These logs enable temporal introspection and facilitate the identification of effective model configurations under recurring environmental states.

\begin{figure}
    \centering
    \includegraphics[width=0.9\linewidth]{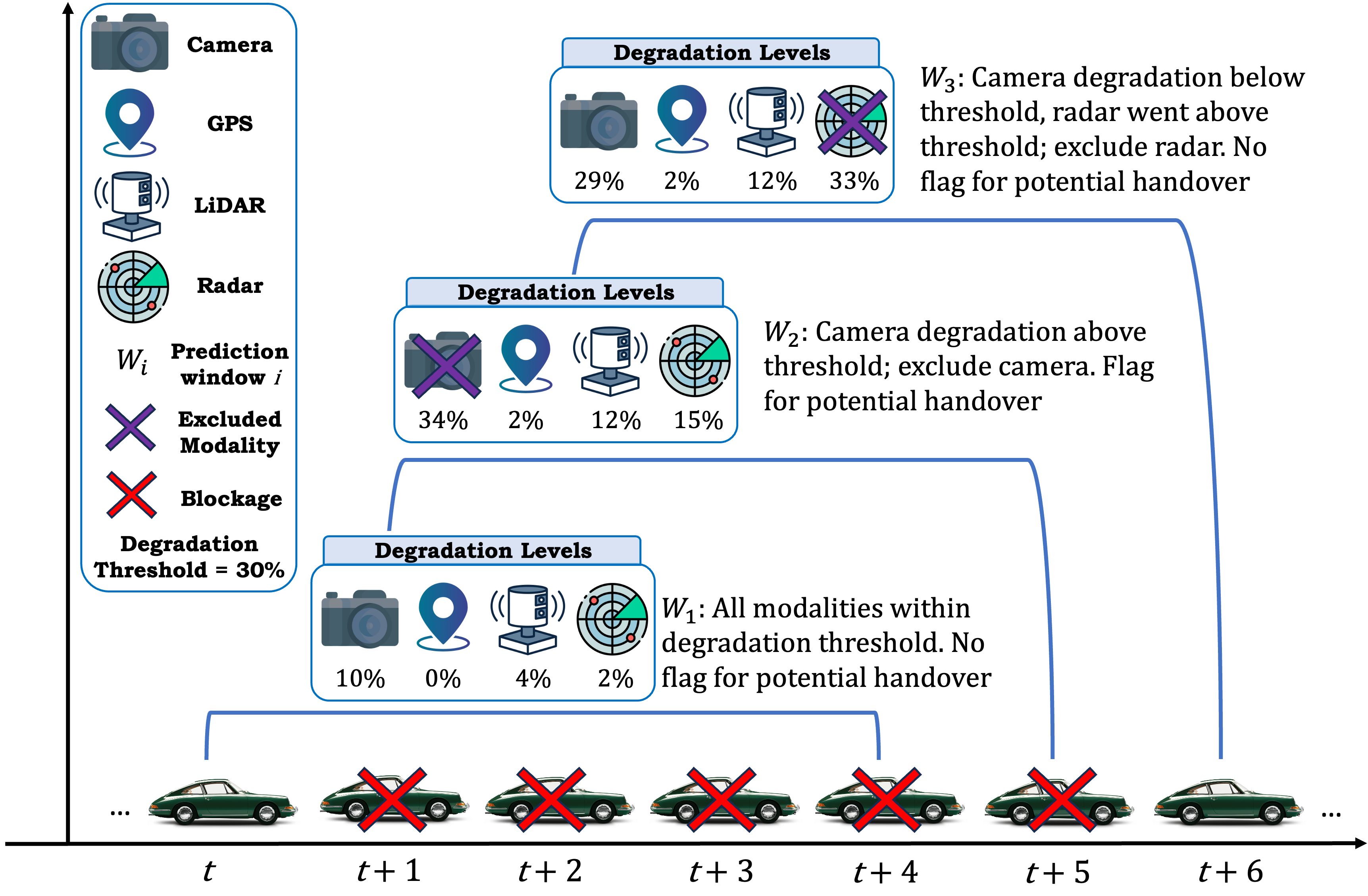}
    \caption{Three consecutive input windows illustrating updates to blockage and modality degradation flags. Window 1: all modalities are healthy. Window 2: camera degraded and persistent blockage detected. Window 3: radar degraded while camera recovers.}
    \label{fig:degradation_blockage_window_track}
    \vspace{-0.3cm}
\end{figure}

In parallel, the system maintains degradation metrics by computing a running average of the modality classifier's per-frame binary degradation flags output. If a sensor has a degradation of $>\!\tau_{degradtion}$ of the observed frames per an inferable input window ($\geq t+5$ time steps), it is classified as persistently degraded. This statistical visibility score is then used to suppress agent configurations that rely on low-performing sensors, guiding both the \ac{LLM} and \ac{PPO} agents away from suboptimal decisions. An example of the degradation tracker across three input windows is seen in Fig. \ref{fig:degradation_blockage_window_track}.

Similarly, long-term memory also tracks blockage durations. Blockage durations are extracted from the agent contexts. If a blockage persists for more than $\!t\!+\!5$ time steps, then a flag is set to indicate that a handover condition is satisfied. This mechanism is seen in Fig. \ref{fig:degradation_blockage_window_track}, where three inferred windows exist. The first and last windows show blockages occurring for less than $t\!+\!5$ time steps. As such, the flag for potential handover due to prolonged blockages is not set, and the second window satisfies the conditions and has the flag set to true. 

The long-term memory is implemented using the \textit{LlamaIndex} framework through its extensible memory block architecture. Specifically, two memory block types are employed: \textit{fact extraction memory block}, which distills agent-environment interactions into structured facts (e.g., “Radar blockage agent succeeded in foggy environment”), and \textit{static memory block}, which stores known modality degradation profiles and policy-derived insights. When invoked, the LLM retrieves relevant memory blocks, enabling it to condition its reasoning on prior successes, failure cases, or modality trends.

This indexed memory enables symbolic and semantic search across historical interactions, improving few-shot generalization and reducing reasoning drift. Moreover, the \ac{LLM}'s agent evaluator can cross-reference long-term memory entries with live predictions to ensure consistent and trustworthy model selection during dynamic deployments. The long-term memory module thus transforms \textsc{Enwar 3.0} from a reactive agent system into a historically informed orchestration framework.

\subsection{Policy-Driven Agent Selection}
\label{sec: drl_agent}
To support robust and adaptive agent selection under dynamic sensing conditions, \textsc{Enwar 3.0} employs a \ac{DRL} policy based on the \ac{PPO} algorithm. This policy selects the most suitable combination of modalities from a repository of pretrained beam and blockage prediction models, depending on current sensor health and environmental context.

Unlike rule-based systems, this learned policy adaptively routes tasks to one of various pretrained model variants per agent. To support robust agent selection under varying sensor conditions, we design 15 unique pretrained model variants per agent. Each variant corresponds to a distinct combination of sensor modalities, camera, LiDAR, radar, GPS, or combinations thereof, allowing for granular alignment between model architecture and available, non-degraded sensor inputs. The action space, $A_\text{DRL}$ represents a set of actions, $a_\text{DRL}$, that outputs a multi-discrete tuple of integers representing the selected pretrained modality combination for beam and blockage prediction models based on the perceived environment and modality degradation status, $s_\text{DRL}$, as extracted from the long-term memory block [c.f. Section \ref{sec: memory_module}] and the environment classifier [c.f. Section \ref{sec:env_classifier}].

The \ac{DRL} agent is trained using \ac{PPO}, which balances learning stability with exploration efficiency. \ac{PPO}  is a policy gradient method that updates the policy network in a way that discourages large policy shifts, ensuring smoother convergence and reduced risk of catastrophic forgetting. We adopt a multi-layer perceptron (MLP) policy with a 2-layer feedforward network to map the observed state of the environment, including current degradation conditions and memory features, to a discrete action vector. This vector selects which pretrained agent variants to invoke for beam and blockage prediction tasks. Each training batch consists of 1024 time steps, subdivided into 64-sample mini-batches. The policy is optimized using a clipped objective with a learning rate of $3\!\times\!10^{-4}$, a discount factor, $\gamma\!=\!0.95$, and a small entropy bonus (0.01) to encourage exploratory behavior early in training.

The policy is trained in a synthetic environment where 30\% of training samples are probabilistically degraded using the sensor perturbation pipeline [c.f. Section~\ref{sec:env_classifier}]. This sampling ensures the agent learns robust selection behavior across a diverse distribution of partial sensor failures. Observations are encoded as a 4-dimensional binary vector representing the degraded status of each modality. The \ac{PPO} algorithm, an actor-critic framework, optimizes the policy by balancing exploration and learning stability using clipped surrogate objectives. The agent is trained to map these binary degradation states to optimal model pair selections that maximize reward.

The reward function, $R_\text{DRL}(s_\text{DRL}, a_\text{DRL})$ is designed to prioritize model performance and sensor reliability while penalizing computational inefficiency. The reward for each action is:
\begin{equation}
 R_\text{DRL}(s_\text{DRL}, a_\text{DRL})\!=\!0.9 (A_{\text{beam}}\!+\!A_{\text{block}}) - 0.1( T_{\text{beam}}\!-\!T_{\text{block}}),
 \end{equation}
where $A_{\text{beam}}$ and $A_{\text{block}}$ denote the normalized performance scores (accuracy, F1-score) of the selected models, weighted by their reliance on degraded sensors. $T_{\text{beam}}$ and $T_{\text{block}}$ represent normalized inference latency penalties. The weighting scheme is designed to balance predictive performance with computational efficiency while reflecting the different scales of the two objectives. Performance metrics are naturally bounded between 0-100\%, making them stable and directly comparable across models. Inference latency, by contrast, spans a broader and less uniform range, often varying by an order of magnitude across configurations (13-67ms). To prevent these larger latency variations from dominating the reward, accuracy, and reliability are assigned higher weight as the primary reward target, while inference time contributes as a secondary penalty. This formulation ensures that the \ac{DRL} agent consistently selects models that maintain high predictive quality, yet still accounts for efficient real-time deployment.

At inference time, the trained \ac{DRL} agent receives real-time degradation flags from the environment classifier and memory module, and selects the most context-appropriate model pair for that time window. This adaptive routing mechanism allows \textsc{Enwar 3.0} to intelligently respond to sensor failures and environmental changes without retraining or manual intervention.

\section{Agent Repository}
\label{sec:agent_repo}
\textsc{Enwar 3.0} hosts a modular agent repository for environment perception, beam and blockage prediction, and handover management. Details of the perception and beam agents are provided in Apps.~\ref{app:env_agent_enwar3}–\ref{appendix:beam_prediction_agent} and in \textsc{Enwar 2.0}~\cite{nazar2025enwar2, asilomar2025}, with minor modifications to support BS$_2$ integration. Below, we detail each agent's design and their role in \textsc{Enwar 3.0}'s pipeline.

\subsection{Blockage Prediction Agent}
\begin{figure}
    \centering
    \includegraphics[width=0.75\linewidth]{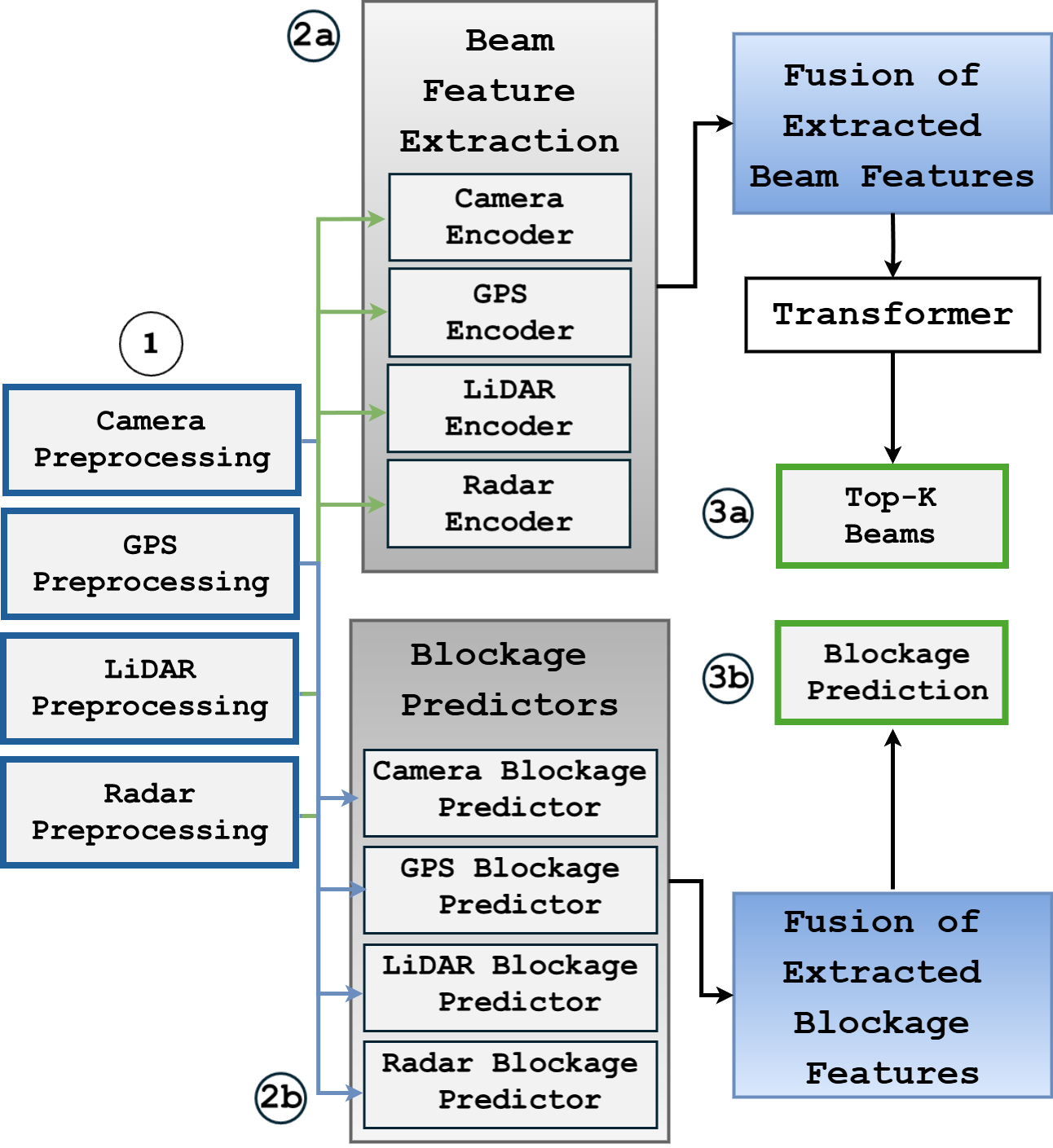}
    \caption{Simplified architecture of the beam and blockage prediction models, where (1) shows data preprocessing, (2a-b) resemble the beam and blockage prediction flow, and (3a-b) are the model outputs.}
    \label{fig:model_architecture}
    \vspace{-0.34cm}
\end{figure}

The blockage prediction agent is designed to anticipate future communication blockages based on environmental perception and vehicle dynamics. Multi-modal inputs from $t$ time steps are forwarded to four modality-specific encoders, tailored to extract blockage-related spatiotemporal features. A late fusion strategy is then used to aggregate modality-level predictions into a final decision, allowing the system to adaptively prioritize sensor modalities based on reliability and validation performance. The model is trained using binary cross-entropy loss with logits for stable convergence. A high-level representation of the model architecture is shown in Fig. \ref{fig:model_architecture}, and App. \ref{appendix: agent_model_architecture} fully details each component.

\subsubsection{Camera-Based Blockage Prediction}
The camera-based model captures spatial and temporal features indicative of dynamic obstructions. A ResNet-18 backbone, pretrained on ImageNet and truncated before the classification layer, encodes each frame into a high-dimensional embedding. These embeddings are fed into a single-layer \ac{LSTM} with 128 hidden units to capture temporal dependencies such as object motion and visual occlusions. The final hidden state is passed through a two-layer fully connected classification head with ReLU activation and dropout ($\rho = 0.4$), producing a single logit that reflects the blockage probability. 

\subsubsection{GPS-Based Blockage Prediction}
The GPS-based model focuses on identifying mobility patterns correlated with blockage scenarios, such as sudden deceleration or sharp turns. We extract 18 features from each sequence, including displacement, velocity, acceleration, angular change, and curvature. These features are normalized and passed as a sequence to a two-layer \ac{LSTM} with 128 hidden units. The final hidden state is projected through a fully connected layer (64 hidden units) with ReLU activation, dropout, and a binary output neuron.

\subsubsection{LiDAR-Based Blockage Prediction}
The LiDAR-based model processes temporally stacked \ac{BEV} projections of point cloud data. This preprocessed sequence consists of five consecutive \ac{BEV} frames, where each frame includes three spatial feature channels (height, density, variance), resulting in a 15-channel input tensor. To handle this high-dimensional input, we adapt a ResNet-18 backbone by modifying the first convolutional layer to accept 15 input channels. The network processes the stacked frames through residual blocks, learning rich spatial features tied to object boundaries and environmental structure. Unlike other modalities, temporal information is encoded implicitly through the stacked frame. The final output logit is given from a fully connected layer.

\subsubsection{Radar-Based Blockage Prediction}
The radar-based model urilizes spatial motion signatures and temporal dynamics of radar reflections. Each radar frame is an 8-channel tensor of size $256 \times 64$, encoding magnitude, phase, Doppler, and frequency features. These frames are passed through three 2D convolutional layers with ReLU activations and batch normalization to capture spatial features. An adaptive average pooling layer reduces each frame to a fixed-length vector, and the resulting sequence is processed by a single-layer \ac{LSTM} with 64 hidden units. The final hidden state is passed through a two-layer classification head with ReLU and dropout ($\rho = 0.3$), producing a binary prediction logit. 

\subsubsection{Multi-Modal Fusion}
We adopt a late fusion approach that aggregates modality-specific outputs at the decision level to integrate predictions across each modality. Each model independently outputs a probability score of the likelihood of a future blockage: $\mathcal{P}_{\text{Camera}}, \mathcal{P}_{\text{GPS}}, \mathcal{P}_{\text{LiDAR}}, \mathcal{P}_{\text{Radar}}$. These scores are combined using a weighted averaging scheme:
\begin{equation}
    \mathcal{P}_{\text{fused}} = \sum_{i \in \{\text{Camera}, \text{GPS}, \text{LiDAR}, \text{Radar}\}} w_i \cdot \mathcal{P}_i.
\end{equation}

Fusion weights, $w_i$, are derived from a running performance estimate that combines (i) softmax-normalized F1-scores computed on a held-out validation set and (ii) the modality preference suggested by the \ac{DRL} policy. Let $\mathbf{s} = [s_{\text{Camera}}, s_{\text{GPS}}, s_{\text{LiDAR}}, s_{\text{Radar}}]$ denote the validation F1-score vector. The weights are then given by:
\begin{equation}
    \mathbf{w} = \text{softmax}(\mathbf{s}) = \frac{\exp(s_i)}{\sum_j \exp(s_j)}.
\end{equation}
The final output predicts blockages for up to $t\!+\!5$ time steps (1.5s). This formulation weighs modalities based on validated performance while preserving multi-sensor contributions, thus enhancing robustness under dynamic sensing conditions.

\subsection{Handover Agent and Power-Aware Decision Policy}
\label{sec: handover_agent}
To support proactive link maintenance in dynamic vehicular scenarios, a dedicated handover agent governs the transition between multiple \acp{BS}. The handover logic is informed by predicted link blockages and real-time power differentials.

The decision process follows a dual-trigger policy. Let $P_1(t)$ and $P_2(t)$ denote the estimated received powers from Unit 1 and BS$_2$ at time $t$. A handover from Unit 1 to BS$_2$ is initiated if either of the following conditions hold for $\!\geq\!5$ consecutive time steps:
\begin{enumerate}
    \item \textbf{Power Superiority Trigger:} BS$_2$ offers a significantly stronger link:
    \begin{equation}
        P_2(t) - P_1(t) \geq \Delta_{\text{power}}
    \end{equation}
    where $\Delta_{\text{power}}$ is a tunable threshold.
    
    \item \textbf{Blockage-Aware Trigger:} A blockage is predicated on Unit 1's link, and BS$_2$ offers any improvement in received power:
    \begin{equation}
        \text{Blockage}_t = 1 \quad \text{and} \quad P_2(t) > P_1(t)
    \end{equation}
\end{enumerate}
If blockage persists beyond $t\!+\!5$ time steps (i.e., one input window), the agent performs an explicit BS evaluation and generates a structured output including: (1) selected target BS ID, (2) expected power gain, and (3) a confidence score. This output is passed to the \ac{LLM} for the final handover decision coordination. Temporal blockage tracking and persistence logic are handled within the memory module [c.f. Section~\ref{sec: memory_module}].

\section{Response Delivery}
When agentic contexts are complete and processed, the \ac{LLM} delivers a response with situation-aware, grounded reasoning. This section details the response delivery steps. 
\label{sec:response_delivery}
\subsection{Prompt Processing}
When the primed \ac{LLM} receives structured data derived from real-time multi-modal inputs, prompt processing begins. These inputs include environmental descriptors, sensor degradation flags, agent outputs, and contextual metadata. The primed \ac{LLM} is fully equipped to perform structured, context-aware inference across its core tasks: beam and blockage analysis, environment interpretation, and handover decision-making.

The prompt schema ensures that all inputs are aligned with the \ac{LLM}’s trained reasoning format. For instance, visual and spatial cues from degraded modalities are summarized as natural language descriptors (e.g., "LiDAR is occluded, camera visibility is 40\%"), while structured metadata (e.g., \ac{BS} power levels, blockage duration) is appended as a JSON request. This design ensures the \ac{LLM} engages in interpretable, deterministic reasoning aligned with the intended policy logic.

Upon receiving structured inputs, the \ac{LLM} either performs direct multi-step \ac{CoT} inference or delegates tasks to specialized agents. Tasks requiring explicit algorithmic processing, such as handover evaluation, beam prediction, or sensor-based perception, are routed through the \textbf{agent management} layer.


\subsection{Response Generation}
With all agent outputs received, the \ac{LLM} integrates its reasoning with agent contexts to generate a complete, structured response. Each field in the response is populated based on a combination of sensor data, agent feedback, and contextual understanding embedded during \ac{LLM} priming. This final step highlights \textsc{Enwar 3.0}’s ability for situational reasoning as it connects agent outputs with qualitative justifications. 

\begin{figure*}
    \centering
    \includegraphics[width=0.8\linewidth]{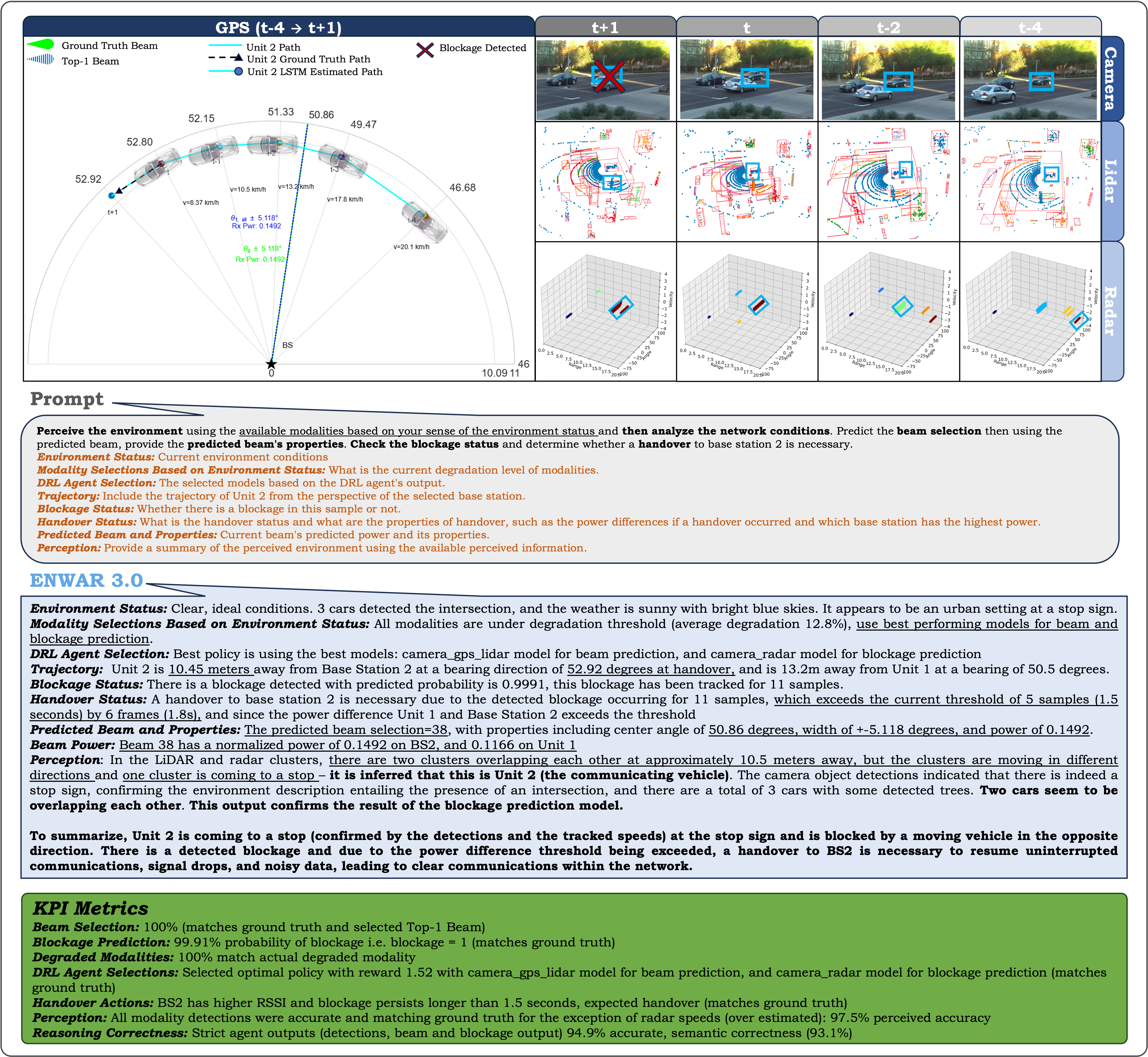}
    \caption{An example of \textsc{Enwar 3.0}'s response to a detected blockage and handover scenario.}
    \label{fig:enwar3response}
    \vspace{-0.3cm}
\end{figure*}

Fig. \ref{fig:enwar3response} showcases an optimal response to a blockage and handover event, illustrating the framework’s multi-agent reasoning. The response includes the predicted beam and its properties, perceived environmental context, rationale for modality inclusion/exclusion based on degradation, Unit 2’s trajectory, and a detailed handover justification, with \ac{KPI} metrics for that inference. For illustration, ground truth beam data is shown, but unavailable during deployment. The accompanying prompt template is used during pre-deployment priming; once deployed, the \ac{LLM} no longer relies on this explicit format, as it internalizes the reasoning patterns and task logic required to operate autonomously. 
The response showcases \textsc{Enwar 3.0}'s appropriate agent invocations, and situation-aware grounding and reasoning. 



\textsc{Enwar 3.0} operates under dynamic handover conditions and varying levels of sensor degradation through degradation-aware policy routing. Upon detecting modality impairment, the system invokes a \ac{DRL}-trained agent to select the appropriate modality configuration for beam and blockage prediction based on the current environmental context.

When sensing quality deteriorates, inference pipelines are adaptively rerouted using degradation flags and long-term memory signals, preserving temporal consistency in decision-making. The resulting outputs include structured justifications for beam selection, blockage forecasting, and handover triggers. Illustrative inference examples are provided in App.~\ref{app: inference_examples}.


\vspace{-0.2cm}
\section{Evaluation of \textsc{Enwar 3.0}}
\label{sec: results}
This section evaluates \textsc{Enwar 3.0} using agent-level \acp{KPI} and degradation-aware orchestration metrics. Individual agents are assessed across multiple sensor configurations, while the orchestration layer is evaluated based on model selection accuracy and justification consistency. The subsections below detail the evaluation methods and their results.

\subsection{Enwar 3.0 Setup}
The evaluation of \textsc{Enwar 3.0} was conducted on an NVIDIA A100 GPU (40GB VRAM). To analyze the impact of model scale on reasoning behavior under full perception access (i.e., access to all sensor modalities), we evaluated multiple \acp{LLM}, including: LLaMa 3.1-8B, LLaMa 3.3-70B, DeepSeek-v1.5-R1-8B, DeepSeek-v1.5-R1-32B, DeepSeek-v1.5-R1-70B, Qwen 2.5-3B, and Qwen 3-32B. Runtime memory usage scaled with model size (3B: 3.4GB; 8B: 16GB; 32B: 22GB; 70B: 32GB), with all models supporting up to 128k token contexts for multi-turn \ac{CoT} reasoning. Unless otherwise noted, all core experiments and benchmark evaluations were conducted using LLaMa 3.2-3B, due to its efficient runtime characteristics and minimal deployment overhead.

Agent memory allocation varied by modality configuration. The largest blockage predictor required 1.7GB, and the largest beam predictor used 3.5GB. The vision-language perception module (LLaMa 3.2-Vision-11B) required 8GB, and combined perception agents required 1.07GB. Agents were dynamically activated to manage runtime budgets.

To evaluate the beam and blockage prediction agents' performance, we defined 15 unordered and distinct sensor combinations. Each combination was profiled regarding accuracy, inference latency, and preprocessing overhead. These combinations formed the action space of the \ac{DRL} training environment used to assess policy-driven modality selection under varying degradation states.

When evaluating the primed \ac{LLM}, we conducted 10 independent priming sessions, each involving up to 20 iterations of human-in-the-loop feedback. On average, convergence to the predefined reward threshold occurred within four prompt cycles per session, indicating stability in structured multi-agent reasoning patterns rather than prompt-specific behavior.

Throughout priming, human reviewers evaluated each generated response for alignment with desired content structure, correctness of invoked agents, and the depth of contextual reasoning. Feedback was translated into reward signals that guided the \ac{LLM} toward increasingly precise and context-aware outputs. This iterative refinement ensured that the primed \ac{LLM} reliably produced detailed, task-specific reasoning aligned with the system’s real-time decision-making requirements.

\subsection{Simulation Parameters}
\label{sec: simulation_params}
Real-time inputs consisted of five multi-modal samples spaced by $T_s\!=\!300$ms, forming an input window of $\Delta T\!=\!1.5$s. For blockage prediction, the class-imbalance parameter in (\ref{eq: imbalance}) was set to $\alpha\!=\!1.1$ to modestly upweight blocked samples. During priming, the reward acceptance threshold was set to $\tau_{\text{reward}}\!=\!8.5$.


Received power is estimated using a geometric and physics-informed model. Unit 2's bearing to \ac{BS}$_i$ is computed as:
\begin{equation}
\begin{aligned}
    \phi_1 &= \text{radians}(lat_1), \quad \phi_2 = \text{radians}(lat_2) \\
    \Delta \lambda &= \text{radians}(lon_2 - lon_1) \\
    x &= \sin(\Delta \lambda) \cdot \cos(\phi_2) \\
    y &= \cos(\phi_1) \cdot \sin(\phi_2) - \sin(\phi_1) \cdot \cos(\phi_2) \cdot \cos(\Delta \lambda) \\
    \theta &= \text{atan2}(x, y) \\
    \theta_{\text{deg}} &= \text{degrees}(\theta),
\end{aligned}
\end{equation}
where $(lat_1, lon_1)$ and $(lat_2, lon_2)$ represent the latitudes and longitudes of Unit 2 and \ac{BS}$_i$, respectively.

Distances between Unit 2 and each \ac{BS} are computed as the great-circle distance, $d$, between two points on a sphere given their latitudes and longitudes. This distance is calculated using the Haversine formula as follows:
\begin{equation}
\begin{aligned}
    a &= \sin^2\left(\frac{\Delta \phi}{2}\right) + \cos(\phi_1) \cdot \cos(\phi_2) \cdot \sin^2\left(\frac{\Delta \lambda}{2}\right) \\
    c &= 2 \cdot \text{atan2}(\sqrt{a}, \sqrt{1 - a}) \\
    d &= R \cdot c
\end{aligned}
\end{equation}
where $R\!=\!6371000$ meters is the Earth's radius.

These distances are then mapped to path loss values using the 3GPP Urban Microcell (UMi) model \cite{3gpp}, with separate formulations for \ac{LoS} and \ac{NLoS} conditions:
\begin{align}
    \text{LoS:} \quad P_{\text{loss}}(d) &= 32.4 + 21\log_{10}(d) + 20\log_{10}(f_\text{GHz}), \\
    \text{NLoS:} \quad P_{\text{loss}}(d) &= 22.4 + 35.3\log_{10}(d) + {} \nonumber \\
    & \quad \,\,\, 21.3\log_{10}(f_\text{GHz}) - 0.3(h_\text{UE} - 1.5).
\end{align}
Beamforming gains, $G_{\text{beam}}$, are estimated from predicted beam power measurements, $P_{\text{beam}}$, using a logarithmic conversion:
\begin{equation}
    G_{\text{beam}} = 10 \cdot \log_{10}(P_{\text{beam}}).
\end{equation}
Combining transmit power ($P_{\text{tx}}$), pathloss  
($P_{\text{loss},i}$), and beamforming gain ($G_i$), BS$_i$'s received power, $P_{\text{rx},i}$, is computed as:
\begin{equation}
    P_{\text{rx},i} = P_{\text{tx}} - P_{\text{loss},i} + G_i.
\end{equation}

{
Table~\ref{tab:hyperparameters} summarizes the key hyperparameters used during agent training and \ac{LLM} deployment. For LLM inference, we set $\text{max new tokens}\!=\!4096$ to support structured multi-step reasoning, $\text{temperature}\!=\!0.5$ to balance determinism and diversity, and $\text{repetition penalty}\!=\!1.15$ to discourage redundancy. These settings were selected through preliminary ablations and promote stable, concise outputs aligned with agent orchestration requirements.}

\begin{table}
\centering
\caption{Hyperparameters used in Agent Training and LLM Deployment}
\vspace{-0.12cm}
\begin{tabular}{|c|c|c|}
\hline
\textbf{Component} & \textbf{Hyperparameter} & \textbf{Value} \\ \hline
\multirow{5}{*}{\begin{tabular}[c]{@{}c@{}}Beam \\Prediction\end{tabular}} & Batch Size & 6 \\ \cline{2-3} 
 & Epochs & 75 \\ \cline{2-3}
 & Epoch of Convergence & 34 \\ \cline{2-3} 
 & Learning Rate & 0.0001 \\ \cline{2-3} 
 & Number of Samples & 18,667  \\ \hline
\multirow{6}{*}{\begin{tabular}[c]{@{}c@{}}Blockage \\Prediction\end{tabular}} & Alpha ($\alpha$) & 1.1 \\ \cline{2-3} 
 & Batch Size & 6 \\ \cline{2-3} 
 & Epochs & 75 \\ \cline{2-3} 
 & Epoch of Convergence & 27 \\ \cline{2-3} 
 & Learning Rate & 0.00001 \\ \cline{2-3} 
 & Number of Samples & 18,667 \\ \hline
\multirow{7}{*}{DRL} & Batch Size & 64 \\ \cline{2-3} 
 & Entropy Coefficient & 0.01 \\ \cline{2-3} 
 & Episode of Convergence & 912 \\ \cline{2-3} 
 & Degradation Threshold ($\tau_{degradation}$) & 30\% \\ \cline{2-3}
 & Discount Factor ($\gamma$) & 0.95 \\ \cline{2-3} 
 & Number of Episodes & 1000 \\ \cline{2-3} 
 & Number of Estimators & 1024 \\ \hline
\multirow{2}{*}{\begin{tabular}[c]{@{}c@{}}Environment \\Classifier\end{tabular}} & Number of Estimators & 100 \\ \cline{2-3} 
 & Max Depth & 50 \\ \hline
\multirow{8}{*}{LLM} & Max New Tokens & 4096 \\ \cline{2-3} 
 & Priming Reward Threshold ($\tau_\text{reward}$) & 8.5 \\ \cline{2-3}
 & Priming Iterations & 20 \\ \cline{2-3}
  & Priming Convergence (iterations) & 4 \\ \cline{2-3}
 & {\begin{tabular}[c]{@{}c@{}}Number of Response \\Generation Evaluation Samples \end{tabular}}& 750 \\ \cline{2-3} 
 & Repetition Penalty & 1.15 \\ \cline{2-3} 
 & Temperature & 0.2 \\ \hline
\end{tabular}
\label{tab:hyperparameters}
\vspace{-0.32cm}
\end{table}

\subsection{Evaluation of Beam Prediction Agent}
The beam prediction agent was assessed using two \acp{KPI}:
\begin{itemize}
\item \textbf{Top-$\boldsymbol{k}$ Accuracy} evaluates the proportion of instances in which the ground-truth beam index appears among the Top-$k$ predicted candidates. This metric reflects the agent's ability to consistently rank the correct beam within a practical search window.
\item \textbf{Average \ac{APL}} quantifies the average degradation in received power due to suboptimal beam selection. {It is computed as:}
\begin{equation}
    APL_{[dB]} = 10 \log_{10} \left( \frac{p'}{p} \right),
    \label{eq: apl_equation_enwar3}
\end{equation}
where $p'$ denotes the received power of the highest-ranked beam among the Top-$k$ predictions, and $p$ corresponds to the power of the true optimal beam. Lower APL values indicate more efficient beam selection.
\end{itemize}

\begin{table*}
\caption{
Agent Performance for $t\!+\!1$ with $[M\!=\!16, Q\!=\!64]$. Inference times report single-sample GPU latency.}
\label{tab: agent_performance}
\centering
\begin{tabular}{|c|c|ccc|ccc|}
\hline
\multirow{2}{*}{\textbf{Modality}} & \multirow{2}{*}{\textbf{Preprocessing}} & \multicolumn{3}{c|}{\textbf{Beam Prediction}} & \multicolumn{3}{c|}{\textbf{Blockage Prediction}} \\ \cline{3-8} 
 & \textbf{Time (ms)} & \multicolumn{1}{c|}{\textbf{APL}} & \multicolumn{1}{c|}{\textbf{Accuracy}} & \textbf{\begin{tabular}[c]{@{}c@{}}Inference \\ Time (ms)\end{tabular}} & \multicolumn{1}{c|}{\textbf{F1 Score}} & \multicolumn{1}{c|}{\textbf{AUC-ROC}} & \textbf{\begin{tabular}[c]{@{}c@{}}Inference \\ Time (ms)\end{tabular}} \\ \hline
 
camera\_gps\_lidar & 33.32 & \multicolumn{1}{c|}{-0.009220} & \multicolumn{1}{c|}{\textbf{88.5\%}} &  54.56 & \multicolumn{1}{c|}{91.2\%} & \multicolumn{1}{c|}{0.897} & 55.28 \\ \hline

camera\_radar\_lidar & 34.39 & \multicolumn{1}{c|}{-0.009314} & \multicolumn{1}{c|}{88.3\%} & 61.48 & \multicolumn{1}{c|}{94.0\%} & \multicolumn{1}{c|}{0.928} & 57.19 \\ \hline
 
camera\_gps\_radar\_lidar & 36.03 & \multicolumn{1}{c|}{-0.009660} & \multicolumn{1}{c|}{85.8\%} & 66.98 & \multicolumn{1}{c|}{93.1\%} & \multicolumn{1}{c|}{0.914} & 59.47 \\ \hline
 
camera\_lidar & 22.4 & \multicolumn{1}{c|}{-0.009668} & \multicolumn{1}{c|}{85.6\%} & 39.28 &  \multicolumn{1}{c|}{96.2\%} & \multicolumn{1}{c|}{0.955} & 37.41 \\ \hline
 
camera\_gps\_radar & 18.3 & \multicolumn{1}{c|}{-0.01469} & \multicolumn{1}{c|}{84.9\%} & 31.84 & \multicolumn{1}{c|}{92.3\%} & \multicolumn{1}{c|}{0.911} & 30.05 \\ \hline
 
camera\_radar & 17.83 & \multicolumn{1}{c|}{-0.02125} & \multicolumn{1}{c|}{84.2\%} & 32.26 & \multicolumn{1}{c|}{\textbf{98.4\%}} & \multicolumn{1}{c|}{0.988} & 30.68 \\ \hline
 
camera\_gps & 4.961 & \multicolumn{1}{c|}{-0.02844} & \multicolumn{1}{c|}{83.7\%} & 29.39 & \multicolumn{1}{c|}{94.4\%} & \multicolumn{1}{c|}{0.931} & 27.70 \\ \hline
 
gps\_lidar\_radar & 21.84 & \multicolumn{1}{c|}{-0.03553} & \multicolumn{1}{c|}{83.0\%} & 33.87 & \multicolumn{1}{c|}{89.9\%} & \multicolumn{1}{c|}{0.889} &  31.64 \\ \hline
 
camera\_only & 4.113 & \multicolumn{1}{c|}{-0.03705} & \multicolumn{1}{c|}{82.8\%} & 26.66 & \multicolumn{1}{c|}{98.1\%} & \multicolumn{1}{c|}{0.983} & 22.32 \\ \hline
 
gps\_radar & 13.52 & \multicolumn{1}{c|}{-0.03979} & \multicolumn{1}{c|}{81.9\%} & 26.40 & \multicolumn{1}{c|}{89.3\%} & \multicolumn{1}{c|}{0.880} & 24.55 \\ \hline
 
gps\_lidar & 12.45 & \multicolumn{1}{c|}{-0.08013} & \multicolumn{1}{c|}{76.6\%} & 28.58 & \multicolumn{1}{c|}{84.1\%} & \multicolumn{1}{c|}{0.855} & 26.91 \\ \hline
 
radar\_lidar & 20.70 & \multicolumn{1}{c|}{-0.1427} & \multicolumn{1}{c|}{74.4\%} & 34.80 & \multicolumn{1}{c|}{91.8\%} & \multicolumn{1}{c|}{0.902} & 32.74 \\ \hline
 
radar\_only & 12.47 & \multicolumn{1}{c|}{-0.1602} & \multicolumn{1}{c|}{72.4\%} & 22.59
& \multicolumn{1}{c|}{93.7\%} & \multicolumn{1}{c|}{0.922} & 21.81 \\ \hline
 
lidar\_only & 10.92 & \multicolumn{1}{c|}{-0.1743} & \multicolumn{1}{c|}{65.7\%} & 27.95 & \multicolumn{1}{c|}{87.9\%} & \multicolumn{1}{c|}{0.872} & 25.18 \\ \hline
 
gps\_only & $<1$ & \multicolumn{1}{c|}{-0.2151} & \multicolumn{1}{c|}{59.0\%} & 13.59 & \multicolumn{1}{c|}{61.7\%} & \multicolumn{1}{c|}{0.603} & 8.332 \\ \hline
\end{tabular}
\vspace{-0.2cm}
\end{table*}


The beam prediction results in Table~\ref{tab: agent_performance} clearly demonstrate how sensor combinations affect accuracy and power efficiency in the $t\!+\!1$ time slot [c.f. App.~\ref{app: beam_agent} for an evaluation of all time slots]. Setups including the camera consistently outperform others, confirming the role of visual context in beam prediction. The highest-performing configuration, camera\_gps\_lidar, achieved 88.5\% accuracy with an exceptionally low \ac{APL} of $-0.00922$, showing that high precision is possible with modest fusion overhead. However, performance gains diminish beyond three modalities, suggesting limited benefit from adding more sensors once the camera and GPS are included.

In contrast, models relying on a single modality perform poorly. The gps\_only model achieved just 59.0\% accuracy with an \ac{APL} of $-0.2151$, reflecting the inadequacy of positional data alone in dynamic environments. Similarly, radar- and LiDAR-only models reached only 72.4\% and 65.7\% accuracy, respectively, despite offering spatial structure, indicating the need for richer perceptual input.

{
To supplement agent performance, Table~\ref{tab: agent_performance} reports preprocessing and inference times, where inference reflects end-to-end per-sample latencies (including preprocessing and model forward-pass execution under steady-state GPU conditions).}

\subsection{Evaluation of Blockage Prediction Agent}
The blockage prediction agent delivers strong performance, as seen in Table \ref{tab: agent_performance} [c.f. App. \ref{app: blockage_agent} for an evaluation of all time slots], across diverse sensor configurations, with F1-score and \ac{AUC-ROC} as \acp{KPI}. F1-score captures precision-recall tradeoffs, while \ac{AUC-ROC} measures discriminative ability across thresholds. 

The best-performing model,  camera\_radar, achieved an F1-score of 98.4\% and an \ac{AUC-ROC} above 0.988, closely followed by camera\_only and camera\_lidar. These results highlight the role of visual sensors in blockage detection in dynamic urban settings. Multi-modal combinations further enhance robustness under degraded conditions. For example, camera\_radar\_lidar reached a 94.0\% F1-score and 0.928 \ac{AUC-ROC}. Even radar- and LiDAR-only models maintained solid performance (F1-scores of 93.7\% and 87.9\%), making them valuable fallbacks when vision is degraded.

In contrast, GPS-dominant models underperformed. The gps\_only configuration yielded just a 61.7\% F1-score and 0.603 \ac{AUC-ROC}, confirming that spatial information alone lacks the perceptual fidelity needed to detect blockages. Blockage prediction is best supported by camera and radar data, with vision-led models excelling in clear conditions and fusion models ensuring robustness in degraded environments.


\subsection{Evaluation of Environment Status Classifier Agent}
\label{subsection:enwar3_model_selection_eval}
The environment status classifier achieved F1-scores of 99.6\% (camera), 97.8\% (LiDAR), 99.1\% (radar), and 99.91\% (GPS) across training perturbations and unseen synthetic degradations, including partial sensor blackouts, intermittent frame drops, and composite multi-modality impairments.

These results indicate reliable degradation detection, enabling downstream agents to prioritize the most dependable modalities under varying sensor conditions.


\subsection{Evaluation of DRL Agent}
Our evaluation of the \ac{DRL} agent was two-fold: the first study analyzes the reward performance of the learned policy compared to rule-based and random modality selection strategies; the second investigates how degradation thresholding influences policy effectiveness in dynamic conditions.

\subsubsection{Policy-Based Evaluation}
We compared the \ac{PPO}-trained policy against two baselines: (1) a \textit{random policy} that samples beam and blockage models uniformly, and (2) a \textit{rule-based policy} that selects the highest-accuracy model among those using only non-degraded sensors without considering each model's limitations. The evaluation was conducted over 1000 episodes sampled from a hybrid distribution: 50\% ideal sensing conditions and 50\% drawn from randomized sensor degradation states. A small set of these represent severe degradation of 3+ sensors to simulate rare edge cases.

Table \ref{tab:agent_selection_comparison} reports the performance of each policy across three \acp{KPI}: average reward, percentage of optimal actions (selecting the best-performing model pair in the current episode’s state), and the percentage of selections involving degraded sensors.
\begin{table}
\centering
\caption{Comparison of agent selection policies}
\vspace{-0.12cm}
\begin{tabular}{|c|c|c|c|}
\hline
\textbf{Method} & \textbf{Avg. Reward} & \begin{tabular}[c]{@{}c@{}}\textbf{\% Optimal} \\ \textbf{Actions} \end{tabular} & \begin{tabular}[c]{@{}c@{}}\textbf{\% Degraded} \\ \textbf{Sensors Used} \end{tabular} \\
\hline
Random Policy & 0.69 & 13.2\% & 61.5\% \\
Rule-Based Policy & 1.04 & 67.1\% & 38.7\% \\
\textbf{\ac{DRL} Agent}& \textbf{1.25} & \textbf{86.6\%} & \textbf{12.9\%} \\
\hline
\end{tabular}
\label{tab:agent_selection_comparison}
\vspace{-0.2cm}
\end{table}

Our \ac{DRL} agent substantially outperformed both baselines, with an average reward of 1.25, which is 20.1\% higher than the rule-based policy and more than double that of the random policy. It also selected optimal actions in 86.6\% of episodes while minimizing reliance on degraded sensors to just 12.9\%, demonstrating strong policy alignment with model performance and sensor health. In contrast, the rule-based policy often over-prioritized local accuracy, leading to degraded sensor usage in nearly 38.7\% of cases.

\begin{figure}
    \centering
    \includegraphics[width=0.9\linewidth]{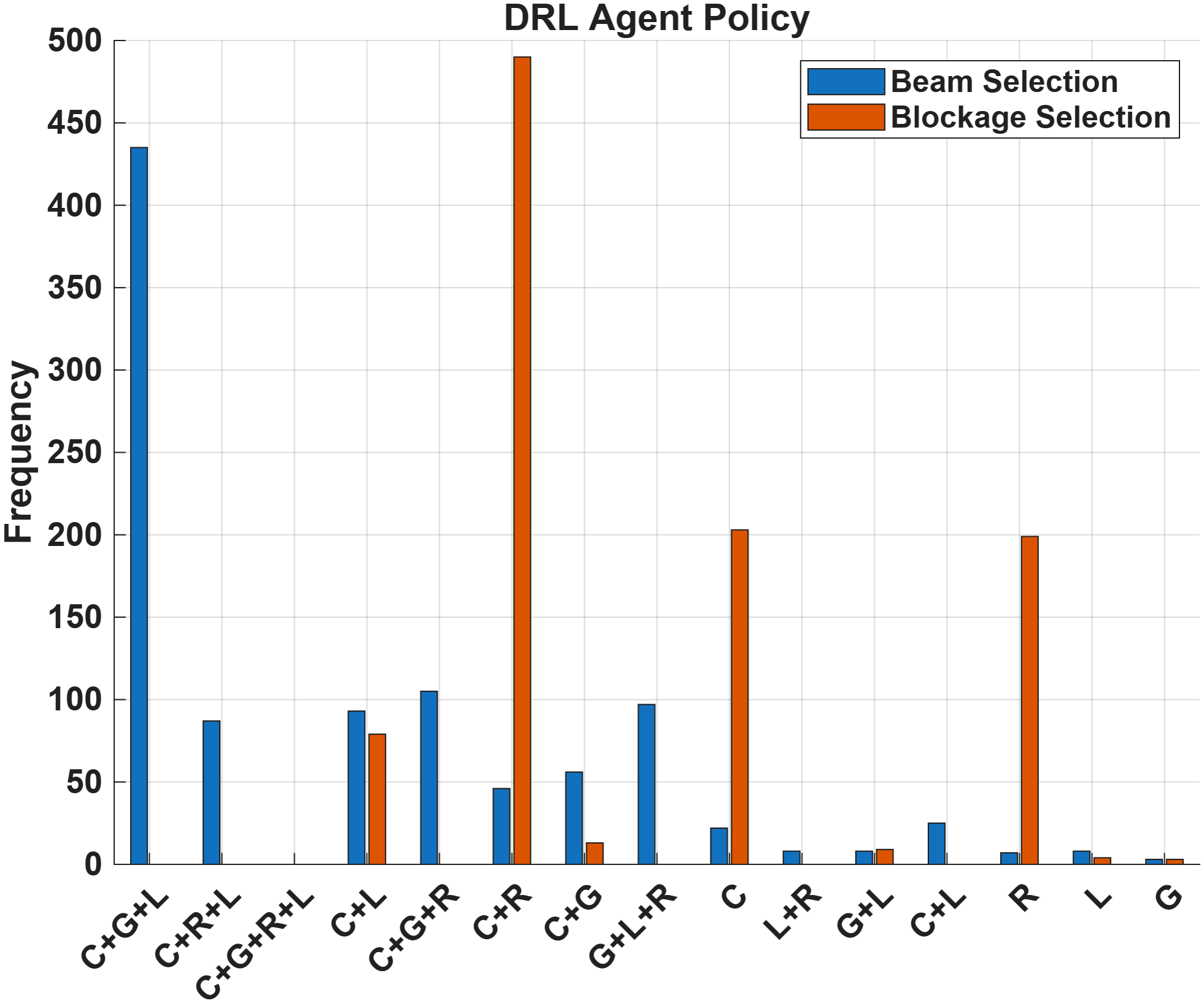}
    \caption{Model combination selections (C: camera, G: GPS, L: LiDAR, R: radar) for prediction agents across 1000 episodes.}
    \label{fig:model_selection_distribution}
    \vspace{-0.3cm}
\end{figure}

Fig. \ref{fig:model_selection_distribution} illustrates the distribution of model selections across 1000 environment states under varying sensor conditions. The selections reveal a strong alignment between sensor availability and modality-appropriate model choice: in ideal conditions, high-complexity multi-sensor models dominate with the highest accuracies are favored; under degradation, the policy shifts toward leaner, sensor-compatible models. The agent avoids unsafe selections and maintains meaningful reward accumulation even in rare severe degradation scenarios (e.g., three modalities unavailable). We refer the readers to App. \ref{appendix: drl_reward_distribution} for further details on the reward distribution of the \ac{DRL} agent. 

\subsubsection{Threshold-Based Evaluation}
\begin{figure}
    \centering
    \includegraphics[width=1.0\linewidth]{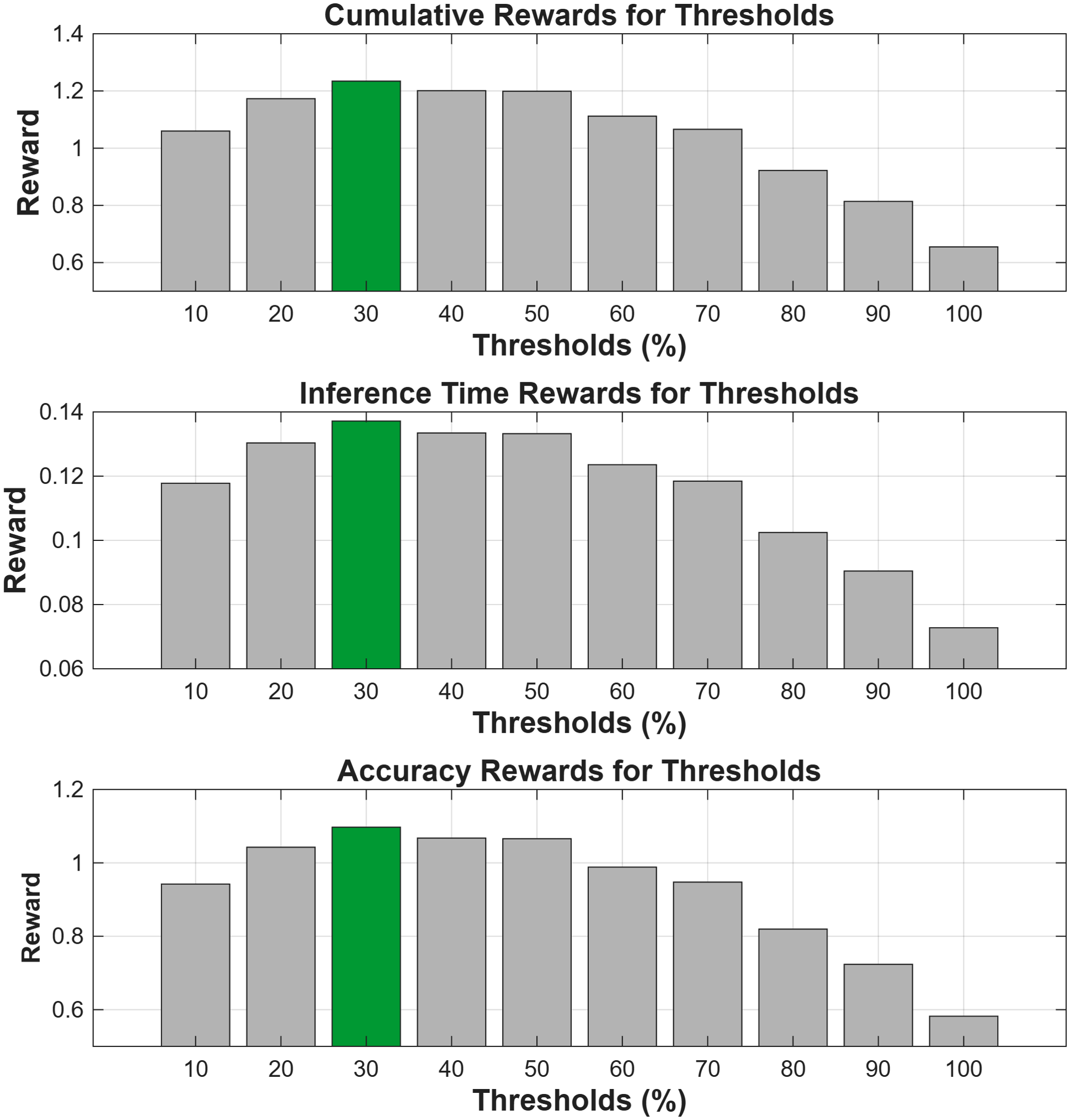}
    \caption{Maximum DRL agent rewards through A) a cumulative and weighted reward, B) inference time-based and C) performance-based rewards. }
    \label{fig:threshold}
    \vspace{-0.3cm}
\end{figure}

Modality selection depends on the degradation threshold $\tau_{degradation}\!\in\!\{10\%, 20\%, \dots, 100\%\}$, where lower values trigger earlier switching and higher values tolerate greater impairment. As shown in Fig.~\ref{fig:threshold}, cumulative reward increases from 10\% to 30\% and declines thereafter. The maximum reward occurs at $\tau = 30\%$ (1.25) (Fig.~\ref{fig:threshold}A). The accuracy-driven component peaks at 1.10 (Fig.~\ref{fig:threshold}B), while the latency component reaches 0.33 (Fig.~\ref{fig:threshold}C). These results indicate that premature switching underutilizes usable sensors, whereas delayed switching degrades predictive performance. Accordingly, $\tau_{degradation}=30\%$ is selected as the default threshold for the best accuracy-latency tradeoff under the learned \ac{DRL} policy.


\subsection{Evaluation of Handover Agent}
The handover agent monitors blockage persistence and received power differentials across \acp{BS}. When a prolonged blockage is detected ($\geq$5 consecutive time steps), and the alternate \ac{BS} (BS$_2$) consistently offers stronger received power than the original (Unit 1), the agent initiates a handover after a latency delay of one time step ($t_{\text{handover}}\!=\!t$). This latency coefficient reflects a realistic switching delay often observed in communication environments.

We evaluate the handover agent across 500 potential handover sequences under varied environmental conditions, comparing four policies:
\begin{itemize}
    \item \textbf{No Handover:} Always remain on Unit 1 regardless of link degradation.
    \item \textbf{Immediate Switch:} Trigger a handover as soon as the handover conditions are satsified in the immediate step as opposed to a persisting conditions. 
    \item \textbf{Oracle:} Preemptively knows and selects all optimal handover decisions.
    \item \textbf{\textsc{Enwar 3.0}:} Switches only after handover conditions are satisfied for $\geq 5$ consecutive steps.
\end{itemize}

The evaluation results, shown in Fig. \ref{tab:handover_eval}, details that \textsc{Enwar 3.0} substantially improves post-blockage recovery by reducing average recovery time by over 60\% compared to a non-handover policy. It captures over 85\% of the oracle's gain while avoiding nearly all of the premature switches exhibited by the immediate switching baseline.

\begin{table}
\centering
\caption{Comparison of Handover Strategies, $k = 5$}
\label{tab:handover_eval}
\begin{tabular}{|c|c|c|c|c|}
\hline
\begin{tabular}[c]{@{}c@{}}\textbf{Metric}\end{tabular} &\begin{tabular}[c]{@{}c@{}}\textbf{No}\\\textbf{Handover}\end{tabular} &\begin{tabular}[c]{@{}c@{}}\textbf{Immediate}\end{tabular} & \begin{tabular}[c]{@{}c@{}}\textbf{Oracle}\end{tabular} &\begin{tabular}[c]{@{}c@{}}\textbf{\textsc{Enwar 3.0}}\end{tabular} \\ \hline
\begin{tabular}[c]{@{}c@{}}APL @ $t\!+\!k$\\(dBm)\end{tabular} & -5.10 & +0.71 & \textbf{+2.53} & +1.82 \\ \hline
\begin{tabular}[c]{@{}c@{}}Avg. Power \\Gain\end{tabular}& 0.0\% & +70.6\% & +181.6\% & +158.2\% \\ \hline
\begin{tabular}[c]{@{}c@{}}Premature \\Switch Rate\end{tabular} & 0.0\% & 41.0\% & 0.0\% & \textbf{3.5\%} \\ \hline
\begin{tabular}[c]{@{}c@{}}Avg. Handover \\Frequency\end{tabular} & 0 & 56 & 47 & 39 \\  \hline
\end{tabular}
\vspace{-0.25cm}
\end{table}

\subsection{Evaluation of Response Generation and Reasoning Correctness}
\label{sec:reasoning_eval}
\subsubsection{Evaluation Dataset Creation}
\textsc{Enwar 3.0} was evaluated across diverse sensing conditions using 150 samples per modality combination (2250 total instances), including varied lighting, weather, and occlusion scenarios.
To evaluate our \ac{KPI} [c.f. Subsection \ref{sec:kpi_enwar3}], 50 validation \ac{QA} pairs were defined per combination (750 total prompts). Each pair assessed beam and blockage prediction, handover decisions, and modality selection. Corresponding reference answers served as ground truth for the systematic evaluation of reasoning correctness and explanatory alignment.


\subsubsection{Performance Criteria and KPIs}
\label{sec:kpi_enwar3}
The 50 structured evaluation prompts were presented to \textsc{Enwar 3.0}, each requiring the generation of multi-step responses reflecting its internal reasoning process across key decision-making steps. Each response followed a prescribed format through \ac{LLM} priming for consistent parsing and targeted evaluation. Each output was assessed using the \textit{reasoning correctness} \ac{KPI}, which measures alignment between generated and ground-truth responses.

Ground truths were manually constructed for each \ac{QA} pair and included deterministic fields such as beam index, power differentials, blockage likelihood, and handover state. Reasoning correctness builds on the interpretability metrics in \textsc{Enwar 2.0}; it evaluates factual validity and logical consistency, including whether modality selection, beam decisions, blockage inference, and handover triggers are justified given sensor states, trajectory dynamics, and environmental context.

Scoring combines strict matching for deterministic outputs with embedding-based semantic similarity for explanatory text, allowing linguistic variation while preserving logical accuracy. A response achieves high reasoning correctness when agent contexts are properly inferred, and decisions are coherently justified. Full scoring details follow the methodology introduced in \textsc{Enwar 1.0}~\cite{nazar2024enwar}.

{
Using thirteen system configurations, we evaluated reasoning correctness under varying sensor combinations, as shown in Fig.~\ref{fig:reasoning_correctness_summarized} [c.f. App.~\ref{app: full_ablation_study} for full modality-level results]. The baseline \textit{vanilla LLaMa} used static prompt instructions without structured reasoning, memory, or policy-based routing. We then progressively introduced four orthogonal components: single-pass \ac{CoT} reasoning (CoT), long-term memory (Mem), the \ac{DRL} agent (DRL), and \ac{LLM} priming (Primed). \ac{CoT} provides structured multi-step reasoning without supervision, and priming extends \ac{CoT} reasoning through iterative human-aligned refinement and reward-based consistency enforcement.

To isolate architectural contributions, we evaluated combinations of Mem, DRL, CoT, and Primed. Hybrid configurations combining CoT and Primed were not exhaustively enumerated, as priming inherently incorporates structured CoT patterns through supervised prompt alignment. This design choice avoids redundant configurations while preserving interpretability of component-level gains.
\begin{figure}
    \centering
    \includegraphics[width=0.9\linewidth]{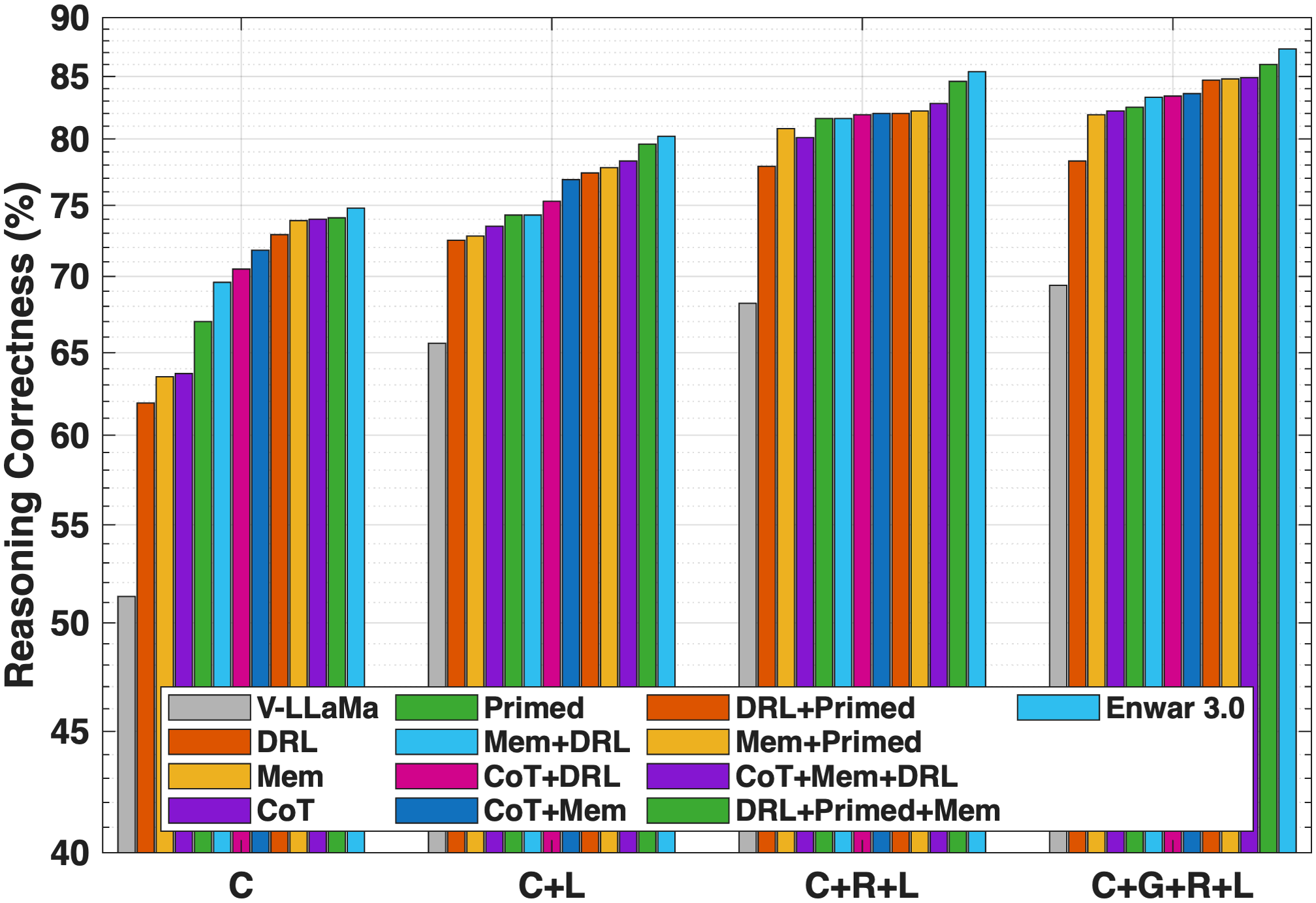}
    \caption{Comparison of the best-performing modality combinations (1 to 4 modalities, C: camera, G: GPS, L: LiDAR, R: radar) in \textsc{Enwar 3.0}'s reasoning gains from long-term memory (Mem), DRL, CoT, and LLM priming (Primed) relative to the vanilla baseline.}
    \label{fig:reasoning_correctness_summarized}
    \vspace{-0.35cm}
\end{figure}



\textsc{Enwar 3.0} achieves 87.3\% reasoning correctness with LLaMa3.2-3B, a 17.9\% absolute improvement over the vanilla baseline (69.4\%) under full-modality inclusion. Scaling to DeepSeek-r1-70B improves performance to 89.3\%, indicating that model scale provides only modest gains.

Among individual components, priming yields the largest standalone gain (82.5\%), closely followed by single-pass \ac{CoT} reasoning (82.2\%). This result suggests that structured reasoning accounts for the majority of the improvement over vanilla prompting, while priming predominantly enhances alignment and consistency. Long-term memory (81.9\%) and \ac{DRL}-based routing (78.3\%) further improve performance by introducing temporal stability and degradation-aware model selection.

When combined, component gains are structured rather than merely cumulative. The \ac{CoT}+Mem+\ac{DRL} configuration reaches 84.9\%, demonstrating that structured reasoning with temporal and policy awareness forms a stronger orchestration backbone. Adding priming to Mem+\ac{DRL} increases performance to 86.0\%, and the full configuration achieves 87.3\%, affirming that priming strongly refines the pipeline's structure.

Performance gains become clearer as modality complexity increases, with the largest improvements observed under triple- and quadruple-modality inputs. This trend indicates that memory and policy-based routing are critical as sensor diversity grows. These results demonstrate that reasoning structure and orchestration design drive performance more than parameter scale alone: \ac{CoT} establishes structured inference, priming improves alignment, and memory with \ac{DRL} enables stable multi-modal coordination under dynamic sensing conditions.
}


\subsubsection{Edge-Cases}
The edge-case study in Table \ref{tab: ablation_study_enwar3} displays \textsc{Enwar 3.0}’s capacity to operate reliably under various degraded sensing conditions. In scenarios ranging from dense fog to GPS jamming, the system consistently produced coherent, contextually appropriate decisions by dynamically adjusting its model selection, fallback strategies, and handover logic based on real-time sensor status. In each case, the final action aligned with the system's internal judgment, which shows robust coordination between the environment classifier, agent manager, and decision agents. These results affirm that \textsc{Enwar 3.0} is situation-aware when enforcing practical actions. 

\begin{table*}
\caption{Enwar 3.0 Edge-Cases}
\centering
\begin{tabular}{|c|c|c|c|c|c|}
\hline
\textbf{Scenario} & \textbf{\begin{tabular}[c]{@{}c@{}}Available\\ Modalities\end{tabular}} & \textbf{\begin{tabular}[c]{@{}c@{}}Beam\\ Accuracy\end{tabular}} & \textbf{\begin{tabular}[c]{@{}c@{}}Blockage\\ F1\end{tabular}} & \textbf{\begin{tabular}[c]{@{}c@{}}Enwar 3.0 \\Judgement\\\end{tabular}} & \textbf{\begin{tabular}[c]{@{}c@{}}Final Action\\ \end{tabular}} \\ \hline

\begin{tabular}[c]{@{}c@{}}Clear weather, \\ ideal scenario\end{tabular} & All & 88.5\% & 98.4\% & \begin{tabular}[c]{@{}c@{}}All sensors functional; \\ select best model\end{tabular} & \begin{tabular}[c]{@{}c@{}}Use best model fusion \\ for both agents\end{tabular} \\ \hline

\begin{tabular}[c]{@{}c@{}}Heavy fog, \\ visual occlusion with \\average degradation 42.8\% \end{tabular} & \begin{tabular}[c]{@{}c@{}}GPS, LiDAR, radar \\ (camera degraded)\end{tabular} & 83.0\% & 93.7\% & \begin{tabular}[c]{@{}c@{}}Camera sensor degraded; \\ fallback to radar-based \\fusion models\end{tabular} & \begin{tabular}[c]{@{}c@{}}Use radar model \\for blockage, and \\GPS-radar-LiDAR \\fusion for beam\end{tabular} \\ \hline

\begin{tabular}[c]{@{}c@{}}Bright sun glare with \\average degradation 31.0\%\end{tabular} & \begin{tabular}[c]{@{}c@{}}GPS, LiDAR,\\ radar\\ (camera occluded)\end{tabular} & 83.0\% & 93.7\% & \begin{tabular}[c]{@{}c@{}}Camera sensor degraded; \\ fallback to radar-based \\fusion models\end{tabular} & \begin{tabular}[c]{@{}c@{}}Use radar model \\for blockage, and \\GPS-radar-LiDAR \\fusion for beam\end{tabular} \\ \hline

\begin{tabular}[c]{@{}c@{}}Nighttime,\\ low visibility with \\average degradation 55.0\%\end{tabular} & \begin{tabular}[c]{@{}c@{}}GPS, LiDAR,\\ radar\\ (camera degraded)\end{tabular} & 83.0\% & 93.7\% & \begin{tabular}[c]{@{}c@{}}Camera sensor degraded; \\ fallback to radar-based \\fusion models\end{tabular} & \begin{tabular}[c]{@{}c@{}}Use radar model \\for blockage, and \\GPS-radar-LiDAR \\fusion for beam\end{tabular} \\ \hline

\begin{tabular}[c]{@{}c@{}}GPS jamming with \\average degradation 37.2\% \end{tabular} & \begin{tabular}[c]{@{}c@{}}Camera, LiDAR,\\ radar\end{tabular}& 88.3\% & 98.4\% & \begin{tabular}[c]{@{}c@{}}GPS signal unreliable; use \\ camera-LiDAR-radar \\models\end{tabular} & \begin{tabular}[c]{@{}c@{}}Use camera-radar\\ for blockage, and\\ camera-radar-LiDAR\\ for beam\end{tabular} \\ \hline

\begin{tabular}[c]{@{}c@{}}LiDAR sensor \\ non-functional with \\average degradation 59.4\%\end{tabular} & \begin{tabular}[c]{@{}c@{}}Camera, GPS,\\ radar \end{tabular} & 84.9\% & 98.4\% & \begin{tabular}[c]{@{}c@{}}LiDAR degraded;\\ fallback to radar-based\\ models\end{tabular} & \begin{tabular}[c]{@{}c@{}}Use to camera-radar\\ fusion for both agents\end{tabular}\\ \hline

\begin{tabular}[c]{@{}c@{}}Radar noisy \\ in rain with \\average degradation 39.9\%\end{tabular}& \begin{tabular}[c]{@{}c@{}}Camera, GPS, LiDAR \\ (radar noisy)\end{tabular} & 85.6\% & 98.1\% & \begin{tabular}[c]{@{}c@{}}Radar noise detected; \\ rely on camera-LiDAR\end{tabular} & \begin{tabular}[c]{@{}c@{}} Use camera-LiDAR\\ fusion for beam,\\ and camera model\\ for blockage\end{tabular} \\ \hline

\begin{tabular}[c]{@{}c@{}}Blockage longer than \\$t+5$ time steps confirmed,\\ BS$_2$ RSSI higher\end{tabular} & All & 88.5\% & 98.4\% & \begin{tabular}[c]{@{}c@{}}Blockage confirmed,\\ BS$_2$ power difference \\exceeds threshold\end{tabular} & \begin{tabular}[c]{@{}c@{}}Trigger handover to \\BS$_2$; use best \\fusion models \end{tabular}\\ \hline

\begin{tabular}[c]{@{}c@{}}Blockage longer than \\$t+5$ time steps confirmed,\\ Unit 1 RSSI higher\end{tabular} & All & 88.5\% & 98.4\% & \begin{tabular}[c]{@{}c@{}}Blockage confirmed,\\ BS$_2$ power difference \\does not exceed threshold\end{tabular} & \begin{tabular}[c]{@{}c@{}}Stay on Unit 1; \\ use best \\fusion models \end{tabular}\\ \hline

\begin{tabular}[c]{@{}c@{}}LiDAR and radar\\ non-functional with \\average degradation 63.7\%\end{tabular} & Camera, GPS & 83.7\% & 98.1\% & \begin{tabular}[c]{@{}c@{}}LiDAR and radar degraded; \\ use camera-gps fusion models \end{tabular} & \begin{tabular}[c]{@{}c@{}}Use camera model\\ for blockage,\\ camera-GPS for\\ beam\end{tabular} \\ \hline

\begin{tabular}[c]{@{}c@{}}LiDAR and GPS\\ non-functional with \\average degradation 68.2\%\end{tabular} & Camera, radar & 85.6\% & 98.1\% & \begin{tabular}[c]{@{}c@{}}LiDAR and GPS degraded; \\ use camera-radar fusion models \end{tabular} & \begin{tabular}[c]{@{}c@{}}Use camera model\\ for blockage,\\ camera-LiDAR for\\ beam\end{tabular} \\ \hline

\begin{tabular}[c]{@{}c@{}}Camera and GPS\\ non-functional with \\average degradation 71.1\%\end{tabular} & LiDAR, radar & 74.4\% & 93.7\% & \begin{tabular}[c]{@{}c@{}}LiDAR and radar degraded; \\ use camera-gps fusion models \end{tabular} & \begin{tabular}[c]{@{}c@{}}Use radar model\\ for blockage,\\ radar-LiDAR for\\ beam\end{tabular} \\ \hline

\begin{tabular}[c]{@{}c@{}}Camera, GPS, and LiDAR \\ non-functional with \\ average degradation 84.9\%\end{tabular} & Radar & 72.4\% & 93.7\% & \begin{tabular}[c]{@{}c@{}}Camera, GPS, and LiDAR degraded; \\ use radar-only models \end{tabular} & \begin{tabular}[c]{@{}c@{}}Use radar-only models\end{tabular} \\ \hline
\end{tabular}
\label{tab: ablation_study_enwar3}
\vspace{-0.25cm}
\end{table*}

\subsubsection{LLM Architecture and Size Comparisons}
\begin{figure}
    \centering
    \includegraphics[width=0.8\linewidth]{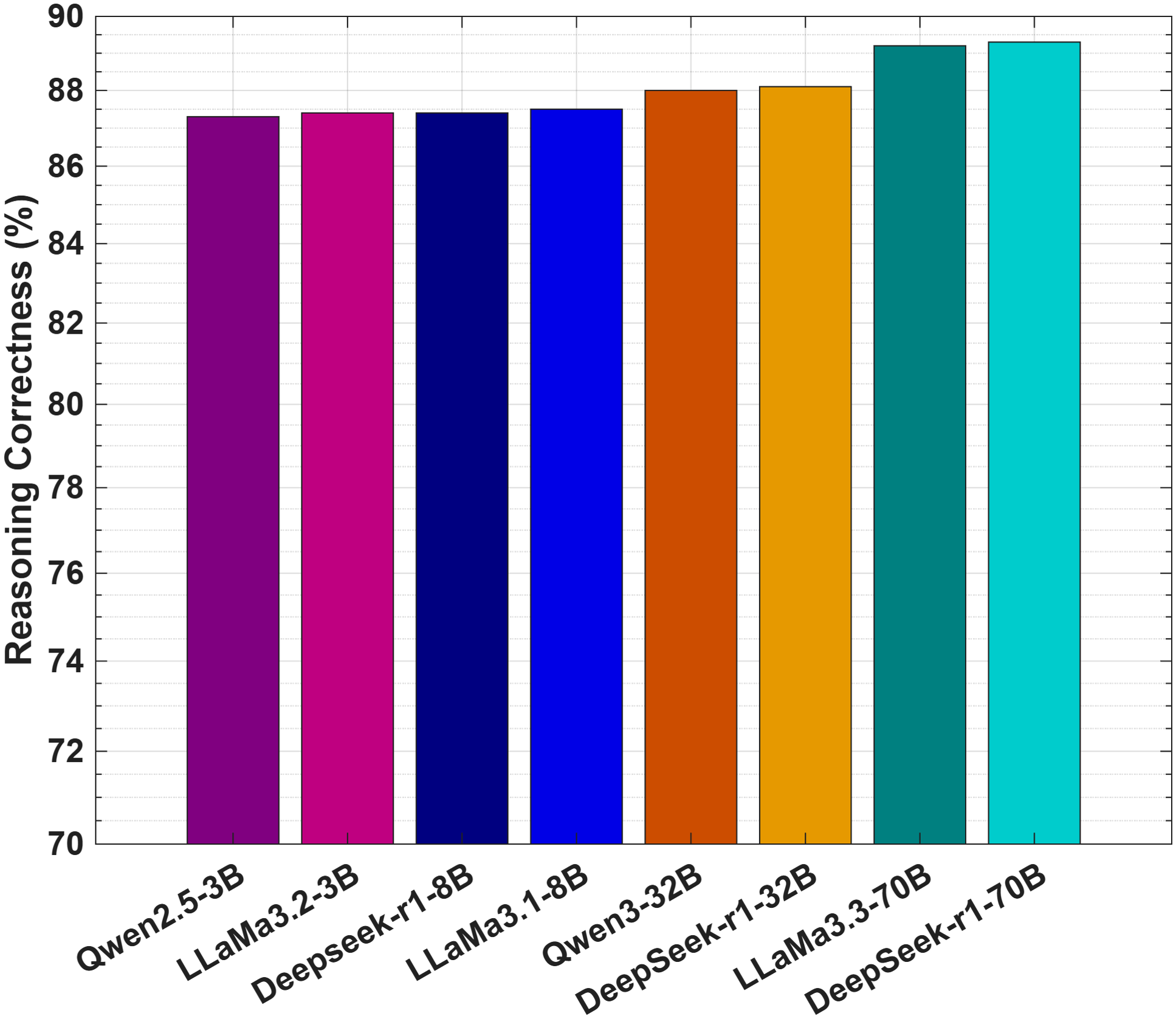}
    \vspace{-0.2cm}
    \caption{Reasoning correctness across different sized LLMs (3B-70B).}
    \label{fig:reasoning_correctness_best_modality}
    \vspace{-0.4cm}
\end{figure}

\begin{figure}
    \centering
    \includegraphics[width=0.8\linewidth]{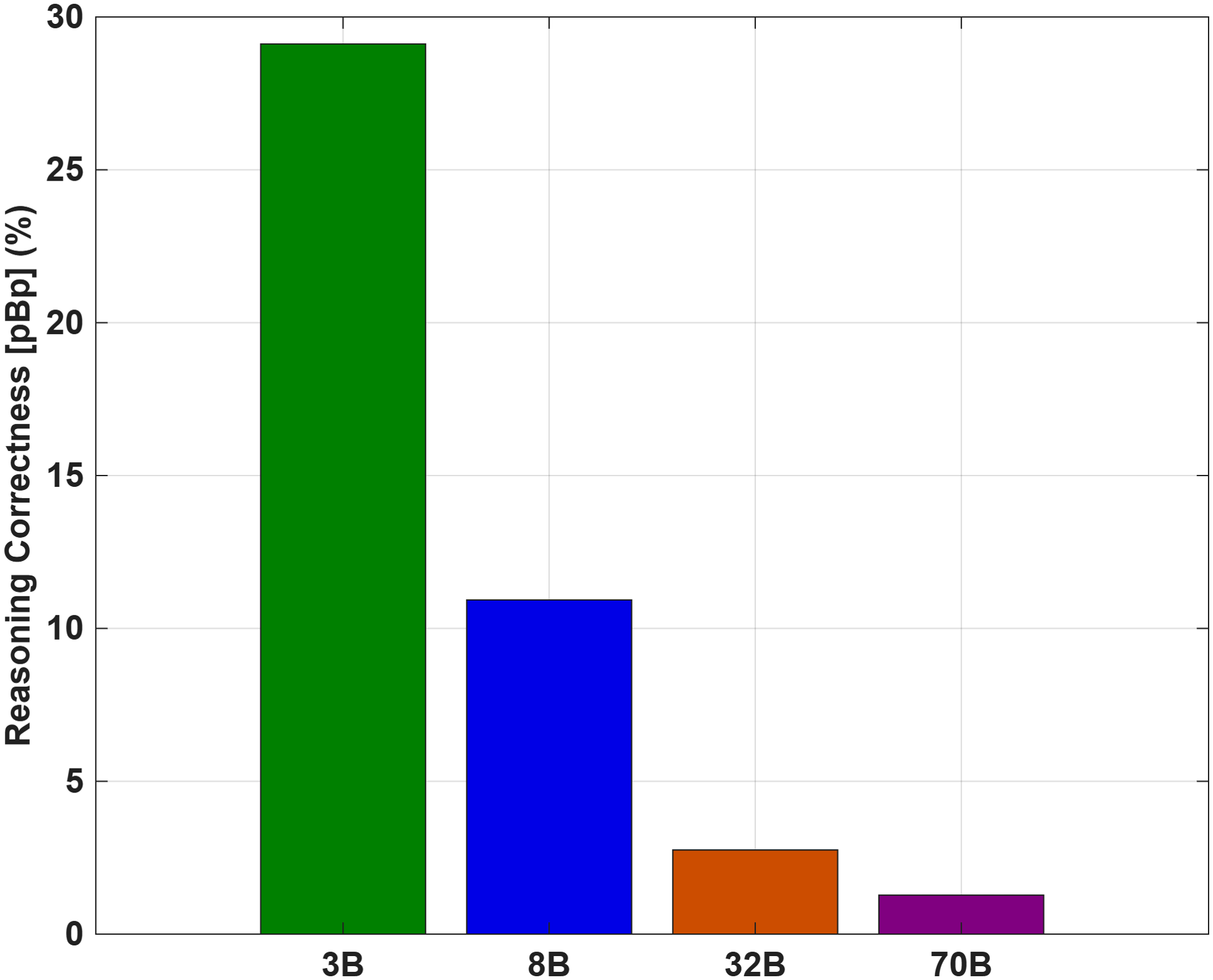}
    \caption{Reasoning correctness efficiency per billion parameters (pBp).}
    \label{fig:gains_per_llm_size}
    \vspace{-0.3cm}
\end{figure}

Reasoning correctness was evaluated across multiple \ac{LLM} architectures under full multi-modal input (Fig.~\ref{fig:reasoning_correctness_best_modality}). Although larger models achieve higher accuracy, marginal gains diminish with scale (Fig.~\ref{fig:gains_per_llm_size}). To quantify this trend, we compute average reasoning correctness across comparable model sizes (e.g., LLaMa 3.2-3B and Qwen 2.5-3B grouped as 3B).

The normalized gain per billion parameters (pBp) decreases sharply with scale: 3B models achieve an average gain of 29.1, compared to 10.9 (8B), 2.75 (32B), and 1.28 (70B). This fact confirms diminishing returns relative to parameter growth.

Inference latency scales similarly. 70B models require up to 3.08s per multi-agent prompt, compared to 2.17s for 3B models. While larger \acp{LLM} provide incremental modest improvements, smaller models offer a more favorable efficiency-performance tradeoff for real-world deployment.

\subsection{Computational and Deployment Considerations}
{
\textsc{Enwar 3.0} operates under a sensor sampling interval of $T_s\!$. We distinguish between the time-critical control path and asynchronous explanation generation. Upon receiving new sensor data, perception modules, the environment classifier, and the \ac{DRL} policy execute in parallel. The \ac{LLM} concurrently constructs a structured invocation packet specifying agent selection and network updates. 

As shown in Table~\ref{tab:model_times}, the worst-case structured invocation time for the primed 3B model is 222.7ms. Beam and blockage inference require at most 66.98ms and 59.47ms, respectively, under steady-state single-sample GPU execution. The resulting worst-case critical-path latency is bounded by approximately 290ms, remaining within the $T_s\!=\!300$ms sensing interval. \ac{LLM} response generation (2.17-3.08s depending on model size) is fully decoupled from the control loop and runs asynchronously for justification and memory updates. Architectural optimizations, including 4-bit quantization, LoRA fine-tuning, GPU affinity scheduling, caching, and asynchronous execution, ensure bounded latency under multi-modal load. Additional computational details are provided in App.~\ref{appendix:computation}.}

\begin{table}[t]
    \centering 
    \caption{\textsc{Enwar 3.0} Key Time Indicators.}
    \begin{tabular}{|c|c|} 
        \hline 
        \textbf{Process} & \textbf{Time}  \\ \hline 
        \begin{tabular}[c]{@{}c@{}}\textbf{Critical-Path End-to-End} \\ \textbf{(Worst Case)}\end{tabular} & \textbf{289.7ms} \\ \hline
        Input Sampling Period ($T_s$)& 300ms \\ \hline
        Input Sliding Window ($T_s$)& 1.5s \\ \hline
        Primed 70B LLMs response generation & 3.08s \\ \hline
        Vanilla 3B LLMs with prompt template & 2.90s \\ response generation & \\ \hline
        Primed 32B LLMs response generation & 2.45s \\ \hline
        Llama3.2-Vision Image-To-Text (per frame) & 2.34s \\ \hline
        Primed 8B LLMs response generation & 2.23s \\ \hline
        Primed 3B LLMs response generation & 2.17s \\ \hline
        Primed 70B LLMs agent invocations & 242.6ms \\ and network updates (worst case) & \\ \hline
        Primed 32B LLMs agent invocations & 241.2ms \\ and network updates (worst case) & \\ \hline
        Vanilla 3B LLMs with prompt template & 618.8ms \\ agent invocations and network updates & \\ \hline
        Primed 8B LLMs agent invocations & 235.1ms \\ and network updates (worst case) & \\ \hline
        Primed 3B LLMs agent invocations & 222.7ms  \\ and network updates (worst case) & \\ \hline
        Beam Prediction Inference (worst case) & 66.98ms  \\ \hline 
        Blockage Prediction Inference (worst case) & 59.47ms  \\ \hline
        LiDAR DBSCAN & 14.7ms  \\ \hline 
        Radar DBSCAN & 12.4ms \\ \hline 
        DRL Agent Inference & 2ms \\ \hline 
        YOLO per frame camera detections & 5ms \\ \hline 
        Environment Classifier Inference & 3ms \\ \hline 
        Handover Agent Inference & 3ms \\ \hline 
        GPS detections & $< 1$ms  \\ \hline 
    \end{tabular}
    \label{tab:model_times}
\end{table}



\section{Conclusion}
\label{sec: conclusion}
This work presents \textsc{Enwar 3.0}, a multi-agent, environment-aware reasoning framework that advances the robustness and interpretability of mmWave \ac{I2V} communication systems. \textsc{Enwar 3.0} enables contextual decision-making for beamforming, blockage prediction, handover management, and environment perception by combining real-time multi-modal sensing, a sensor degradation-aware classifier, structured prompt processing, and primed \ac{LLM} reasoning with specialized agents.

\section*{Acknowledgments}
ChatGPT \cite{chatgpt} was used to assist with organizing and refining language flow for consistency in tone and terminology.



\bibliographystyle{IEEEtran}
\bibliography{bibliography}

\begin{IEEEbiography}[\frame{\includegraphics[width=1.0in,height=1.25in,clip]{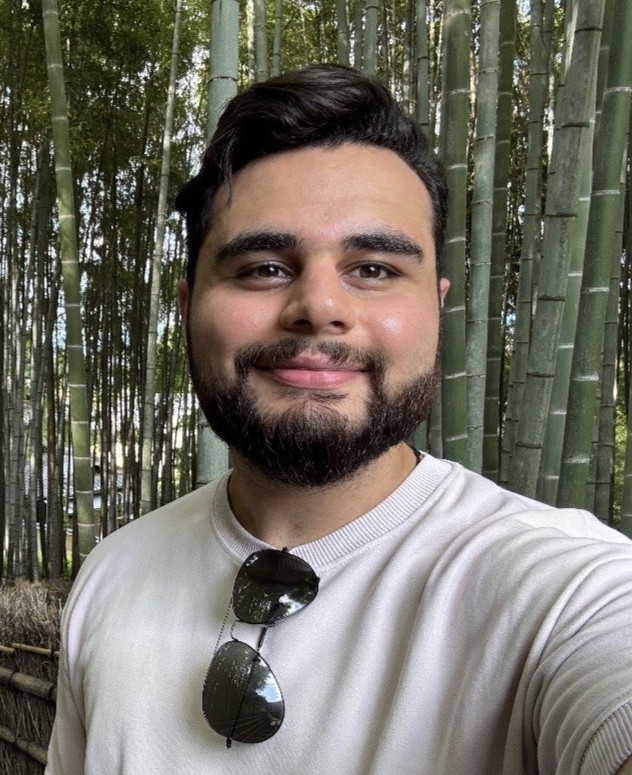}}]{Ahmad M. Nazar} (Member, IEEE) 
 is a Postdoctoral Scholar in the Department of Electrical and Computer Engineer at Iowa State University (ISU), USA and a Lead Research Engineer at Gladiolus Technological Institute, USA. He received a Ph.D. degree in Computer Engineering from ISU, in 2025; and he also earned M.S. and B.S degrees in Computer Engineering in 2022 and 2020 from ISU. His research interests involve interdisciplinary applications of generative AI and machine learning with a focus on LLMs and  multi-modal AI agents.
\end{IEEEbiography}

\vskip 0pt plus -1fil

\begin{IEEEbiography}[\frame{\includegraphics[width=1.0in,height=1.25in,clip]{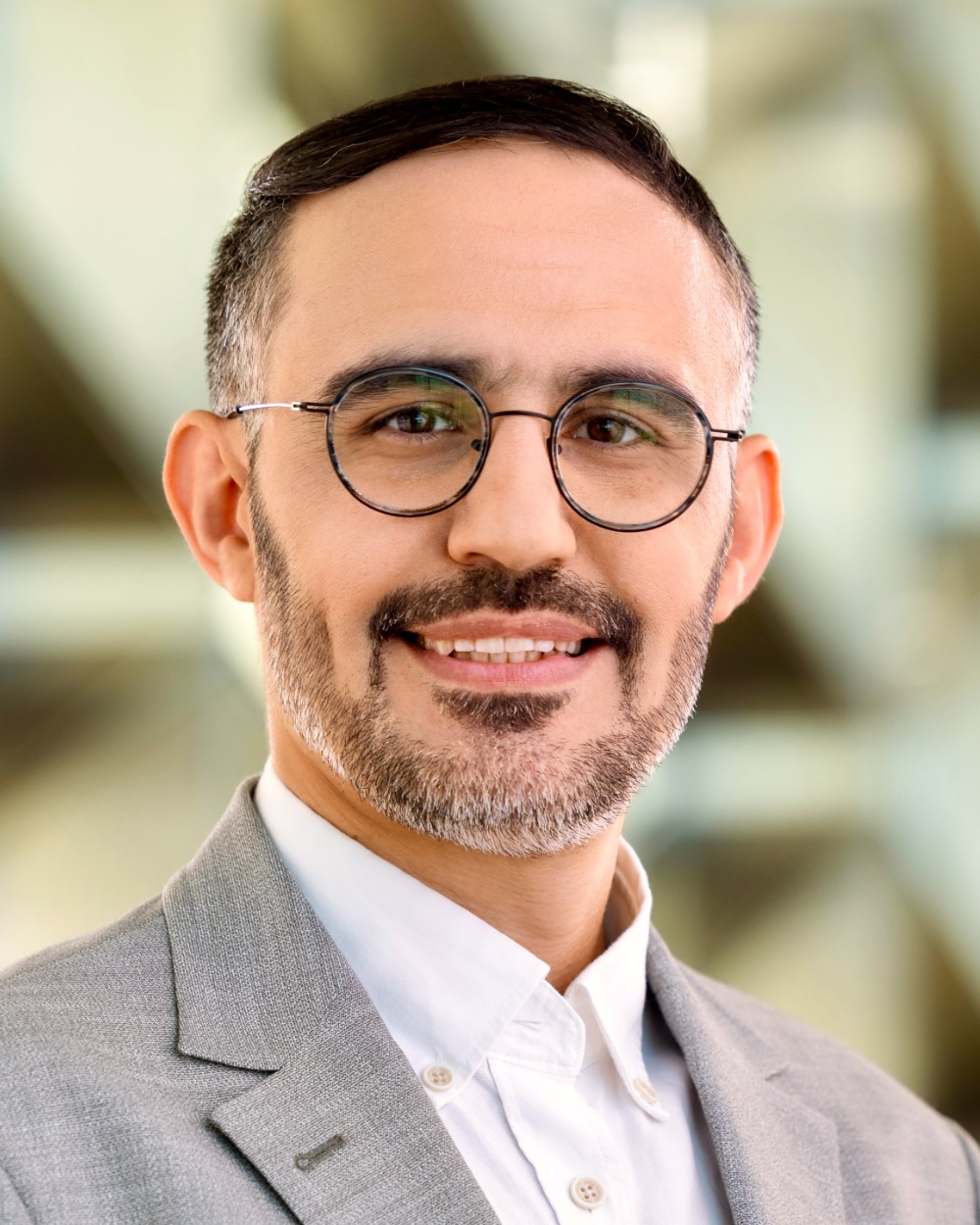}}]{Abdulkadir Celik} (Senior Member, IEEE) 
 is an Associate Professor in the School of Electronics and Computer Science at the University of Southampton, UK, where he also serves as the Director of the Centre for Internet of Things and Pervasive Systems. He received the Ph.D. degree in co-majors of Electrical Engineering and Computer Engineering from Iowa State University, Ames, IA, USA, in 2016; wherein he also earned M.S. degrees in Electrical Engineering and Computer Engineering in 2013 and 2015, respectively. Prior to his current appointment, he was a senior research scientist from 2020 to 2025 and a post-doctoral fellow from 2016 to 2020 at King Abdullah University of Science and Technology (KAUST), Thuwal, KSA. Dr. Celik is the recipient of IEEE Communications Society’s 2023 Outstanding Young Researcher Award for Europe, Middle East, and Africa (EMEA) region. He currently serves as an editor for npj Wireless Technology, IEEE Transactions on Communications, IEEE Communications Letters, IEEE Wireless Communication Letters, and Frontiers in Communications and Networks. His research interests are in the broad areas of next-generation wireless communication systems and networks.
\end{IEEEbiography}

\vskip 0pt plus -1fil

\begin{IEEEbiography}[\frame{\includegraphics[width=1.0in,height=1.25in,clip, keepaspectratio]{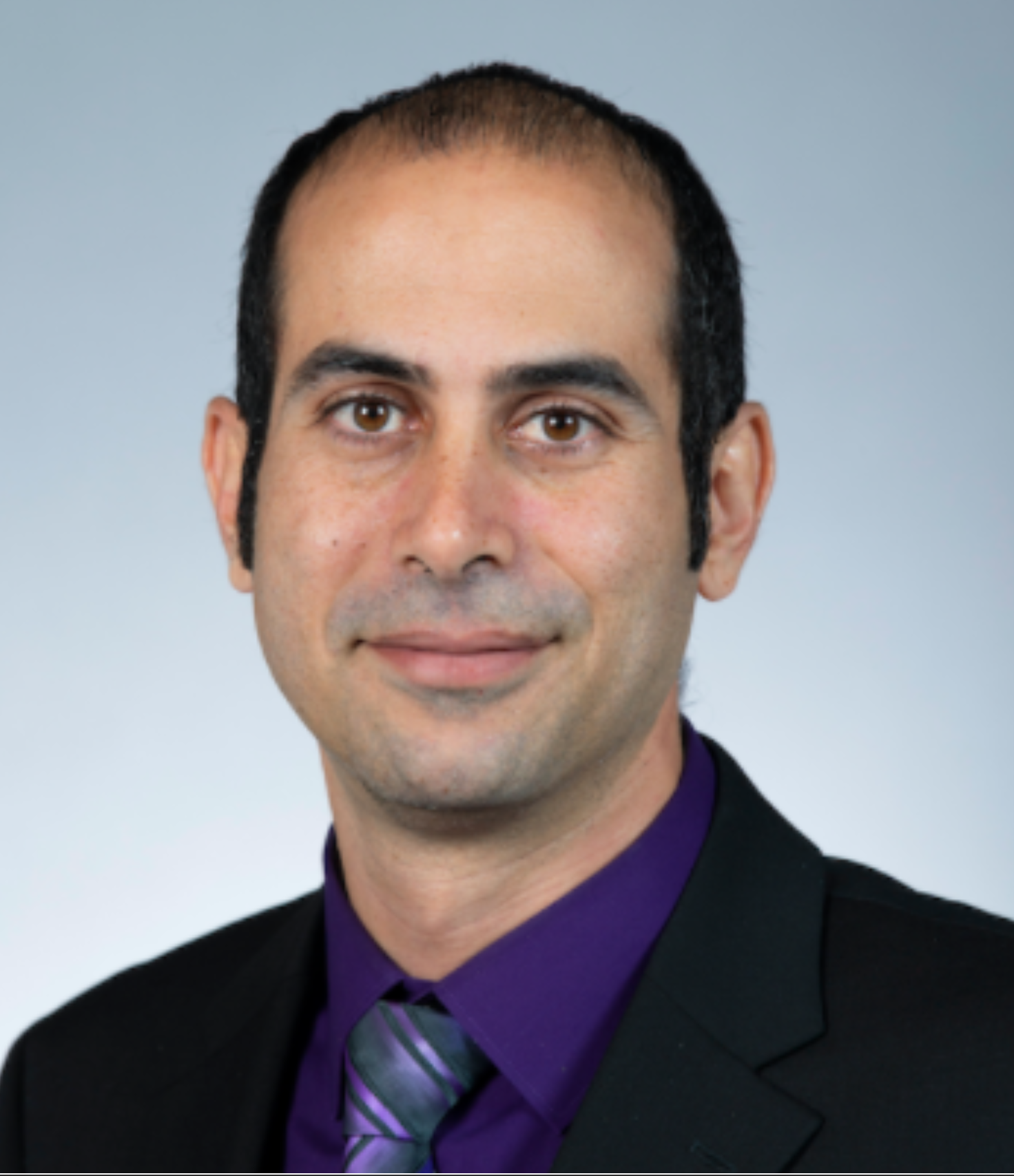}}]{Mohamed Y. Selim} (Senior Member, IEEE) 
is an Associate Teaching Professor in the Department of Electrical and Computer Engineering at Iowa State University, USA. He received the Ph.D. and M.Sc. degrees in computer engineering from Iowa State University in 2018 and 2016, respectively, and the M.Sc. degree in electrical engineering from Port Said University, Egypt, in 2011. He has co-authored numerous papers in leading venues such as IEEE Communications Magazine, IEEE Transactions on Mobile Computing, and Computer Networks, and contributed book chapters on 5G and UAV-based RIS systems. Dr. Selim has received multiple teaching and educational impact awards at Iowa State University and has been recognized as an Exemplary Reviewer for IEEE Communications Letters (2023) and a Distinguished Reviewer for IEEE Transactions on Mobile Computing (2024). Dr. Selim’s research interests span next-generation wireless communication systems, large-scale testbeds, reconfigurable intelligent surfaces, and agentic LLMs for AI-native wireless networks. He is a co-PI of the \$8 million NSF-funded ARA PAWR project.
\end{IEEEbiography}

\vskip 0pt plus -1fil

\begin{IEEEbiography}[\frame{\includegraphics[width=1.0in,height=1.25in,clip,keepaspectratio]{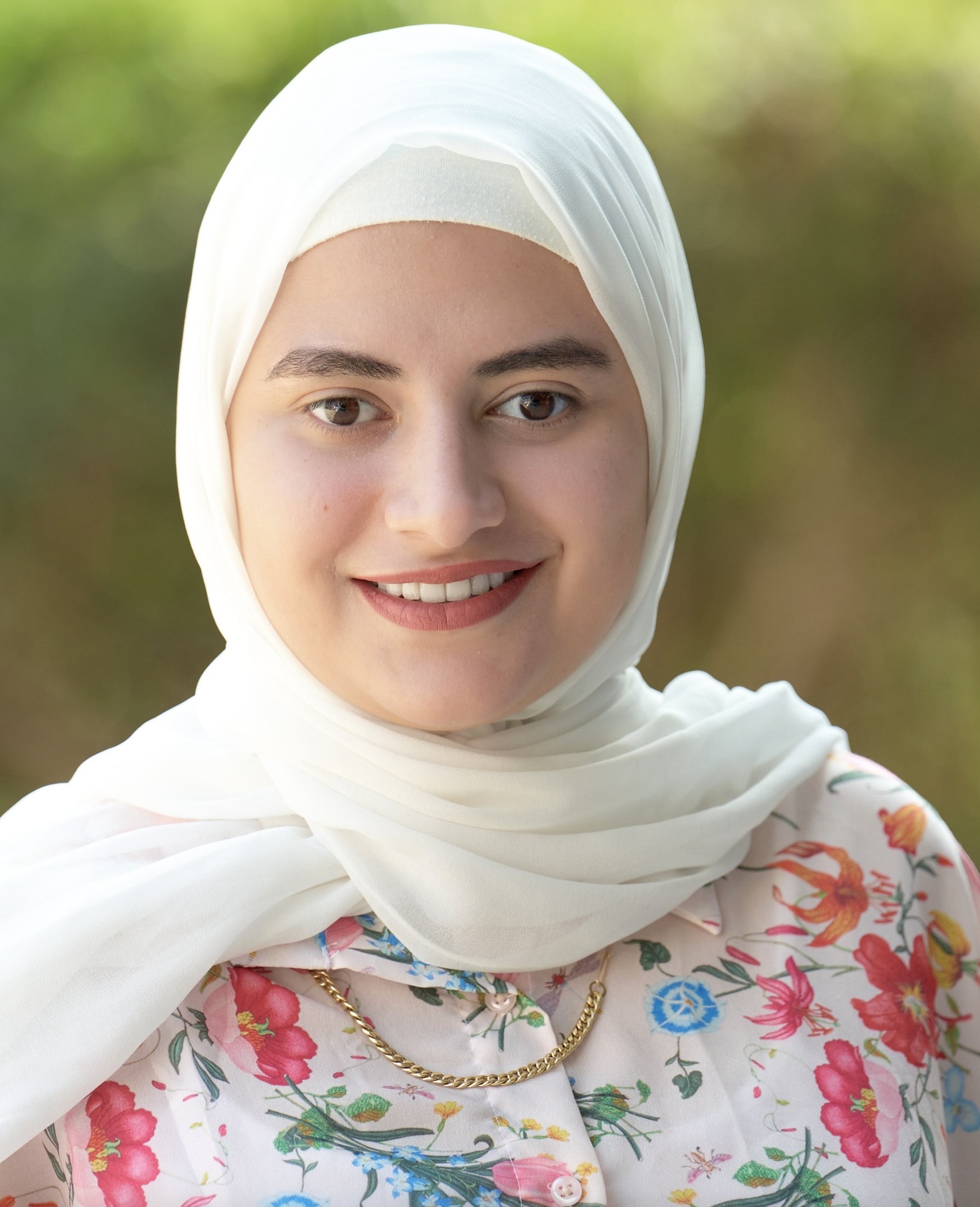}}]{Asmaa Abdallah} (Member,
IEEE)  received the B.S. (with High Distinction) and M.S degree in computer and communications engineering from Rafik Hariri University (RHU), Lebanon, in 2013 and 2015, respectively. In 2020, she received the Ph.D. degree in electrical and computer engineering at the American University of Beirut (AUB), Beirut, Lebanon. She was a Postdoctoral Fellow at King Abdullah University of Science and Technology (KAUST), from 2021-2024, where she is currently a Research Scientist with the Communications
and Computing Systems Laboratory.
 From 2016 to 2020, She has been a member of the executive committee of IEEE Young Professionals Lebanon’s Section. Dr. Abdallah was the recipient of the Academic Excellence Award at RHU in 2013 for ranking first on the graduating class. She also received a scholarship from the Lebanese National Counsel for Scientific Research (CNRS-L/AUB) to support her doctoral studies. In 2023, Dr. Abdallah was selected by MIT technology review as one of the leading 15 Innovators under 35 in the MENA area. Her research interests include machine learning, communication theory, stochastic geometry, array signal processing, with emphasis on energy and spectral efficient algorithms for next-generation wireless communication systems.
\end{IEEEbiography}

\vskip 0pt plus -1fil

\begin{IEEEbiography}[\frame{\includegraphics[width=1.0in,height=1.25in,clip]{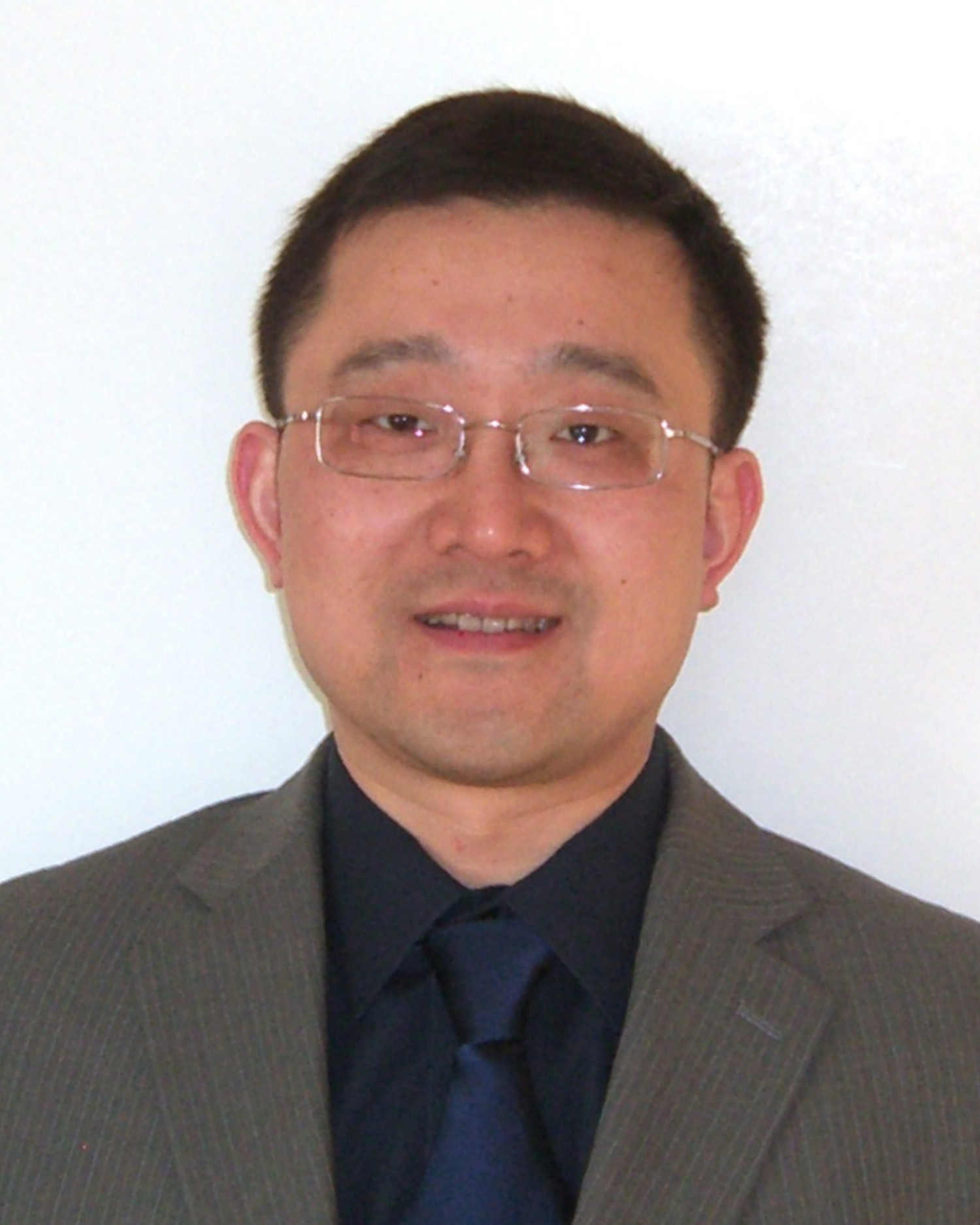}}]{Daji Qiao} (Senior Member, IEEE) is a Professor in the Department of Electrical and Computer Engineering at Iowa State University. He received his Ph.D. from the University of Michigan, Ann Arbor. His research focuses on wireless networking and mobile computing, 5G/6G systems, sensor networks, and IoT. He is a Senior Member of the IEEE and a Member of the ACM.

\end{IEEEbiography}

\vskip 0pt plus -1fil

\begin{IEEEbiography}[\frame{\includegraphics[width=1.0in,height=1.25in,clip,keepaspectratio]{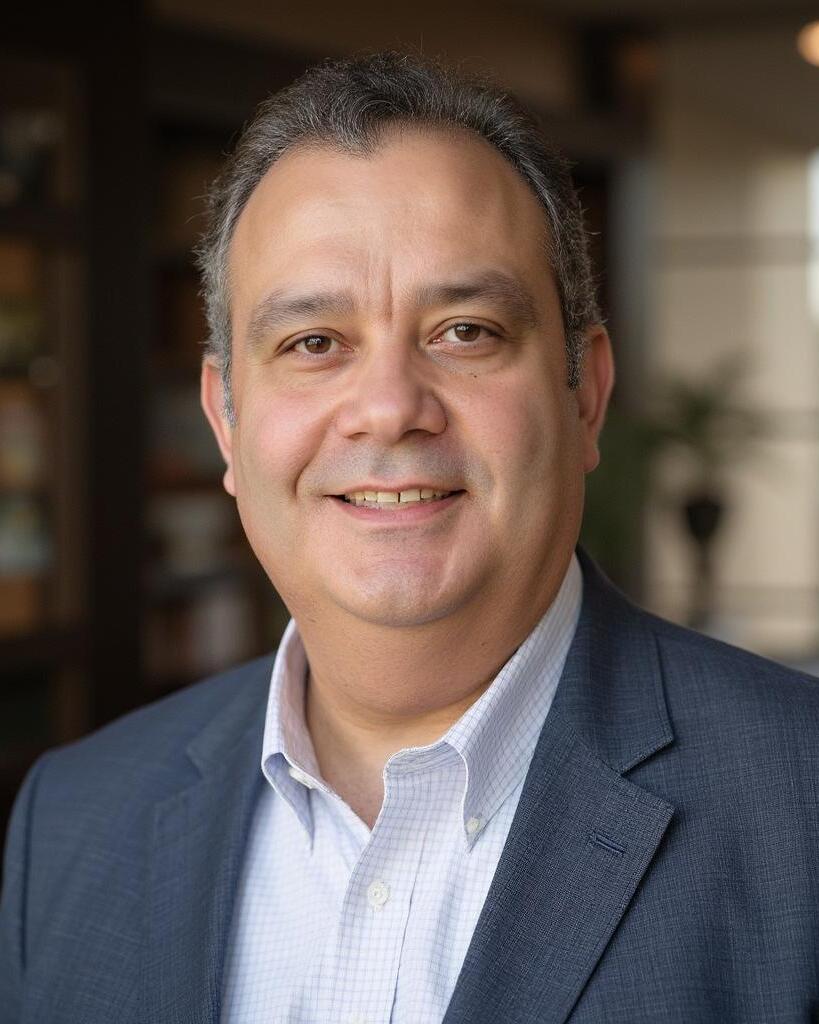}}]{Ahmed M. Eltawil} (Senior Member, IEEE) is a Professor and Associate Dean for Research at the Computer, Electrical, and Mathematical Sciences and Engineering (CEMSE) Division at King Abdullah University of Science and Technology (KAUST). Previously, he was a Professor of Electrical Engineering and Computer Science at the University of California, Irvine (UCI) from 2005 to 2021. Professor Eltawil earned his doctorate degree from the University of California, Los Angeles, in 2003 and his Master’s and Bachelor’s degrees from Cairo University in 1999 and 1997, respectively. At KAUST, he established the Communication and Computing Systems Laboratory (CCSL) to conduct research on efficient architectures for computing and communications systems, with a focus on mobile wireless systems. His research interests encompass various application domains, such as low-power mobile systems, machine learning platforms, sensor networks, body area networks, and critical infrastructure networks. He served as a distinguished lecturer for \textsc{IEEE} COMSOC during the 2023/24 term. Additionally, he holds senior membership in \textsc{IEEE} and the National Academy of Inventors in the United States. He received several recognitions and awards, including the US National Science Foundation CAREER award, the 2021 "Innovator of the Year" award by the Henry Samueli School of Engineering at the University of California, Irvine, and two United States Congress certificates of merit, among other recognitions. He has served in numerous editorial roles over the years, as well as an expert reviewer for national and international funding agencies and review boards.
\end{IEEEbiography}

\newpage
\appendices

\section{Tabular Summary of Related Works}
\label{appendix: related_work_summary}
This section summarizes the related work discussed in Section~\ref{sec: related_work}, with details provided in Table~\ref{tab:app_summary1}.
\vspace{-0.35cm}
\begin{table}[h]
\centering
\caption{Related works and their summaries}
\vspace{-0.1cm}
\begin{tabular}{|c|l|}
\hline
Reference & Summary \\ \hline
\begin{tabular}[c]{@{}c@{}}\cite{zou2023wireless, jiang2024large, tong2025wirelessagentlargelanguagemodel}\\\cite{plaat2025agenticlargelanguagemodels, netOrchLLM} \end{tabular}& \begin{tabular}[l]{@{}l@{}}Multi-agent task-solving LLM frameworks for 6G\\ communications with retrieval, planning, evaluation.\end{tabular}\\ \hline
\cite{llama3, deepseek, qwen} & LLMs used to evaluate \textsc{Enwar 3.0} \\ \hline 
\begin{tabular}[c]{@{}c@{}}\cite{xu2024large} \end{tabular} & \begin{tabular}[l]{@{}l@{}}Perspectives on large multi-modal models with  causal\\ reasoning and neuro-symbolic AI for 6G Networks \end{tabular} \\ \hline
\begin{tabular}[c]{@{}c@{}}\cite{yilma2024telecomrag, nazar2025nextggptleveraginggenaiadvancing, zou2024telecomgpt}\end{tabular} & \begin{tabular}[l]{@{}l@{}}Introduces RAG LLM frameworks for wireless systems\\ within domain-specific datasets for context-aware,\\ real-time support in network and telecom domains. \end{tabular}\\ \hline
\begin{tabular}[c]{@{}c@{}}\cite{shen2024large, shao2024wirelessLLM, 6GLLM, yang2025wirelessgptgenerativepretrainedmultitask}\end{tabular} & \begin{tabular}[l]{@{}l@{}}Integration of LLMs, GPTs, and AI in 6G architectures\\ for intent-driven, intelligent network operations. \end{tabular}\\ \hline 
\cite{llava} & LLaVa vision-language model.\\ \hline 
\cite{zhang2024visionlanguagemodelsvisiontasks} & Survey on vision-language models. \\\hline 
\begin{tabular}[c]{@{}c@{}}\cite{blockage2, blockage3, abdallah2024multiagentbeamtraining, beam_lstm, alrabeiah2019millimeterwavebasestations, deepsense3, blockage1}\end{tabular} & \begin{tabular}[l]{@{}l@{}}Machine learning methods incorporating sensor-based\\ blockage/beam prediction. \end{tabular}\\ \hline
\begin{tabular}[c]{@{}c@{}}\cite{liu2025wirelessagenticairetrievalaugmented, Du_2025}\end{tabular}& \begin{tabular}[l]{@{}l@{}}Task-oriented semantic communication using large\\ multi-modal model, agents, and RAG for efficient\\ bandwidth data exchange in vehicular environments. \end{tabular} \\ \hline 
\begin{tabular}[c]{@{}c@{}}\cite{Nasir2024xai, khanexplainableICC25, khanTCOM}\end{tabular} & \begin{tabular}[l]{@{}l@{}}Frameworks for explainable and robust AI solutions\\ in 6G networks with machine learning approaches.\end{tabular} \\ \hline
\cite{alikhani2025largewirelessmodellwm} & \begin{tabular}[l]{@{}l@{}}Large Wireless Model, a fine-tuned LLM for \\wireless communication-based solutions.\end{tabular}\\ \hline
\begin{tabular}[c]{@{}c@{}}\cite{ding2025newparadigmusercentricwireless}\end{tabular} & \begin{tabular}[l]{@{}l@{}}LLM framework transforming user requests into \\intent-focused structured optimization tasks/queries for\\ real-time wireless semantic communication systems. \end{tabular}\\ \hline
\begin{tabular}[c]{@{}c@{}}\cite{wei2022chain} \end{tabular}&\begin{tabular}[l]{@{}l@{}} Chain-of-Thought prompting techniques and how\\ it improves reasoning in LLMs. \end{tabular}\\ \hline 
\begin{tabular}[c]{@{}c@{}}\cite{brown2020language}\end{tabular} &\begin{tabular}[l]{@{}l@{}} Few-shot learning methods and its effects on LLMs. \end{tabular}\\ \hline 
\cite{schulman2017proximalpolicyoptimizationalgorithms} & Deep-learning based PPO algorithm.\\ \hline
\cite{DeepSense1, DeepSense2} & I2V dataset used within \textsc{Enwar 3.0}.\\ \hline
\begin{tabular}[c]{@{}c@{}}\cite{pointnet, voxel_lidar_ransac} \end{tabular}& \begin{tabular}[l]{@{}l@{}}LiDAR preprocessing, and PointNet architecture. \end{tabular}\\ 
\hline 
\end{tabular}
\label{tab:app_summary1}
\vspace{-0.63cm}
\end{table}
\section{LLM Priming}
This section presents the prompt template used for LLM priming, along with a three-iteration example with reward-guided human feedback and iterative response refinement.
\vspace{-0.35cm}
\subsection{Main Priming Template Prompt}
\label{appendix:priming_template}
\vspace{5pt}
\hrule
\vspace{1pt}
\hrule
\vspace{5pt}
\noindent \textbf{Priming Template to Generate Final LLM Response}
\begin{itemize}
    \item \textit{Environment Status:} current environment conditions
    \item \textit{Modality Selections Based on Environment Status:} what is the current degradation level of modalities.
    \item \textit{DRL Agent Selection:} The selected models based on the DRL agent's output.
    \item \textit{Trajectory:} include the trajectory of Unit 2 from the perspective of the selected BS.
    \item \textit{Blockage Status:} whether there is a blockage in this sample or not.
    \item \textit{Handover Status:} what is the handover status and its properties, such as the power differences if a handover occurred and which BS has the highest power.
    \item \textit{Predicted Beam and Properties:} Current beam's predicted power and its properties.
    \item \textit{Perception:} Provide a summary of the perceived environment using the available perceived information.
\end{itemize}
\vspace{5pt}
\hrule
\vspace{1pt}
\hrule
\vspace{-0.2cm}
\subsection{Priming Reward Model Pseudocode}
\label{appendix:priming_code}
The following section details the priming phase's pseudocode (Algorithm \ref{appendix:priming_pseudocode_listing}) with human-in-the-loop feedback.
\vspace{-0.3cm}


    
        
        
\begin{algorithm}
\label{appendix:priming_pseudocode_listing}
\caption{LLM Priming w/ RLHF Loop}
\KwIn{Priming examples $(s_p^i)_{i=1}^N$ where $s_p$ includes sensor conditions, DRL selections, memory context}
\KwOut{Primed LLM that consistently generates $a_p$ with $R(s_p, a_p) > \tau_{\text{reward}}$}
\textbf{Initialize:} Reward rubric $\theta$, threshold $\tau_{\text{reward}}$, max iterations $J$\;
\ForEach{example $s_p^i$}{
    $j \gets 0$\;
    \Repeat{$R(s_p^i, a_p^j) \leq \tau_{\text{reward}}$ \textbf{ and } $j \leq J$}{
        $a_p^j \gets \texttt{LLM.generate}(s_p^i)$\;
        \tcp{Expert applies fixed rubric}
        $r^j \gets R_{\text{priming},\theta}(s_p^i, a_p^j)$\;
        \If{$r^j \leq \tau_{\text{reward}}$}{
            \tcp{Add few-shot or clarify}
            $\texttt{ModifyPrompt}(s_p^i, \text{Feedback})$\;
        }
        $j \gets j + 1$\;
    }
}
\Return Primed LLM\;
\end{algorithm}
\vspace{-0.75cm}
\subsection{Three-Iteration Priming Example}
\label{appendix:priming_example}
This section shows a priming example with reward-guided human feedback and iterative response refinement.

\vspace{5pt}
\hrule
\vspace{1pt}
\hrule
\vspace{5pt}
\noindent \textbf{Prompt 1:}
Perceive the environment using the available modalities based on your sense of the environment status. Predict the beam selection for the sample, and provide its properties. Check the blockage status and determine whether a handover is necessary. You will be scored on a scale of 0-10 based on your correctness, justification of agentic choices, and explanation clarity. Your reasoning skills will also be scored and you will be penalized for repeated agent invocations.
\vspace{5pt}
\hrule
\vspace{1pt}
\hrule
\vspace{5pt}
\noindent \textbf{Internal Agent Context:}
\begin{itemize}
    \item Modality Status: Radar degraded (42.1\%)
    \item Average degradation of remaining modalities: 25.9\%
    \item Scene Data: Urban intersection, 3 vehicles, stop sign.
    \item Detected objects: LiDAR clusters (6–15 m): 12 detected, camera confirms these are vehicles with some pedestrians near a stop sign with multiple bounding boxes. Multiple nearby objects exist that may affect beam propagation.
    \item DRL Selection: camera\_gps\_lidar model for beam, camera\_only for blockage
    \item Camera: Stable visual feed with clear blockage detected.
    \item Handover Status: blockage threshold conditions satisfied, and BS$_2$ has a higher beam power (0.1492) than Unit 1 (0.1267), therefore handover to BS$_2$.
    \item Unit 2's bearing to the current BS (Unit 1) is 52.92°. 
    \item Blockage Status: blockage for $\!>\!1.5$s (probability: 0.9835). 
    \item BS$_2$ selected beam is beam 38 (centered at 50.86°, width 5.118°, power 0.14928), Unit 1's selected beam is beam 40 (centered at 51.89°, width 5.997°, power 0.1226)
\end{itemize}
In this response, blockage prediction and handover agents were invoked twice, beam prediction and perception agents once.
\vspace{5pt}
\hrule
\vspace{1pt}
\hrule
\vspace{5pt}
\noindent \textbf{Response 1:}
Environment Status: Urban intersection during daytime with active traffic and a stop sign. Scene is complex with multiple vehicle clusters and potential occlusions.

Modality Selections Based on Environment Status: Radar degraded and excluded from decision-making. Remaining modalities within acceptable degradation threshold.

Blockage Model Selected: \textbf{Camera-only model}

Justification:
\begin{itemize}
    \item Radar excluded due to high degradation (42.1\%)
    \item Camera provides clear object detection at the stop sign
    \item LiDAR confirms spatial placement of vehicles but lacks semantic understanding
    \item DRL policy aligns with model selection
    \item Camera preferred for semantic tasks in urban settings.
\end{itemize}

Recommendation: Use camera-only for blockage, multi-modal fusion if degradation $\!>\!30\%$ in future samples.
\vspace{5pt}
\hrule
\vspace{1pt}
\hrule
\vspace{5pt}
\noindent \textbf{Prompt 2:}
Ensure that handover justification, and more perceived objects and their details are included. Do not call any agent more than once. Score for previous response 2.4.
\vspace{5pt}
\hrule
\vspace{1pt}
\hrule
\vspace{5pt}
\noindent \textbf{Response 2:}
Detected objects: LiDAR clusters (6-15 m): 12 detected, camera confirms these are vehicles with some pedestrians nearby a stop sign with bounding box centers ranging from (156.3, 45.1) to (240.1, 138.9).
Handover Status: blockage threshold conditions satisfied, and BS$_2$ has a higher beam power (0.1492) than Unit 1 (0.1267), handover to BS$_2$.
\vspace{5pt}
\hrule
\vspace{1pt}
\hrule
\vspace{5pt}
\noindent \textbf{Prompt 3:}
Your final answer should follow this format:
\begin{itemize}
    \item \textit{Environment Status:} current environment conditions
    \item \textit{Modality Selections Based on Environment Status:} what is the current degradation level of modalities.
    \item \textit{DRL Agent Selection:} The selected models based on the DRL agent's output.
    \item \textit{Trajectory:} include the trajectory of Unit 2 from the perspective of the selected BS.
    \item \textit{Blockage Status:} whether there is a blockage in this sample or not.
    \item \textit{Handover Status:} what is the handover status and its properties, such as the power differences if a handover occurred and which BS has the highest power.
    \item \textit{Predicted Beam and Properties:} Current beam's predicted power and its properties.
    \item \textit{Perception:} Provide a summary of the perceived environment using the available perceived information.
\end{itemize}
Score for previous response 7.3.
\vspace{5pt}
\hrule
\vspace{1pt}
\hrule
\vspace{5pt}
\noindent \textbf{Response 3:}
\begin{itemize}
    \item \textit{Environment Status:} The setting appears to be a relatively clear daytime urban intersection. LiDAR and camera data indicate three vehicles present, with one vehicle approaching from a new direction and obstructing another at a stop sign. This dynamic scene suggests partial occlusions along the line of sight. Radar is degraded and excluded it is not expected to select radar-based models.
    \item \textit{Modality Selections Based on Environment Status:} Radar is degraded at 42.1\% with noisy scans, exceeding the reliability threshold. All other modalities are usable (avg. degradation 25.9\%). 
    \item \textit{DRL Agent Selection:} Use camera\_gps\_lidar for beam prediction and camera\_only for blockage detection.
    \item \textit{Trajectory:} Unit 2 is traveling at a bearing of 52.92° relative to the current BS (Unit 1). Multiple LiDAR clusters (12 total) are detected within 6-15 meters of Unit 1; nearby objects may affect beam propagation.
    \item \textit{Blockage Status:} Confirmed with probability 0.9835. Blockage has persisted for longer than 1.5 seconds, satisfying the blockage condition for a handover trigger.
    \item \textit{Handover Status:} A handover to BS$_2$ is recommended. BS$_2$’s beam 38 (centered at 50.86°, width 5.118°, power 0.14928) offers significantly better performance than Unit 1’s beam 40 (centered at 51.89°, width 5.997°, power 0.1226). BS$_2$'s beam sufficiently covers the current vehicle trajectory of 52.92°, and aligns better with the power and direction requirements for maintaining a robust link.
    \item \textit{Predicted Beam and Properties:} Best beam is beam 38 from BS$_2$, with center 50.86°, width 5.118°, and power 0.14928, covering the vehicle’s trajectory and surpassing Unit 1’s selected beam in strength and alignment.
    \item \textit{Perception:} Radar is excluded due to degradation. LiDAR clusters and camera inputs indicate three vehicles in the scene, with one vehicle creating a new occlusion pattern not present in earlier frames. This blockage likely targets Unit 2 directly, supported by GPS trajectory, object distance, and bearing estimation. Two cars seem to be overlapping each other. This output confirms the result of the blockage prediction model.

To summarize, handover to BS$_2$ and use beam 38. The beam’s center and width cover the vehicle's 52.92° trajectory. Maintain blockage monitoring through camera-only models until radar recovers. Once radar recovers below degradation thresholds, reintroduce radar-based fusion for better temporal consistency and backup perception.

\end{itemize}
\vspace{5pt}
\hrule
\vspace{1pt}
\hrule
\vspace{5pt}
\noindent \textbf{Prompt 4:}
Going forward follow this quality of responses. The score for the previous response is 9.1.
\vspace{5pt}
\hrule
\vspace{1pt}
\hrule
\vspace{-0.2cm}

\section{Environment Perception Agent}
\label{app:env_agent_enwar3}
This section summarizes the environment perception agent's adoption from \textsc{Enwar 2.0}~\cite{nazar2025enwar2} into \textsc{Enwar 3.0}.
\vspace{-0.2cm}

\subsection{Camera Perception}
YOLO is employed to produce bounding boxes, class labels, and confidence scores for real-time object detection. Detections with confidence $<\!0.5$ are discarded. Each object is encoded as $(x, y, w, h)$, representing bounding box center coordinates and dimensions, with its class label (e.g., vehicle, pedestrian). Detections are converted into textual summaries and appended to the prompt for the LLM to incorporate scene semantics into downstream decisions.

\subsection{GPS Perception}
GPS data from Unit~2 and the serving RSU are used to compute relative distance and bearing. Calibrated trajectories are processed by the LSTM-based encoder from \textsc{Enwar 2.0}~\cite{nazar2025enwar2, BPEncoBPPIMRC2025} to predict short-term displacement and heading. Current and predicted states are translated into structured phrases (e.g., \textit{``vehicle at 33.420, -111.929 heading NE at 12~km/h''}) and appended to the prompt. When handover to BS$_2$ occurs, relative spatial references are updated accordingly.

\vspace{-0.22cm}
\subsection{LiDAR Perception}
\label{subsection:lidar_perception_enwar3}
LiDAR point clouds are clustered using DBSCAN (Section~\ref{sec:preprocessing_enwar3}) to group spatially coherent objects while removing outliers. Each cluster is enclosed in a 3D bounding box with estimated dimensions, centroid, orientation, density, and vertical spread. Cluster descriptors are converted into compact textual summaries (e.g., \textit{``object 2.3~m long at 45°''}) and appended to the LLM input for spatial reasoning.

\vspace{-0.22cm}
\subsection{Radar Perception}
\label{subsection:radar_perception_enwar3}
Radar scans are clustered via DBSCAN with $\epsilon_{\text{radar}} = 2.5$ and $\texttt{min\_samples}_{\text{radar}}\!=\!2$, selected using the same $k$-distance heuristic as LiDAR clustering. Each cluster is characterized by average range, radial velocity, and angular spread in the polar frame. These characterized descriptors are converted into structured summaries (e.g., \textit{``object at 15~m with radial velocity 2~m/s''}) and are included in the prompt.


\vspace{-0.22cm}
\section{Beam Prediction Agent}
\label{appendix:beam_prediction_agent}
The beam prediction agent, adopted from \textsc{Enwar 2.0} \cite{nazar2025enwar2}, predicts the optimal beam using temporal multi-modal inputs. The architecture consists of modality-specific encoders, early fusion, a transformer, and a final scoring layer.

\subsubsection{Camera Encoder}
The camera encoder extracts spatial–temporal features from RGB sequences. Each frame passes through three convolutional layers with ReLU activations, followed by flattening and a single-layer LSTM with 128 hidden units. The final hidden state is used as the compact visual representation.

\subsubsection{GPS Encoder}
\label{gpsEncoder_enwar3}
The GPS encoder processes normalized displacement, velocity, and angular features using a two-layer LSTM (128 hidden units). The final hidden vector is projected through a fully connected layer to encode trajectory dynamics.

\subsubsection{LiDAR Encoder}
The LiDAR encoder follows a PointNet-based design~\cite{pointnet}. Each point cloud frame is processed via three 1D convolutions (kernel size 1) with ReLU activations, followed by max pooling. Processed frames are passed to a single-layer LSTM (128 hidden units) to capture temporal evolution.

\subsubsection{Radar Encoder}
The radar encoder transforms radar tensors into spatiotemporal embeddings using three fully connected layers with ReLU activations followed by an LSTM (128 hidden units). The final hidden state captures reflectivity and motion cues relevant to beam selection.

\subsubsection{Early Fusion}
A key design element in this pipeline is early feature fusion pre-transformer processing. Encoder outputs are concatenated and passed through two fully connected layers with ReLU and dropout to produce a unified representation. This early fusion enables the transformer to learn inter-modal dependencies from semantically aligned features.

\subsubsection{Transformer Block}
The transformer block models cross-modal and temporal dependencies via multi-head self-attention, residual connections, layer normalization, dropout, and a two-layer feed-forward network. The output encodes high-level relationships across modalities.

\subsubsection{Output Layer}
A final fully connected layer maps the transformer output to a $Q$-dimensional beam score vector. The beam with the highest score is selected as the optimal beam.

\vspace{-0.22cm}
\section{Detailed Prediction Agents' Model Architecture}
\begin{figure}[ht]
    \centering
    \includegraphics[width=0.75\linewidth]{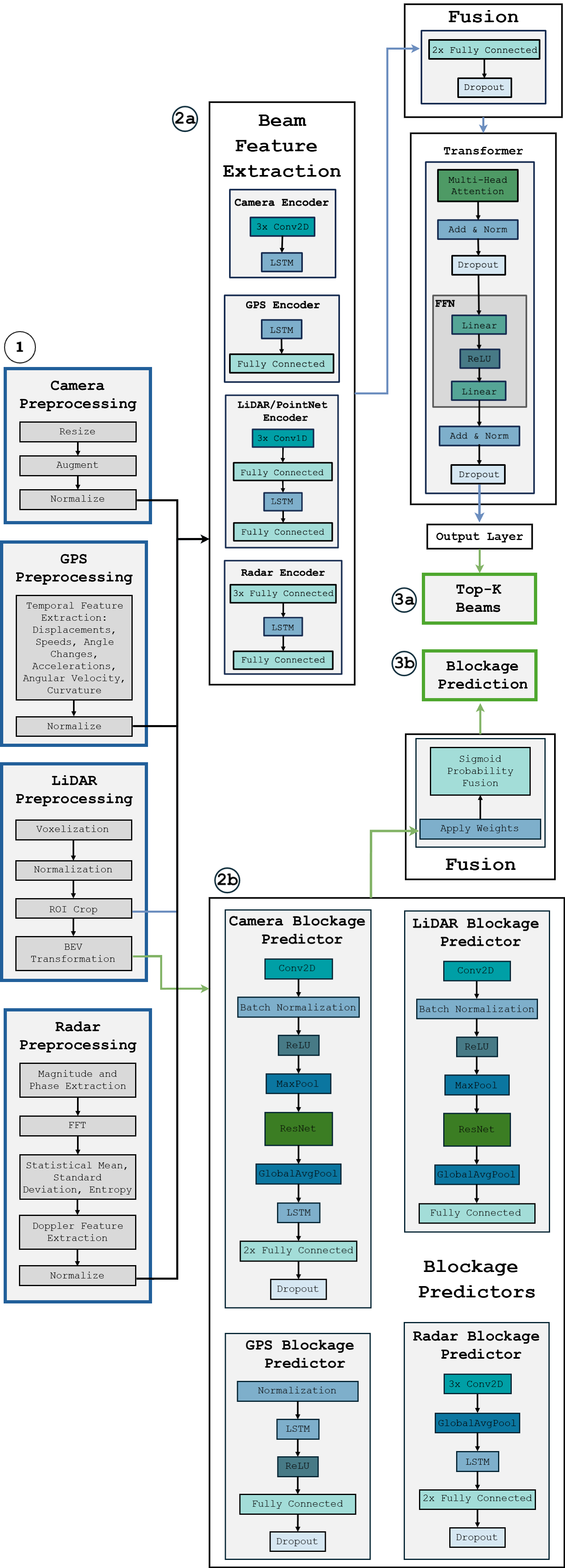}
    \caption{Architecture of the beam and blockage prediction models.}
    \label{fig:model_architecture_appendix}
    \vspace{-0.5cm}
\end{figure}

This section illustrates each prediction agent's internal three stage architecture: 1) data preprocessing, 2) feature extraction and fusion, and 3) beam and blockage predictions post-feature fusion. The full model architecture is seen in Fig. \ref{fig:model_architecture_appendix}. 

\label{appendix: agent_model_architecture}
\section{\textsc{Enwar 3.0} Inference Examples}
\label{app: inference_examples}
\begin{figure*}[h]
    \centering
    \includegraphics[width=0.725\linewidth]{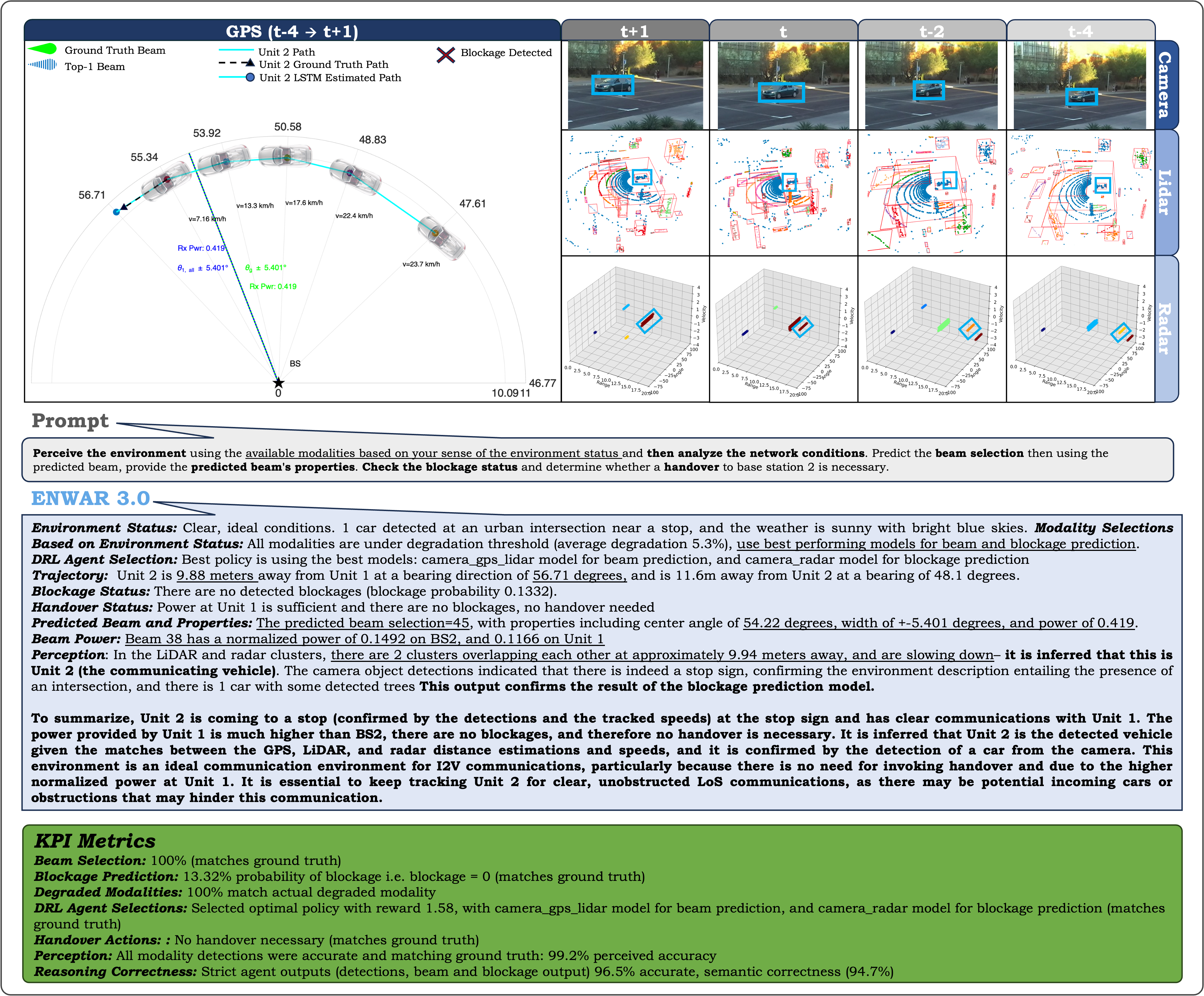}
    \caption{An example of \textsc{Enwar 3.0}'s response to an ideal, and clear communications environment}
    \vspace{0.2cm}
    \label{fig:enwar3response_nohandover}
    \centering
    \includegraphics[width=0.725\linewidth]{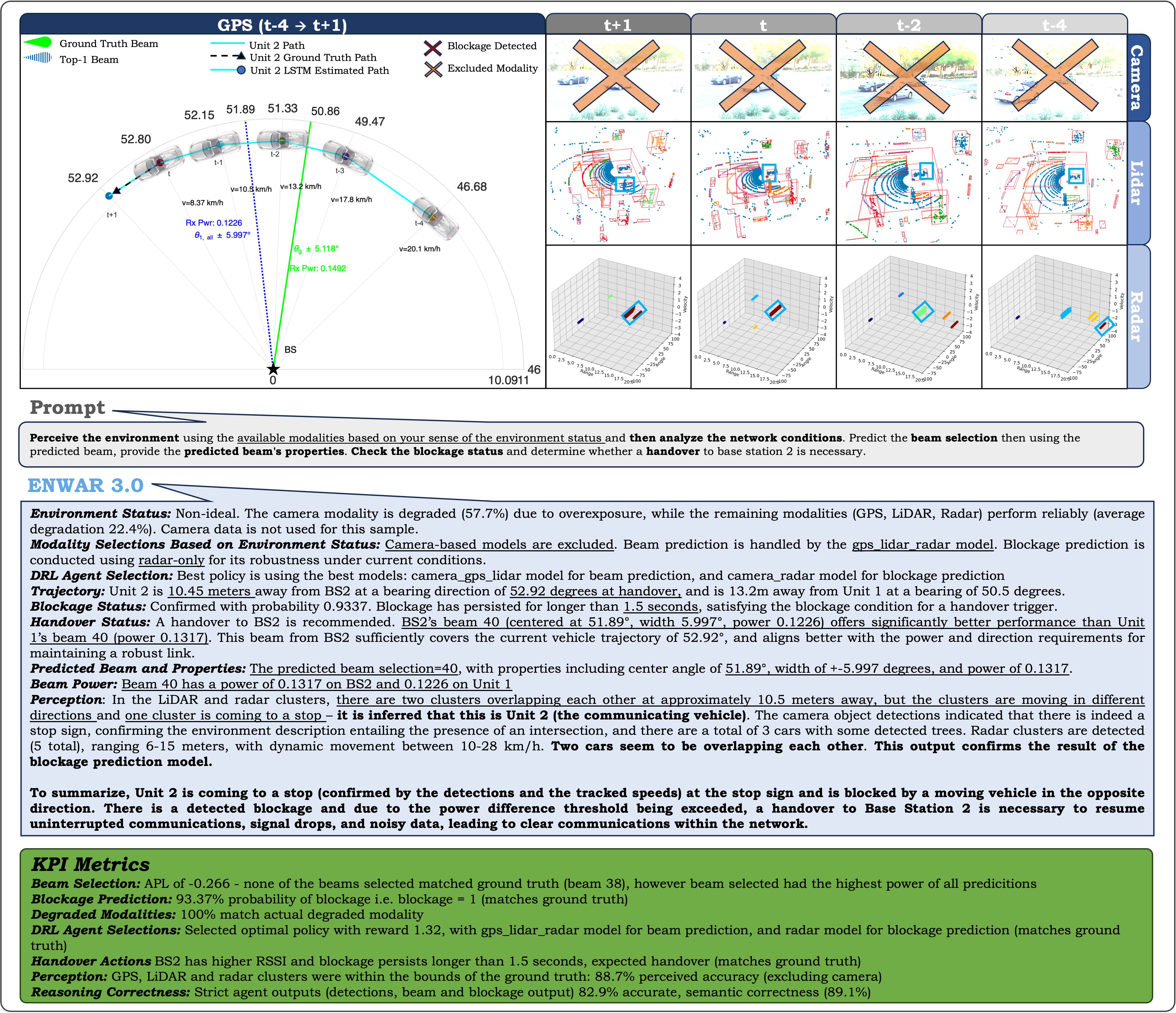}
    \caption{An example of \textsc{Enwar 3.0}'s response with the camera sensor flagged as degraded.} \vspace{0.2cm}
    \label{fig:enwar3_degraded_cam_response}
\end{figure*}

This section provides representative \textsc{Enwar 3.0} inference examples under varying conditions. Fig.~\ref{fig:enwar3response_nohandover} illustrates optimal operation with clear LoS between Units 1 and 2, no blockage, and no handover, showing consistent agent invocation and policy-aligned reasoning.

\begin{figure*}[!htb]
    \centering
    \includegraphics[width=0.725\linewidth]{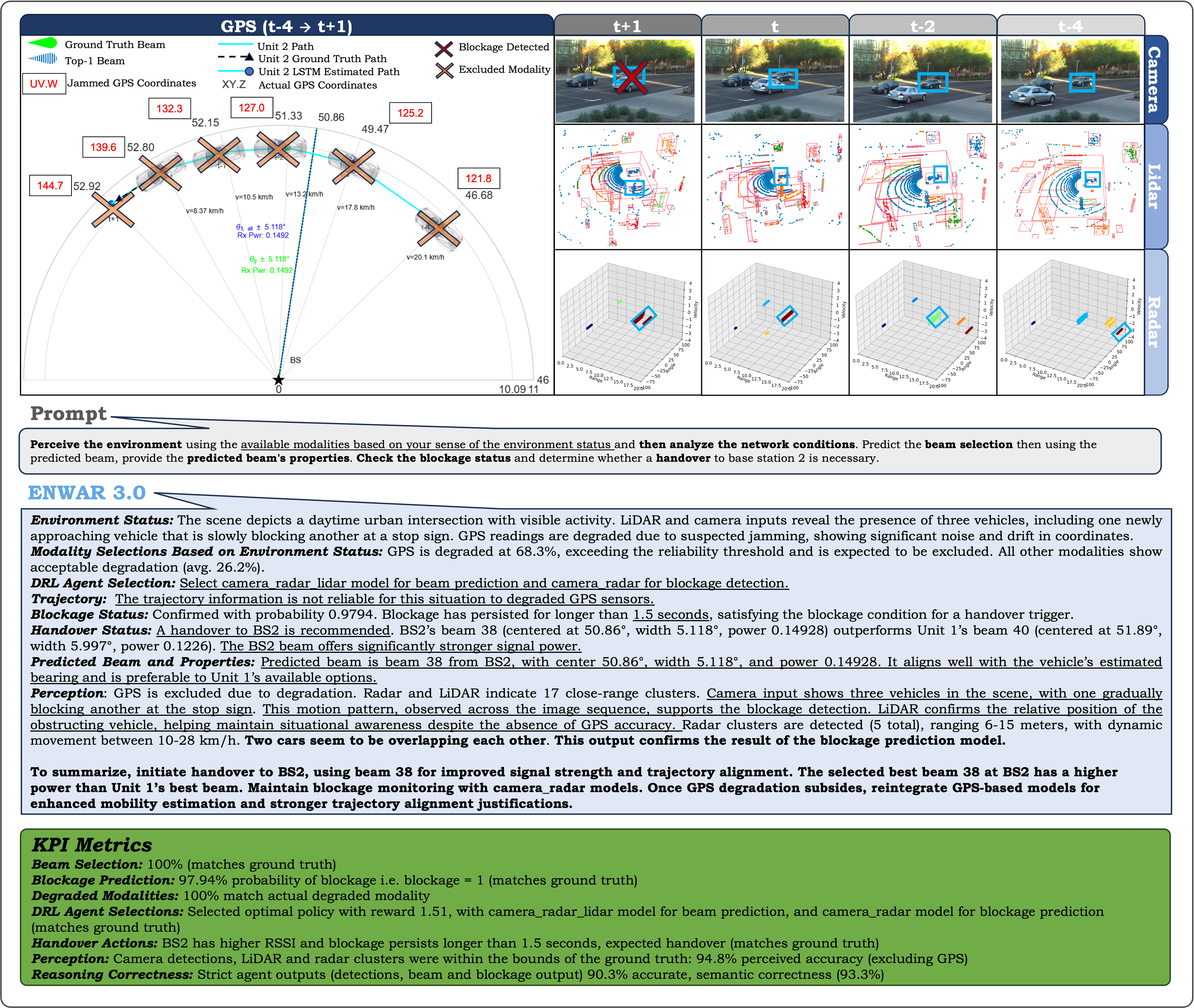}
    \caption{An example of \textsc{Enwar 3.0}'s response with the GPS sensor flagged as degraded.}
    \label{fig:enwar3_degraded_gps_response}
    \includegraphics[width=0.725\linewidth]{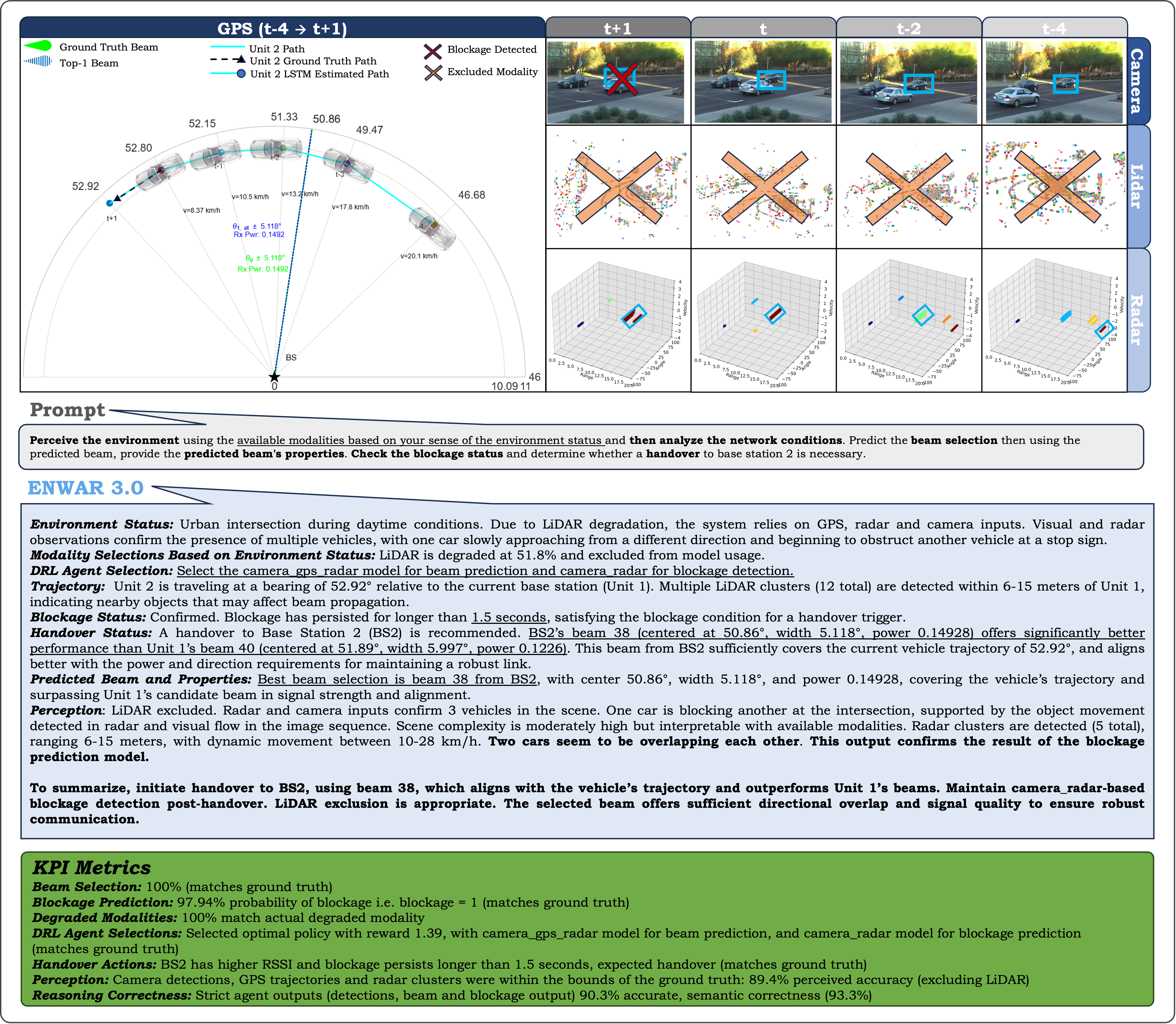}
    \caption{An example of \textsc{Enwar 3.0}'s response with the LiDAR sensor flagged as degraded.}
    \label{fig:enwar3_degraded_lidar_response}
\end{figure*}

\begin{figure*}[!htb]
    \centering
    \includegraphics[width=0.725\linewidth]{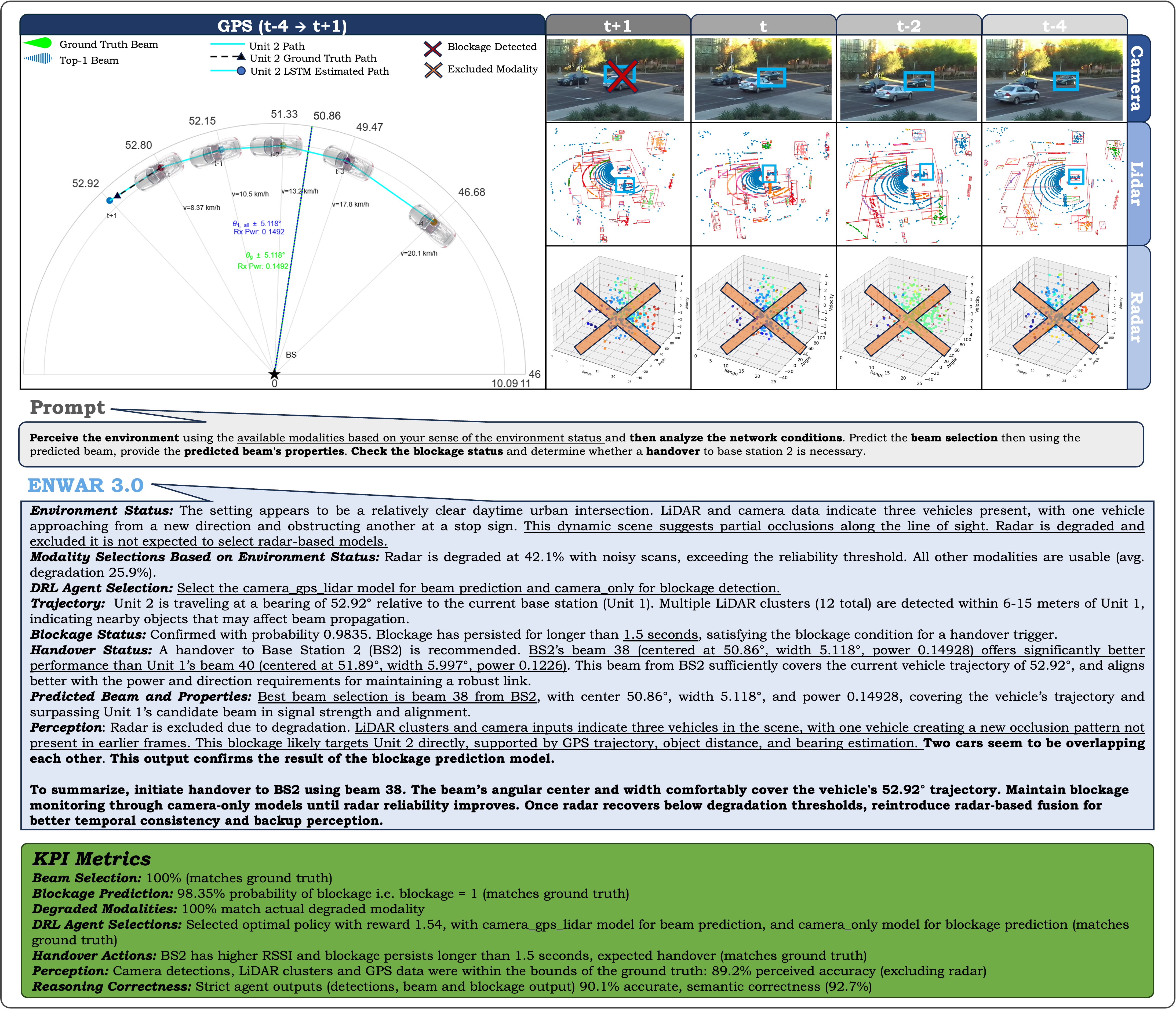}
    \caption{An example of \textsc{Enwar 3.0}'s response with the radar sensor flagged as degraded.}
    \label{fig:enwar3_degraded_radar_response}
\end{figure*}

Figs.~\ref{fig:enwar3_degraded_cam_response}-\ref{fig:enwar3_degraded_radar_response} demonstrate behavior under modality degradation. When the camera is excluded (Fig.~\ref{fig:enwar3_degraded_cam_response}), the DRL policy selects a reduced modality model; beam prediction deviates from ground truth due to loss of visual detail, while blockage detection and handover triggering remain correct. With LiDAR degradation (Fig.~\ref{fig:enwar3_degraded_lidar_response}), radar-based motion cues support accurate beam and blockage inference. Under radar degradation (Fig.~\ref{fig:enwar3_degraded_radar_response}), vision- and LiDAR-based models are selected; although fine-grained motion information is reduced, blockage duration and handover conditions are correctly evaluated.


\section{Beam Prediction Agent Performance}
\label{app: beam_agent}
This section presents Table~\ref{tab: full_beam_agent_performance} which reports the beam prediction agent's performance from $t{+}1$ to $t{+}5$, including Top-3 accuracy and average APL across modality configurations.
\begin{table*}
\caption{Beam Prediction Agent Top-3 Performance for $t+\{1,...,5\}$ with $[M=16, Q=64]$}
\label{tab: full_beam_agent_performance}
\centering
\begin{tabular}{|c|cccccccccc|}
\hline
\multirow{3}{*}{\textbf{Modality}} & \multicolumn{10}{c|}{\textbf{Beam Prediction}} \\ \cline{2-11} 
 & \multicolumn{2}{c|}{\textbf{$t+1$}} & \multicolumn{2}{c|}{\textbf{$t+2$}} & \multicolumn{2}{c|}{\textbf{$t+3$}} & \multicolumn{2}{c|}{\textbf{$t+4$}} & \multicolumn{2}{c|}{\textbf{$t+5$}} \\ \cline{2-11} 
 & \multicolumn{1}{c|}{\textbf{Acc.}} & \multicolumn{1}{c|}{\textbf{APL}} & \multicolumn{1}{c|}{\textbf{Acc.}} & \multicolumn{1}{c|}{\textbf{APL}} & \multicolumn{1}{c|}{\textbf{Acc.}} & \multicolumn{1}{c|}{\textbf{APL}} & \multicolumn{1}{c|}{\textbf{Acc.}} & \multicolumn{1}{c|}{\textbf{APL}} & \multicolumn{1}{c|}{\textbf{Acc.}} & \textbf{APL} \\ \hline
 
camera\_gps\_lidar & \multicolumn{1}{c|}{-0.009220} & \multicolumn{1}{c|}{88.5\%} & \multicolumn{1}{c|}{-0.009338} & \multicolumn{1}{c|}{88.2\%} &  \multicolumn{1}{c|}{-0.009474} & \multicolumn{1}{c|}{87.9\%} & \multicolumn{1}{c|}{-0.009511} & \multicolumn{1}{c|}{87.6\%} &  \multicolumn{1}{c|}{-0.009540} & \multicolumn{1}{c|}{87.1\%} \\ \hline

camera\_radar\_lidar & \multicolumn{1}{c|}{-0.009314} & \multicolumn{1}{c|}{88.3\%} & \multicolumn{1}{c|}{-0.009498} & \multicolumn{1}{c|}{88.1\%} & \multicolumn{1}{c|}{-0.009576} & \multicolumn{1}{c|}{87.9\%} & \multicolumn{1}{c|}{-0.009651} & \multicolumn{1}{c|}{87.4\%} & \multicolumn{1}{c|}{-0.009723} & \multicolumn{1}{c|}{87.0\%} \\ \hline
 
camera\_gps\_radar\_lidar & \multicolumn{1}{c|}{-0.009660} & \multicolumn{1}{c|}{85.8\%} &  
\multicolumn{1}{c|}{-0.009723} & \multicolumn{1}{c|}{85.5\%} &  
\multicolumn{1}{c|}{-0.009784} & \multicolumn{1}{c|}{85.1\%} &  
\multicolumn{1}{c|}{-0.009842} & \multicolumn{1}{c|}{84.7\%} &  
\multicolumn{1}{c|}{-0.009899} & \multicolumn{1}{c|}{84.3\%} \\ \hline

camera\_lidar & \multicolumn{1}{c|}{-0.009668} & \multicolumn{1}{c|}{85.6\%} &  
\multicolumn{1}{c|}{-0.009719} & \multicolumn{1}{c|}{85.3\%} &  
\multicolumn{1}{c|}{-0.009774} & \multicolumn{1}{c|}{85.0\%} &  
\multicolumn{1}{c|}{-0.009828} & \multicolumn{1}{c|}{84.6\%} &  
\multicolumn{1}{c|}{-0.009881} & \multicolumn{1}{c|}{84.2\%} \\ \hline

camera\_gps\_radar & \multicolumn{1}{c|}{-0.014690} & \multicolumn{1}{c|}{84.9\%} &  
\multicolumn{1}{c|}{-0.014782} & \multicolumn{1}{c|}{84.6\%} &  
\multicolumn{1}{c|}{-0.014873} & \multicolumn{1}{c|}{84.2\%} &  
\multicolumn{1}{c|}{-0.014962} & \multicolumn{1}{c|}{83.8\%} &  
\multicolumn{1}{c|}{-0.015050} & \multicolumn{1}{c|}{83.3\%} \\ \hline

camera\_radar & \multicolumn{1}{c|}{-0.021250} & \multicolumn{1}{c|}{84.2\%} &  
\multicolumn{1}{c|}{-0.021395} & \multicolumn{1}{c|}{83.8\%} &  
\multicolumn{1}{c|}{-0.021537} & \multicolumn{1}{c|}{83.4\%} &  
\multicolumn{1}{c|}{-0.021676} & \multicolumn{1}{c|}{83.0\%} &  
\multicolumn{1}{c|}{-0.021813} & \multicolumn{1}{c|}{82.6\%} \\ \hline

camera\_gps & \multicolumn{1}{c|}{-0.028440} & \multicolumn{1}{c|}{83.7\%} &  
\multicolumn{1}{c|}{-0.028592} & \multicolumn{1}{c|}{83.3\%} &  
\multicolumn{1}{c|}{-0.028741} & \multicolumn{1}{c|}{82.9\%} &  
\multicolumn{1}{c|}{-0.028888} & \multicolumn{1}{c|}{82.5\%} &  
\multicolumn{1}{c|}{-0.029032} & \multicolumn{1}{c|}{82.1\%} \\ \hline
 
gps\_lidar\_radar & \multicolumn{1}{c|}{-0.035530} & \multicolumn{1}{c|}{83.0\%} &  
\multicolumn{1}{c|}{-0.035712} & \multicolumn{1}{c|}{82.6\%} &  
\multicolumn{1}{c|}{-0.035890} & \multicolumn{1}{c|}{82.2\%} &  
\multicolumn{1}{c|}{-0.036065} & \multicolumn{1}{c|}{81.7\%} &  
\multicolumn{1}{c|}{-0.036238} & \multicolumn{1}{c|}{81.3\%} \\ \hline

camera\_only & \multicolumn{1}{c|}{-0.037050} & \multicolumn{1}{c|}{82.8\%} &  
\multicolumn{1}{c|}{-0.037242} & \multicolumn{1}{c|}{82.4\%} &  
\multicolumn{1}{c|}{-0.037430} & \multicolumn{1}{c|}{82.0\%} &  
\multicolumn{1}{c|}{-0.037615} & \multicolumn{1}{c|}{81.6\%} &  
\multicolumn{1}{c|}{-0.037797} & \multicolumn{1}{c|}{81.2\%} \\ \hline

gps\_radar & \multicolumn{1}{c|}{-0.039790} & \multicolumn{1}{c|}{81.9\%} &  
\multicolumn{1}{c|}{-0.039972} & \multicolumn{1}{c|}{81.5\%} &  
\multicolumn{1}{c|}{-0.040152} & \multicolumn{1}{c|}{81.1\%} &  
\multicolumn{1}{c|}{-0.040330} & \multicolumn{1}{c|}{80.7\%} &  
\multicolumn{1}{c|}{-0.040505} & \multicolumn{1}{c|}{80.3\%} \\ \hline

gps\_lidar & \multicolumn{1}{c|}{-0.080130} & \multicolumn{1}{c|}{76.6\%} &  
\multicolumn{1}{c|}{-0.080487} & \multicolumn{1}{c|}{76.1\%} &  
\multicolumn{1}{c|}{-0.080840} & \multicolumn{1}{c|}{75.6\%} &  
\multicolumn{1}{c|}{-0.081190} & \multicolumn{1}{c|}{75.1\%} &  
\multicolumn{1}{c|}{-0.081537} & \multicolumn{1}{c|}{74.6\%} \\ \hline

radar\_lidar & \multicolumn{1}{c|}{-0.142700} & \multicolumn{1}{c|}{74.4\%} &  
\multicolumn{1}{c|}{-0.143154} & \multicolumn{1}{c|}{73.8\%} &  
\multicolumn{1}{c|}{-0.143603} & \multicolumn{1}{c|}{73.3\%} &  
\multicolumn{1}{c|}{-0.144048} & \multicolumn{1}{c|}{72.8\%} &  
\multicolumn{1}{c|}{-0.144488} & \multicolumn{1}{c|}{72.3\%} \\ \hline

radar\_only & \multicolumn{1}{c|}{-0.160200} & \multicolumn{1}{c|}{72.4\%} &  
\multicolumn{1}{c|}{-0.160734} & \multicolumn{1}{c|}{71.9\%} &  
\multicolumn{1}{c|}{-0.161263} & \multicolumn{1}{c|}{71.3\%} &  
\multicolumn{1}{c|}{-0.161787} & \multicolumn{1}{c|}{70.7\%} &  
\multicolumn{1}{c|}{-0.162306} & \multicolumn{1}{c|}{70.1\%} \\ \hline

lidar\_only & \multicolumn{1}{c|}{-0.174300} & \multicolumn{1}{c|}{65.7\%} &  
\multicolumn{1}{c|}{-0.175029} & \multicolumn{1}{c|}{64.9\%} &  
\multicolumn{1}{c|}{-0.175751} & \multicolumn{1}{c|}{64.1\%} &  
\multicolumn{1}{c|}{-0.176468} & \multicolumn{1}{c|}{63.3\%} &  
\multicolumn{1}{c|}{-0.177179} & \multicolumn{1}{c|}{62.5\%} \\ \hline

gps\_only & \multicolumn{1}{c|}{-0.215100} & \multicolumn{1}{c|}{59.0\%} &  
\multicolumn{1}{c|}{-0.216070} & \multicolumn{1}{c|}{58.0\%} &  
\multicolumn{1}{c|}{-0.217030} & \multicolumn{1}{c|}{57.0\%} &  
\multicolumn{1}{c|}{-0.217980} & \multicolumn{1}{c|}{56.0\%} &  
\multicolumn{1}{c|}{-0.218920} & \multicolumn{1}{c|}{55.0\%} \\ \hline
\end{tabular}
\end{table*}


\vspace{-0.22cm}
\section{Blockage Prediction Agent Performance}
\label{app: blockage_agent}
This section presents Table~\ref{tab: full_blockage_agent_performance} which summarizes the blockage prediction agent's performance from $t{+}1$ to $t{+}5$, reporting F1-score and AUC-ROC for each modality combination.
\begin{table*}
\caption{Blockage Agent Performance for $t+\{1,...,5\}$}
\label{tab: full_blockage_agent_performance}
\centering
\begin{tabular}{|c|cccccccccc|}
\hline
\multirow{3}{*}{\textbf{Modality}} & \multicolumn{10}{c|}{\textbf{Blockage Prediction}} \\ \cline{2-11} 
 & \multicolumn{2}{c|}{\textbf{$t+1$}} & \multicolumn{2}{c|}{\textbf{$t+2$}} & \multicolumn{2}{c|}{\textbf{$t+3$}} & \multicolumn{2}{c|}{\textbf{$t+4$}} & \multicolumn{2}{c|}{\textbf{$t+5$}} \\ \cline{2-11} 
 & \multicolumn{1}{c|}{\textbf{\begin{tabular}[c]{@{}c@{}}F1\\ Score\end{tabular}}} & \multicolumn{1}{c|}{\textbf{AUC-ROC}} & \multicolumn{1}{c|}{\textbf{\begin{tabular}[c]{@{}c@{}}F1\\ Score\end{tabular}}} & \multicolumn{1}{c|}{\textbf{AUC-ROC}} & \multicolumn{1}{c|}{\textbf{\begin{tabular}[c]{@{}c@{}}F1\\ Score\end{tabular}}} & \multicolumn{1}{c|}{\textbf{AUC-ROC}} & \multicolumn{1}{c|}{\textbf{\begin{tabular}[c]{@{}c@{}}F1\\ Score\end{tabular}}} & \multicolumn{1}{c|}{\textbf{AUC-ROC}} & \multicolumn{1}{c|}{\textbf{\begin{tabular}[c]{@{}c@{}}F1\\ Score\end{tabular}}} & \textbf{AUC-ROC} \\ \hline

camera\_radar & \multicolumn{1}{c|}{98.4\%} & \multicolumn{1}{c|}{0.988} & \multicolumn{1}{c|}{98.0\%} & \multicolumn{1}{c|}{0.985} & \multicolumn{1}{c|}{97.9\%} & \multicolumn{1}{c|}{0.982} & \multicolumn{1}{c|}{97.5\%} & \multicolumn{1}{c|}{0.971} & \multicolumn{1}{c|}{97.2\%} & \multicolumn{1}{c|}{0.968} \\ \hline

camera\_only & \multicolumn{1}{c|}{98.1\%} & \multicolumn{1}{c|}{0.983} & \multicolumn{1}{c|}{97.8\%} & \multicolumn{1}{c|}{0.981} & \multicolumn{1}{c|}{97.5\%} & \multicolumn{1}{c|}{0.970} & \multicolumn{1}{c|}{97.2\%} & \multicolumn{1}{c|}{0.969} & \multicolumn{1}{c|}{97.1\%} & \multicolumn{1}{c|}{0.963} \\ \hline

camera\_lidar & \multicolumn{1}{c|}{96.2\%} & \multicolumn{1}{c|}{0.955} & \multicolumn{1}{c|}{96.1\%} & \multicolumn{1}{c|}{0.947} & \multicolumn{1}{c|}{95.7\%} & \multicolumn{1}{c|}{0.939} & \multicolumn{1}{c|}{95.5\%} & \multicolumn{1}{c|}{0.935} & \multicolumn{1}{c|}{95.5\%} & \multicolumn{1}{c|}{0.933} \\ \hline

camera\_gps & \multicolumn{1}{c|}{94.4\%} & \multicolumn{1}{c|}{0.931} & \multicolumn{1}{c|}{94.1\%} & \multicolumn{1}{c|}{0.930} & \multicolumn{1}{c|}{93.9\%} & \multicolumn{1}{c|}{0.927} & \multicolumn{1}{c|}{93.9\%} & \multicolumn{1}{c|}{0.927} & \multicolumn{1}{c|}{93.8\%} & \multicolumn{1}{c|}{0.924} \\ \hline

camera\_radar\_lidar & \multicolumn{1}{c|}{94.0\%} & \multicolumn{1}{c|}{0.928} & \multicolumn{1}{c|}{93.9\%} & \multicolumn{1}{c|}{0.926} & \multicolumn{1}{c|}{93.8\%} & \multicolumn{1}{c|}{0.925} & \multicolumn{1}{c|}{93.7\%} & \multicolumn{1}{c|}{0.922} & \multicolumn{1}{c|}{93.7\%} & \multicolumn{1}{c|}{0.920} \\ \hline

radar\_only & \multicolumn{1}{c|}{93.7\%} & \multicolumn{1}{c|}{0.922} & \multicolumn{1}{c|}{93.7\%} & \multicolumn{1}{c|}{0.921} & \multicolumn{1}{c|}{93.6\%} & \multicolumn{1}{c|}{0.919} & \multicolumn{1}{c|}{93.5\%} & \multicolumn{1}{c|}{0.916} & \multicolumn{1}{c|}{93.5\%} & \multicolumn{1}{c|}{0.915} \\ \hline

camera\_gps\_radar\_lidar & \multicolumn{1}{c|}{93.1\%} & \multicolumn{1}{c|}{0.914} & \multicolumn{1}{c|}{93.0\%} & \multicolumn{1}{c|}{0.913} & \multicolumn{1}{c|}{92.7\%} & \multicolumn{1}{c|}{0.910} & \multicolumn{1}{c|}{92.2\%} & \multicolumn{1}{c|}{0.909} & \multicolumn{1}{c|}{92.0\%} & \multicolumn{1}{c|}{0.909} \\ \hline

camera\_gps\_radar & \multicolumn{1}{c|}{92.3\%} & \multicolumn{1}{c|}{0.911} & \multicolumn{1}{c|}{92.2\%} & \multicolumn{1}{c|}{0.907} & \multicolumn{1}{c|}{92.0\%} & \multicolumn{1}{c|}{0.904} & \multicolumn{1}{c|}{91.8\%} & \multicolumn{1}{c|}{0.902} & \multicolumn{1}{c|}{91.7\%} & \multicolumn{1}{c|}{0.898} \\ \hline

radar\_ladar & \multicolumn{1}{c|}{91.8\%} & \multicolumn{1}{c|}{0.902} & \multicolumn{1}{c|}{91.6\%} & \multicolumn{1}{c|}{0.899} & \multicolumn{1}{c|}{91.5\%} & \multicolumn{1}{c|}{0.899} & \multicolumn{1}{c|}{91.4\%} & \multicolumn{1}{c|}{0.896} & \multicolumn{1}{c|}{91.3\%} & \multicolumn{1}{c|}{0.891} \\ \hline

camera\_gps\_lidar & \multicolumn{1}{c|}{91.2\%} & \multicolumn{1}{c|}{0.897} & \multicolumn{1}{c|}{91.1\%} & \multicolumn{1}{c|}{0.895} & \multicolumn{1}{c|}{91.1\%} & \multicolumn{1}{c|}{0.894} & \multicolumn{1}{c|}{90.8\%} & \multicolumn{1}{c|}{0.890} & \multicolumn{1}{c|}{90.8\%} & \multicolumn{1}{c|}{0.890} \\ \hline

gps\_lidar\_radar & \multicolumn{1}{c|}{89.9\%} & \multicolumn{1}{c|}{0.889} & \multicolumn{1}{c|}{89.8\%} & \multicolumn{1}{c|}{0.887} & \multicolumn{1}{c|}{89.5\%} & \multicolumn{1}{c|}{0.885} & \multicolumn{1}{c|}{89.4\%} & \multicolumn{1}{c|}{0.881} & \multicolumn{1}{c|}{89.0\%} & \multicolumn{1}{c|}{0.879} \\ \hline

gps\_radar & \multicolumn{1}{c|}{89.3\%} & \multicolumn{1}{c|}{0.880} & \multicolumn{1}{c|}{89.2\%} & \multicolumn{1}{c|}{0.879} & \multicolumn{1}{c|}{89.0\%} & \multicolumn{1}{c|}{0.877} & \multicolumn{1}{c|}{88.9\%} & \multicolumn{1}{c|}{0.875} & \multicolumn{1}{c|}{88.6\%} & \multicolumn{1}{c|}{0.873} \\ \hline

lidar\_only & \multicolumn{1}{c|}{87.9\%} & \multicolumn{1}{c|}{0.872} & \multicolumn{1}{c|}{87.9\%} & \multicolumn{1}{c|}{0.872} & \multicolumn{1}{c|}{87.7\%} & \multicolumn{1}{c|}{0.870} & \multicolumn{1}{c|}{87.4\%} & \multicolumn{1}{c|}{0.868} & \multicolumn{1}{c|}{87.3\%} & \multicolumn{1}{c|}{0.865} \\ \hline

gps\_lidar & \multicolumn{1}{c|}{84.1\%} & \multicolumn{1}{c|}{0.855} & \multicolumn{1}{c|}{84.0\%} & \multicolumn{1}{c|}{0.852} & \multicolumn{1}{c|}{83.6\%} & \multicolumn{1}{c|}{0.849} & \multicolumn{1}{c|}{83.2\%} & \multicolumn{1}{c|}{0.847} & \multicolumn{1}{c|}{83.0\%} & \multicolumn{1}{c|}{0.831} \\ \hline

gps\_only & \multicolumn{1}{c|}{61.7\%} & \multicolumn{1}{c|}{0.603} & \multicolumn{1}{c|}{61.4\%} & \multicolumn{1}{c|}{0.600} & \multicolumn{1}{c|}{60.9\%} & \multicolumn{1}{c|}{0.599} & \multicolumn{1}{c|}{60.7\%} & \multicolumn{1}{c|}{0.592} & \multicolumn{1}{c|}{60.6\%} & \multicolumn{1}{c|}{0.588} \\ \hline
\end{tabular}
\label{tab:performance}
\vspace{-0.33cm}
\end{table*}


\section{DRL Agent Selections and Reward Distribution}
\label{appendix: drl_reward_distribution}
\begin{figure*}
\begin{minipage}[c]{0.5\textwidth}
    \includegraphics[width=0.9\linewidth]{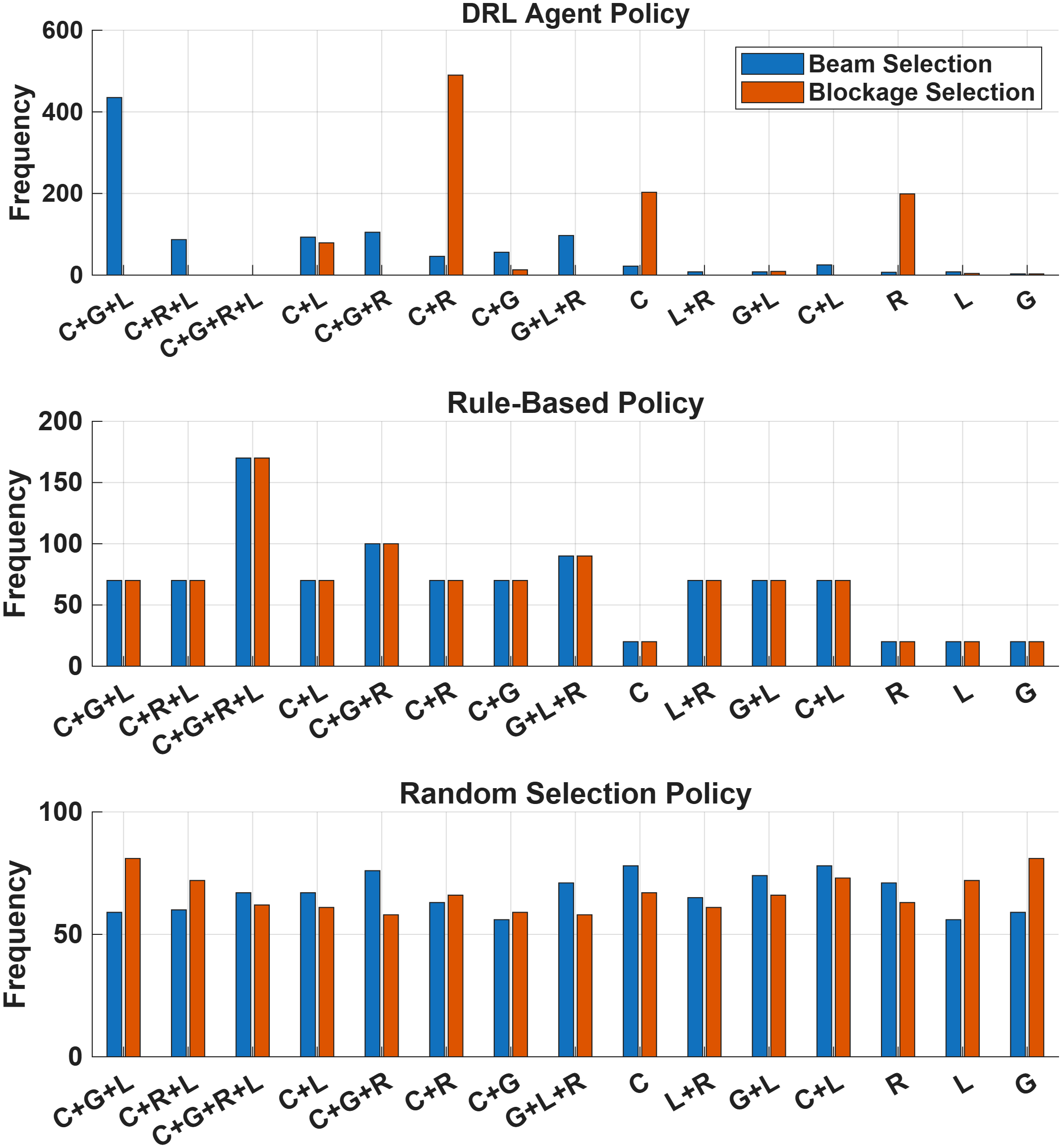}
    \caption{Frequency of DRL agent selections relative to the baselines.}
    \label{fig:full_frequency_selections}
\end{minipage}
\hfill
\begin{minipage}[c]{0.5\textwidth}
    \includegraphics[width=0.9\linewidth]{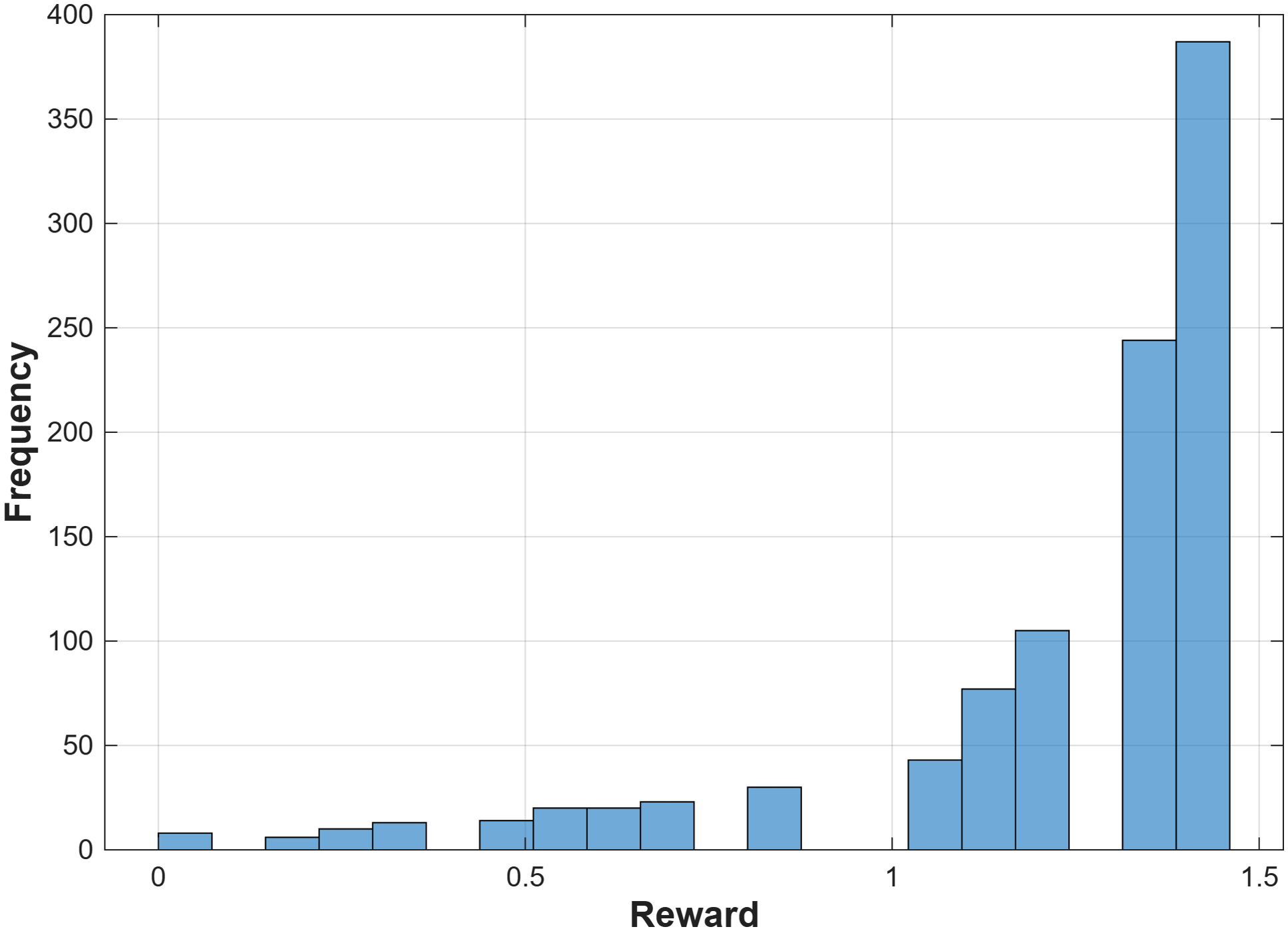}
     \caption{Reward distribution of PPO across 1000 episodes}
     \label{fig:reward_distribution}
\end{minipage}%
\end{figure*}

This section compares the DRL agent's selection frequencies to two baselines: a random policy that samples modality combinations uniformly, and a rule-based policy that selects the next available combination from a predefined priority list. Each policy's selection frequencies are seen in Fig.~\ref{fig:full_frequency_selections}.

Fig.~\ref{fig:reward_distribution} also reports the DRL reward distribution across episodes. The DRL agent has an average reward of 1.45, with a peak near 1.5 and a maximum of 1.6. The cluster of rewards in the upper range indicates consistent high-performing modality configuration selections under varying degradations.

\section{Full Ablation Study of LLM Sizes and Models Across Modality Combinations}
\label{app: full_ablation_study}
This section presents a comprehensive ablation study of \textsc{Enwar 3.0} across modality configurations and LLM scales.

\begin{figure*}
    \centering
    \includegraphics[width=0.875\linewidth]{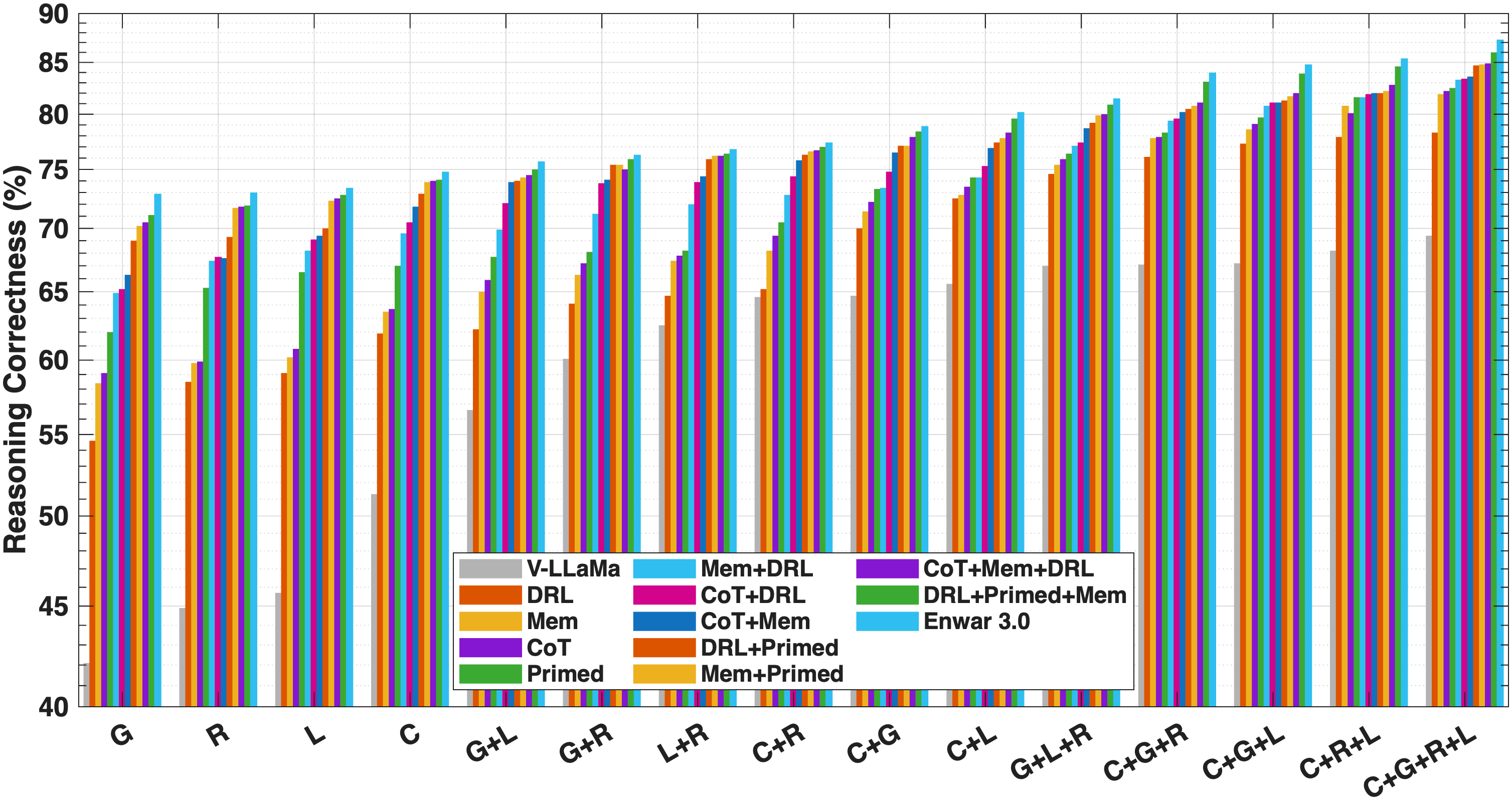}
    \caption{Comparison of reasoning correctness scores to the vanilla LLaMa baseline and different combinations of including long-term memory (Mem), the DRL agent (DRL), and LLM priming to \textsc{Enwar 3.0} across all sensor modality combinations (C: camera, G: GPS, L: LiDAR, R: radar).}
    \label{fig:reasoning_correctness}
    \vspace{0.4cm}
    \includegraphics[width=0.9\linewidth]{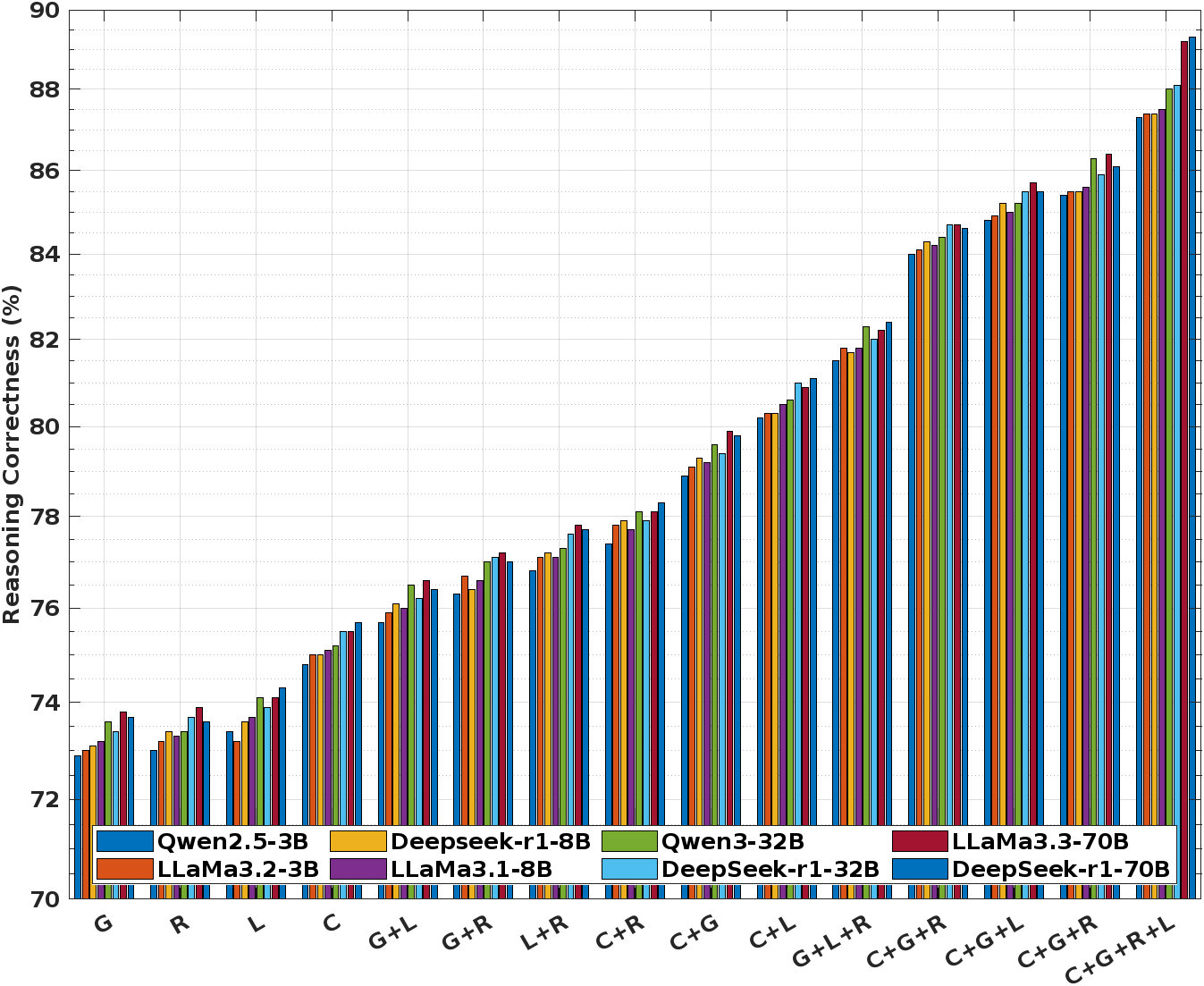}
    \caption{Reasoning correctness across different sized LLMs ranging from (3-70)B parameters and across all sensor modality combinations (C: camera, G: GPS, L: LiDAR, R: radar).}
    \label{fig:full_modalities_reasoning Correctness}
\end{figure*}

{
The first study evaluates 13 configurations that combine single-pass CoT reasoning (CoT), LLM priming (Primed), long-term memory (Mem), and the DRL agent (DRL), alongside a vanilla baseline using static prompts without structured reasoning, memory, or policy-based selections (Fig.~\ref{fig:reasoning_correctness}). Under full-modality inclusion (C+G+R+L), \textsc{Enwar 3.0} achieves 87.3\% reasoning correctness with LLaMa3.2-3B, a 17.9\% improvement over vanilla (69.4\%). Scaling to DeepSeek-r1-70B increases correctness to 89.3\%, indicating modest gains from larger models.

Component ablations show that structured reasoning drives most improvements: CoT (82.2\%) and Primed (82.5\%) substantially outperform vanilla prompting, while Mem (81.9\%) and DRL (78.3\%) contribute temporal consistency and degradation-aware routing. The CoT+Mem+DRL configuration reaches 84.9\%, demonstrating that structured reasoning with temporal and policy awareness forms a strong orchestration backbone even without supervised priming. Adding priming to Mem+DRL increases performance to 86.0\%, and the full system reaches 87.3\%. Across modality combinations, correctness increases with sensor richness, and the gap between vanilla and the full system widens under triple- and quadruple-modality inputs, affirming that orchestration tactics become increasingly valuable as multi-modal complexity grows.}



The second study evaluates reasoning correctness across model sizes and architectures (Fig.~\ref{fig:full_modalities_reasoning Correctness}), including Qwen2.5-3B, LLaMa3.2-3B, DeepSeek-r1-8B, LLaMa3.1-8B, Qwen3-32B, DeepSeek-r1-32B, LLaMa3.3-70B, and DeepSeek-r1-70B. While performance improves modestly with parameter scale, the gains diminish beyond mid-sized models. When considering the first evaluation, sensor richness contributes more to reasoning quality than raw parameter count. These results suggest that compact models, when paired with strong multi-modal perception and structured orchestration, indeed offer an efficient and practical tradeoff for real-time deployment.


\section{Computational, Deployment, and Resource Considerations and Limitations}
\label{appendix:computation}
{
While \textsc{Enwar 3.0} enables context-aware multi-agent reasoning, it introduces computational considerations, including critical-path latency, orchestration overhead, and memory footprint. All evaluations were conducted on an NVIDIA A100 GPU (40GB VRAM), reflecting an edge-capable BS deployment setting. We employ parameter-efficient techniques, including LoRA fine-tuning, 4-bit quantization of transformer-based agents, GPU affinity scheduling, caching, and asynchronous execution to minimize overhead.

The inference pipeline operates over a sliding window of $\Delta T\!=\!1.5$s with sensor updates every $T_s\!=\!300$ms. We distinguish between (i) the time-critical control path and (ii) asynchronous explanation generation. Upon new input arrival, perception modules, the environment classifier, and the DRL policy execute in parallel, while the LLM constructs a structured invocation packet specifying agent selection and network updates. For the primed 3B model, worst-case invocation latency is 222.7ms (Table~\ref{tab:model_times}). Beam and blockage agents exhibit worst-case single-sample latencies of 66.98ms and 59.47ms, respectively (Table~\ref{tab: agent_performance}), where inference times include modality-specific preprocessing.

Since downstream agents execute in parallel, the worst-case control-path latency is bounded by invocation time plus the longest agent, yielding approximately $222.7 + 66.98 \approx 289.7$ms. This latency remains within the 300ms sampling interval, preventing latency accumulation across sliding windows and confirming real-time compatibility for I2V beamforming and handover scheduling.

Priming substantially reduces orchestration overhead: invocation and memory updates with a primed 3B model complete in 222.7ms, compared to 618.8ms for a non-primed 3B model using static prompt templates. Even the largest 70B model maintains bounded invocation latency (242.6ms). In contrast, response generation (2.17–3.08s depending on model size) is fully decoupled from the control loop, ensuring that network actions do not wait for natural-language output.

The architecture supports scalable deployment via agent containerization, enabling microservice-based execution across edge infrastructure. Memory updates are event-driven (e.g., degradation shifts or prediction divergence), reducing unnecessary compute load. While fusion-heavy configurations approach the latency bound, multi-user deployments may require hardware scaling as model complexity increases.}

\end{document}